\documentclass[conference,compsoc,10pt]{IEEEtran}
\usepackage{enumerate,paralist}
\usepackage{comment}
\usepackage{parskip}
\usepackage{multirow}
\usepackage{graphicx}
\usepackage{color}
\usepackage[T1]{fontenc}

\usepackage{tikz,pgffor}
\usetikzlibrary{arrows}
\usetikzlibrary{shapes}
\usetikzlibrary{calc}
\usetikzlibrary{automata}
\usetikzlibrary{positioning}

\tikzstyle{proglabel}=[shape=circle,draw,inner sep=0pt,minimum size=5mm]
\tikzstyle{tran}=[draw,->,>=stealth, rounded corners]

\usepackage{amsmath}
\usepackage{amssymb}

\usepackage{stmaryrd}
\usepackage{url}                  

\ifCLASSOPTIONcompsoc
\usepackage[nocompress]{cite}
\else
\usepackage{cite}
\fi

\usepackage{listings}
\lstdefinelanguage{prog}
{
morekeywords={prob, if, then, else, fi, while, do, od, true, false, and, or, skip},
sensitive = false
}

\newtheorem{lemma}{Lemma}

\newtheorem{proposition}{Proposition}
\newtheorem{definition}{Definition}
\newtheorem{corollary}{Corollary}
\newtheorem{theorem}{Theorem}
\newtheorem{example}{Example}
\newtheorem{remark}{Remark}

\newcommand{\Rset}{\mathbb{R}}
\newcommand{\Nset}{\mathbb{N}}
\newcommand{\Zset}{\mathbb{Z}}

\newcommand{\fnames}{\mathit{F}}
\newcommand{\pvars}[1]{V^{\mathsf{#1}}_\mathrm{p}}
\newcommand{\rvars}{V_{\mathrm{r}}}
\newcommand{\fn}[1]{\mathsf{#1}}

\newcommand{\loc}{\ell}
\newcommand{\locs}[1]{\mathit{L}^{\mathsf{#1}}}
\newcommand{\clocs}[1]{\mathit{L}_{\mathrm{b}}^{\mathsf{#1}}}
\newcommand{\alocs}[1]{\mathit{L}_{\mathrm{a}}^{\mathsf{#1}}}
\newcommand{\flocs}[1]{\mathit{L}_{\mathrm{c}}^{\mathsf{#1}}}
\newcommand{\dlocs}[1]{\mathit{L}_{\mathrm{d}}^{\mathsf{#1}}}
\newcommand{\transitions}[1]{{\rightarrow}_{#1}}

\newcommand{\assgn}[2]{\left[#1/#2\right]}
\newcommand{\lin}[1]{\loc_\mathrm{in}^{#1}}
\newcommand{\lout}[1]{\loc_\mathrm{out}^{#1}}
\newcommand{\val}[1]{\mbox{\sl Val}_{#1}}
\newcommand{\samples}{\val{}^\mathrm{r}}
\newcommand{\id}{\mbox{\sl id}}

\newcommand{\dpd}{q}
\newcommand{\supp}[1]{{\mathrm{supp}}{\left(#1\right)}}

\newcommand{\expvt}{\overline{T}}
\newcommand{\tertime}{T}

\newcommand{\probm}{\mathbb{P}}
\newcommand{\expv}{\mathbb{E}}
\newcommand{\condexpv}[2]{{\expv}{\left({#1}{\mid}{#2}\right)}}

\newcommand{\MDPkernel}{\mathbf{P}}
\newcommand{\MDPstates}{S}
\newcommand{\MDPactions}{\mbox{\sl Act}}
\newcommand{\MDP}[1]{\mathcal{M}_{#1}}
\newcommand{\infruns}{\Lambda}

\newcommand{\last}[1]{{#1}{\downarrow}}
\newcommand{\stopping}[1]{Z_{#1}}

\newcommand{\rvlen}{\mathsf{len}}
\newcommand{\rvfn}{\mathsf{fn}}
\newcommand{\rvlb}{\mathsf{lb}}
\newcommand{\rvval}{\mathsf{val}}
\newcommand{\rvsam}{\mathsf{samp}}
\newcommand{\rvtop}{\mathsf{top}}

\newcommand{\configs}{\mathcal{C}}
\newcommand{\stackelems}{\mathcal{E}}

\newcommand{\sampdpd}{\overline{\Upsilon}}

\newcommand{\enabled}[1]{\mathrm{En}(#1)}
\newcommand{\cyl}{\mbox{\sl Cyl}}

\newcommand{\Dirac}[1]{\Delta_{#1}}

\newcommand{\wbsch}[1]{{\langle {#1}\rangle}_\mathbf{0}}
\newcommand{\stksch}[2]{\left[{#1}{\uparrow}{#2}\right]}
\newcommand{\probht}[1]{\mbox{\sl pr}\left(#1\right)}

\renewcommand{\baselinestretch}{0.96}
\begin{document}

\title{Termination of Nondeterministic Recursive Probabilistic Programs}

\author{\IEEEauthorblockN{Krishnendu Chatterjee}\IEEEauthorblockA{IST Austria}\and \IEEEauthorblockN{Hongfei Fu} \IEEEauthorblockA{Institute of Software, State Key Laboratory of Computer Science\\ Chinese Academy of Sciences, Beijing, P.R. China}}

\maketitle

\begin{abstract}
We study the termination problem for nondeterministic recursive probabilistic programs.
First, we show that a ranking-supermartingales-based approach is both sound and complete
for bounded terminiation (i.e., bounded expected termination time over all schedulers).
Our result also clarifies previous results which claimed that ranking supermartingales are not a complete approach even for
nondeterministic probabilistic programs without recursion.
Second, we show that conditionally difference-bounded ranking supermartingales provide a sound approach for lower bounds of expected termination time.
Finally, we show that supermartingales with lower bounds on conditional absolute difference provide a sound approach for almost-sure termination,
along with explicit bounds on tail probabilities of nontermination within a given number of steps.
We also present several illuminating counterexamples that establish the necessity of certain prerequisites
(such as conditionally difference-bounded condition).
\end{abstract}


\section{Introduction}\label{sec:introduction}

\noindent{\em Probabilistic programs.}
The extension of classical imperative programs with \emph{random value
generators} that produce random values according to some desired probability
distribution gives rise to the class of probabilistic programs.
Probabilistic programs provide the appropriate model for a wide
variety of applications, such as analysis of stochastic network
protocols~\cite{BaierBook,prism}, robot planning~\cite{kaelbling1998planning}, etc.
The formal analysis of probabilistic systems and probabilistic programs is
an active research topic across different disciplines, such as
probability theory and statistics~\cite{Durrett,Howard,Kemeny,Rabin63,PazBook},
formal methods~\cite{BaierBook,prism},
artificial intelligence~\cite{LearningSurvey,kaelbling1998planning},
and programming languages~\cite{SriramCAV,HolgerPOPL,SumitPLDI,EGK12,DBLP:conf/popl/ChatterjeeFNH16}.

\smallskip\noindent{\em Termination questions.}
The most basic and fundamental notion of liveness for programs is the
\emph{termination} problem.
For nonprobabilistic programs, the proof of termination coincides with
the construction of \emph{ranking functions}~\cite{rwfloyd1967programs},
and many different approaches exist for such construction~\cite{DBLP:conf/cav/BradleyMS05,DBLP:conf/tacas/ColonS01,DBLP:conf/vmcai/PodelskiR04,DBLP:conf/pods/SohnG91}.
For probabilistic programs the most natural and basic extensions of the
termination problem are:
First, the \emph{almost-sure} termination question asks whether the program
terminates with probability~1.
Second, the \emph{bounded} termination question asks whether the expected
termination time is bounded.
While the bounded termination implies almost-sure termination, the converse
is not true in general.

\smallskip\noindent{\em Two key aspects: Nondeterminism and Recursion.}
Nondeterminism plays a fundamental role in program analysis.
A classic example is abstraction: for efficient static analysis of large programs,
it is infeasible to track all variables of the program.
Abstraction ignores certain variables and replaces them with worst-case
behavior modeled as nondeterminism.
Another fundamental aspect in program analysis is the role of recursion.
Thus probabilistic recursive programs with nondeterminism is a fundamental model
in program analysis.
We study the termination questions for this model.
We first present the previous results, and then our contributions.

\smallskip\noindent{\em Previous results: Nonrecursive probabilistic programs.}
We describe the most relevant previous results for termination of
probabilistic nonrecursive programs.
\begin{compactitem}
\item \emph{Finite probabilistic choices.}
First, in~\cite{MM04,MM05} quantitative invariants were used to establish termination
for probabilistic programs with nondeterminism, but restricted only to finite probabilistic choices.

\item \emph{Infinite probabilistic choices without nondeterminism.}
The approach of~\cite{MM04,MM05} was extended in~\cite{SriramCAV} to \emph{ranking}
supermartingales to obtain a sound (but not complete) approach
for almost-sure termination for infinite-state probabilistic programs with
infinite-domain random variables.
The above approach was for probabilistic programs without nondeterminism.
The connection between termination of probabilistic programs without nondeterminism and
\emph{Lyapunov ranking functions} was considered in~\cite{BG05}.
For probablistic programs with countable state space and without
nondeterminism, the Lyapunov ranking functions provide a sound and complete
method to prove bounded termination~\cite{BG05,Foster53}.

\item \emph{Infinite probabilistic choices with nondeterminism.}
In the presence of nondeterminism, the Lyapunov-ranking-function method
as well as the ranking-supermartingale method are sound but not complete,
and completeness was established for a subclass~\cite{HolgerPOPL}.
A martingale-based approach for high probability termination and nontermination
has also been considered~\cite{ChatterjeeNZ2017}.

\end{compactitem}

\smallskip\noindent{\em Previous results: Recursive probabilistic programs.}
Probabilistic recursive programs with bounded-domain variables or equivalently
recursive MDPs have been studied extensively~\cite{DBLP:journals/jacm/EtessamiY15,DBLP:journals/iandc/BrazdilBFK08,DBLP:journals/jcss/BrazdilKKV15}
for decidability and complexity results. In contrast we consider probabilistic programs with integer variables.
The notion of proof rules for probabilistic recursive programs has also been studied~\cite{DBLP:conf/lics/OlmedoKKM16}.
However, a sound and complete approach for probabilistic recursive programs with nondeterminism
has not been studied.

\smallskip\noindent{\em Important open questions.}
Given the many important results established in the literature, there
are still several fundamental open questions.
First, a sound and complete approach for recursive probabilistic programs
with nondeterminism is an important open question.
Second, a thorough understanding of the inability of the ranking-supermartingale
approach for the bounded termination problem in the presence of nondeterminism
is also missing.
We address these fundamental questions in this work.

\smallskip\noindent{\em Our results.}
We consider probabilistic recursive programs with nondeterminism where all variables are integer-valued,
which leads to countable state-space MDPs.
Our main contributions are as follows.
\begin{compactitem}
\item {\em Bounded termination.}
We show that a ranking supermartingales based approach is both sound and complete
for the bounded termination problem for probabilistic recursive programs
with nondeterminism.
Note that this is in contrast to~\cite[Theorem~5.7]{HolgerPOPL}
which states that ranking supermartingales are not complete even for
probabilistic nonrecursive programs with nondeterminism.
Our semantics follows the standard MDP semantics and is different
from the one of~\cite{HolgerPOPL} (see Remark~\ref{rmk:semantics}),
and our ranking supermartingales (when applied to programs) are also different from~\cite{HolgerPOPL}
(see Remark~\ref{rmk:differencersm}).
A closer look at the claim of~\cite[Theorem~5.7]{HolgerPOPL} reveals that the counterexample
used for nonexistence of ranking supermartingales (in standard setting) admit a ranking supermartingale 
(see Example~\ref{ex:refutation}).
The significance of our result is as follows:
\begin{compactitem}
\item First, our result presents both a sound and complete approach for bounded
termination for probabilistic recursive programs with nondeterminism, which settles
an important open question.

\item Second, it clarifies the understanding of the ranking-supermartingale approach in the
presence of nondeterminism for probabilistic programs. In particular,
quite surprisingly we show that with the standard MDP semantics and the appropriate notion
of ranking supermartingales, we obtain a sound and complete approach.
\end{compactitem}

\item {\em Lower bound and tail probabilities.}
We show that conditionally difference-bounded ranking supermartingales provide
a sound approach for lower bounds on expected termination time.
We show that bounds on tail probabilities can be obtained
for difference-bounded ranking supermartingales in the presence of recursion,
and without the difference-boundedness condition optimal bounds can
be obtained from Markov's inequality.

\item {\em Almost-sure termination.}
We show that supermartingales with lower bounds on conditional absolute difference
present a sound approach for almost-sure termination.
Note that our results for almost-sure termination use supermartingales (i.e., 
not necessarily ranking supermartingales).
Moreover, with supermartingales no previous works present explicit bounds on
tail probabilities of nontermination within a given number of steps.
We present the first method to explicitly obtain bounds on tail probabilities
 from supermartingales.

\end{compactitem}
Besides the main results above we present several illuminating examples which show
the necessity of several prerequisites (e.g., (conditionally) difference-boundedness):
once they are dropped the desired results no longer hold.

\smallskip\noindent{\em Technical contributions.}
The key novelties of our results are as follows:
first, is the construction of ranking supermartingales over configurations
(which correspond to states of the infinite-state MDP)
of probabilistic programs with nondeterminism and recursion;
second, is an elaborate construction of martingales that allows to derive
almost-sure termination with bounds on tail probabilities from supermartingales.
Our proofs require delicate handling of integrability conditions as well as
clever use of Optional Stopping Theorem.
Detailed proofs are presented in the appendix.


\section{Recursive Probabilistic Programs}\label{sect:preliminaries}

We consider a simple programming language for nondeterministic recursive probabilistic programs extended from C programming language, with basic capabilities for recursion, demonic nondeterminism and probabilistic features.
Moreover, all variables hold integers in our language.
We first introduce some basic notations and concepts, then illustrate the
syntax of the language and finally the semantics.

\subsection{Basic Notations and Concepts}\label{sect:basicnotations}

We denote by $\Nset$, $\Nset_0$, $\Zset$, and $\Rset$ the sets of all positive integers, nonnegative integers, integers, and real numbers, respectively.

\smallskip\noindent{\bf Arithmetic Expressions.} An \emph{arithmetic expression} $\mathfrak{e}$ over a finite set $V$ of variables is an expression built from variables in $V$, integer constants and
arithmetic operations namely addition, subtraction, multiplication, exponentiation, etc.
Since our results are theoretical,
we consider a general setting for arithmetic expressions.

\noindent{\bf Valuations.} Let $V$ be a finite set of variables. A \emph{valuation} over $V$ is a function $\nu$ from $V$ into $\Zset$.
The set of valuations over $V$ is denoted by $\val{V}$.
Given an arithmetic expression $\mathfrak{e}$ and a valuation $\nu$, we denote by $\mathfrak{e}(\nu)$ the integer obtained by evaluating $\mathfrak{e}$ through replacing every $x\in V$ with $\nu(x)$.
In this paper, we assume that $\mathfrak{e}(\nu)$ be always well-defined and integer (e.g., no zero denominator, $2^{-1}$ or $0^0$).

\noindent{\bf Propositional Arithmetic Predicates.}
Let $V$ be a finite set of variables.
A \emph{propositional arithmetic predicate} (over $V$) is a logical formula $\phi$ built from (i) atomic formulae of the form $\mathfrak{e}\Join\mathfrak{e}'$ where $\mathfrak{e},\mathfrak{e}'$ are arithmetic expressions over $V$ and $\Join\in\{<,\le, >,\ge\}$, and (ii) logical connectives namely $\vee,\wedge,\neg$.
The satisfaction relation $\models$ between a valuation $\nu$ and a propositional arithmetic predicate $\phi$ is defined through evaluation and standard semantics of logical connectives such that (i) $\nu\models \mathfrak{e}\Join\mathfrak{e}'$ iff $\mathfrak{e}(\nu)\Join \mathfrak{e}'(\nu)$, (ii) $\nu\models\neg\phi$ iff $\nu\not\models\phi$ and (iii) $\nu\models\phi_1\wedge\phi_2$ (resp. $\nu\models\phi_1\vee\phi_2$) iff $\nu\models\phi_1$ and (resp. or) $\nu\models\phi_2$.

\noindent{\bf Probability Space.} A \emph{probability space} is a triple $(\Omega,\mathcal{F},\probm)$, where $\Omega$ is a nonempty set (so-called \emph{sample space}), $\mathcal{F}$ is a \emph{$\sigma$-algebra} over $\Omega$ (i.e., a collection of subsets of $\Omega$ that contains the empty set $\emptyset$ and is closed under complementation and countable union), and $\probm$ is a \emph{probability measure} on $\mathcal{F}$, i.e., a function $\probm\colon \mathcal{F}\rightarrow[0,1]$ such that (i) $\probm(\Omega)=1$ and
(ii) for all set-sequences $A_1,A_2,\dots \in \mathcal{F}$ that are pairwise-disjoint
(i.e., $A_i \cap A_j = \emptyset$ whenever $i\ne j$)
it holds that $\sum_{i=1}^{\infty}\probm(A_i)=\probm\left(\bigcup_{i=1}^{\infty} A_i\right)$.
Elements $A\in\mathcal{F}$ are usually called \emph{events}.
An event $A\in\mathcal{F}$ is said to hold \emph{almost surely} (a.s.) if $\probm(A)=1$.

\noindent{\bf Random Variables.} A \emph{random variable} $X$ from a probability space $(\Omega,\mathcal{F},\probm)$
is an $\mathcal{F}$-measurable function $X\colon \Omega \rightarrow \Rset \cup \{-\infty,+\infty\}$, i.e.,
a function satisfying the condition that for all $d\in \Rset \cup \{-\infty,+\infty\}$, the set $\{\omega\in \Omega\mid X(\omega)<d\}$ belongs to $\mathcal{F}$; $X$ is \emph{bounded} if there exists a real number $M>0$ such that for all $\omega\in\Omega$, we have $X(\omega)\in\Rset$ and $|X(\omega)|\le M$.
By convention, we abbreviate $+\infty$ as $\infty$.

\noindent{\bf Expectation.} The \emph{expected value} of a random variable $X$ from a probability space $(\Omega,\mathcal{F},\probm)$, denote by $\expv(X)$, is defined as the Lebesgue integral of $X$ w.r.t $\probm$, i.e.,
$\expv(X):=\int X\,\mathrm{d}\probm$~;
the precise definition of Lebesgue integral is somewhat technical and is
omitted  here (cf.~\cite[Chapter 5]{probabilitycambridge} for a formal definition).
In the case that $\mbox{\sl range}~X=\{d_0,d_1,\dots,d_k\dots,\}$ is countable with distinct $d_k$'s, we have
$\expv(X)=\sum_{k=0}^\infty d_k\cdot \probm(X=d_k)$.

\noindent{\bf Characteristic Random Variables.} Given random variables $X_0,\dots,X_n$ from a probability space $(\Omega,\mathcal{F},\probm)$ and a predicate $\Phi$ over $\Rset \cup \{-\infty,+\infty\}$, we denote by $\mathbf{1}_{\Phi(X_0,\dots,X_n)}$ the random variable such that
\[
\mathbf{1}_{\Phi(X_0,\dots,X_n)}(\omega)=
\begin{cases}
1 & \mbox{if }\Phi\left(X_0(\omega),\dots,X_n(\omega)\right)\mbox{ holds} \\
0 & \mbox{otherwise}
\end{cases}\enskip.
\]
By definition, $\expv\left(\mathbf{1}_{\Phi(X_0,\dots,X_n)}\right)=\probm\left(\Phi(X_0,\dots,X_n)\right)$.
Note that if $\Phi$ does not involve any random variable, then $\mathbf{1}_{\Phi}$ can be deemed as a constant whose value depends only on whether $\Phi$ holds or not.

\noindent{\bf Filtrations and Stopping Times.} A \emph{filtration} of a probability space $(\Omega,\mathcal{F},\probm)$ is an infinite sequence $\{\mathcal{F}_n \}_{n\in\Nset_0}$ of $\sigma$-algebras over $\Omega$ such that $\mathcal{F}_n \subseteq \mathcal{F}_{n+1} \subseteq\mathcal{F}$ for all $n\in\Nset_0$.
A \emph{stopping time} (from $(\Omega,\mathcal{F},\probm)$) w.r.t $\{\mathcal{F}_n\}_{n\in\Nset_0}$ is a random variable $R:\Omega\rightarrow \Nset_0\cup\{\infty\}$ such that for every $n\in\Nset_0$, the event $R\le n$ belongs to $\mathcal{F}_n$.

\noindent{\bf Conditional Expectation.}
Let $X$ be any random variable from a probability space $(\Omega, \mathcal{F},\probm)$) such that $\expv(|X|)<\infty$.
Then given any $\sigma$-algebra $\mathcal{G}\subseteq\mathcal{F}$, there exists a random variable (from $(\Omega, \mathcal{F},\probm)$), conventionally denoted by $\condexpv{X}{\mathcal{G}}$, such that
\begin{compactitem}
\item[(E1)] $\condexpv{X}{\mathcal{G}}$ is $\mathcal{G}$-measurable, and
\item[(E2)] $\expv\left(\left|\condexpv{X}{\mathcal{G}}\right|\right)<\infty$, and
\item[(E3)] for all $A\in\mathcal{G}$, we have $\int_A \condexpv{X}{\mathcal{G}}\,\mathrm{d}\probm=\int_A {X}\,\mathrm{d}\probm$.
\end{compactitem}
The random variable $\condexpv{X}{\mathcal{G}}$ is called the \emph{conditional expectation} of $X$ given $\mathcal{G}$.
The random variable $\condexpv{X}{\mathcal{G}}$ is a.s. unique in the sense that if $Y$ is another random variable satisfying (E1)--(E3), then $\probm(Y=\condexpv{X}{\mathcal{G}})=1$.
Some properties of conditional expectation are listed in Appendix~\ref{app:condexpv}.
We refer to~\cite[Chapter~9]{probabilitycambridge} for more details.


\noindent{\bf Discrete-Time Stochastic Processes.}
A \emph{discrete-time stochastic process} is a sequence $\Gamma=\{X_n\}_{n\in\Nset_0}$ of random variables where $X_n$'s are all from some probability space (say, $(\Omega,\mathcal{F},\probm)$);
and $\Gamma$ is \emph{adapted to} a filtration $\{\mathcal{F}_n\}_{n\in\Nset_0}$ of sub-$\sigma$-algebras of $\mathcal{F}$ if for all $n\in\Nset_0$, $X_n$ is $\mathcal{F}_n$-measurable.

\noindent{\bf (Conditionally) Difference-boundedness.}
A discrete-time stochastic process $\Gamma=\{X_n\}_{n\in\Nset_0}$ when adapted to a filtration $\{\mathcal{F}_n\}_{n\in\Nset_0}$, is
\begin{compactitem}
\item \emph{conditionally difference-bounded}, if there exists $c\in(0,\infty)$ such that for all $n\in\Nset_0$, we have $\expv\left(|X_{n+1}-X_n|\right)< \infty$, $\condexpv{|X_{n+1}-X_n|}{\mathcal{F}_n}\le c$ a.s.;
\item \emph{difference-bounded}, if there exists $c\in(0,\infty)$ such that for every $n\in\Nset_0$, we have $|X_{n+1}-X_n|\le c$ a.s.
\end{compactitem}

\noindent{\bf Stopping time $\stopping{\Gamma}$.}
Given a discrete-time stochastic process $\Gamma=\{X_n\}_{n\in\Nset_0}$ adapted to a filtration $\{\mathcal{F}_n\}_{n\in\Nset_0}$, we define the random variable $\stopping{\Gamma}$ by
$\stopping{\Gamma}(\omega):=\min\{n\mid X_{n}(\omega)\le 0\}$
where $\min\emptyset:=\infty$.
Note that by definition, $\stopping{\Gamma}$ is a stopping time w.r.t $\{\mathcal{F}_n\}_{n\in\Nset_0}$.
Moreover,
\begin{eqnarray*}
\expv(\stopping{\Gamma}) &=& \infty\cdot \probm(\stopping{\Gamma}=\infty)+\sum_{k=0}^\infty k\cdot \probm(\stopping{\Gamma}=k) \\
&=& \infty\cdot \probm(\stopping{\Gamma}=\infty)+\sum_{k=0}^\infty \probm(k<\stopping{\Gamma}<\infty)\enskip.
\end{eqnarray*}
We follow the convention that $0\cdot\infty:=0$ as in the setting of Lebesgue Integral.

\noindent{\bf Martingales.} A discrete-time stochastic process $\Gamma=\{X_n\}_{n\in\Nset}$ adapted to a filtration $\{\mathcal{F}_n\}_{n\in\Nset_0}$ is a \emph{martingale} (resp. \emph{supermartingale}, \emph{submartingale})
if for every $n\in\Nset_0$, $\expv(|X_n|)<\infty$ and it holds a.s. that
$\condexpv{X_{n+1}}{\mathcal{F}_n}=X_n$ (\mbox{resp. } $\condexpv{X_{n+1}}{\mathcal{F}_n}\le X_n$, $\condexpv{X_{n+1}}{\mathcal{F}_n}\ge X_n$).
We refer to ~\cite[Chapter~10]{probabilitycambridge} for more details.

\noindent{\bf Discrete Probability Distributions over Countable Support.} A \emph{discrete probability distribution} over a countable set $U$ is a function $\dpd:U\rightarrow[0,1]$ such that $\sum_{z\in U}\dpd(z)=1$.
The \emph{support} of $q$, is defined as $\supp{q}:=\{z\in U\mid q(z)>0\}$.

\subsection{The Syntax}

Due to page limit, we only present an informal description for the syntax of our programming language.

Informally, our program language involves two types of variables: \emph{program variables} and \emph{sampling variables}. Program variables are normal variables, while each sampling variable will be bound to a discrete probability distribution later (cf. Section~\ref{sect:semantics}).
Statements in our language are similar to C programming language: assignment statements are indicated by `$:=$', while `\textbf{skip}' is the special assignment statement that does not change values;
if-branches (resp. while-loops) are indicated by `\textbf{if}' (resp. `\textbf{while}') together with a propositional arithmetic predicate and possibly `\textbf{then}' and `\textbf{else}' branches;
function-calls takes the same form as in C and are call-by-value;
demonic branches are indicated by `\textbf{if}' and the special symbol `$\star$'.
For the sake of simplicity, we do not allow return statements.
Moreover, we do not specify a $\mathsf{main}$ function body since we consider termination analysis of all function entities. See Appendix~\ref{app:syntax} for details.

\smallskip\noindent{\bf Statement Labelling.}
Given a nondeterministic recursive probabilistic program, we assign a distinct natural number (called \emph{label} in our context)
to every single assignment/skip statement, function call, if/while-statement and terminal line
within each function body in the program.
Informally, each label serves as a program counter which indicates the next statement to be executed.

We illustrate an example with labelling in Example~\ref{ex:runningexample}.

\lstset{language=prog}
\lstset{tabsize=3}
\newsavebox{\progrunningexamplea}
\begin{lrbox}{\progrunningexamplea}
\begin{lstlisting}[mathescape]
$\mathsf{f}(n)~\{$
$1$: if $n\ge 1$ then
$2$:   if $\star$ then
$\underline{3;4}$:     $\mathsf{f}\left(\left\lfloor \frac{n}{2}\right\rfloor\right)$; $\mathsf{f}\left(\left\lfloor \frac{n}{2}\right\rfloor\right)$
    else
$5$:     $\fn{g}(n-1)$
    fi
   else
$6$:   skip
   fi
$7$: $\}$
\end{lstlisting}
\end{lrbox}

\lstset{language=prog}
\lstset{tabsize=3}
\newsavebox{\progrunningexampleb}
\begin{lrbox}{\progrunningexampleb}
\begin{lstlisting}[mathescape]
|
|
|$\mathsf{g}(n)~\{$
|$1$: if $n\ge 1$ then
|$2$:   $n:=n+r$;
|$3$:   $\fn{f}(n)$
|   else
|$4$:   skip
|   fi
|$5$: $\}$
|
|
\end{lstlisting}
\end{lrbox}

\begin{figure}
\centering
\usebox{\progrunningexamplea}
\usebox{\progrunningexampleb}
\caption{An Example: Nondeterministic Recursive Probabilistic Programs}
\label{fig:runningexample}
\end{figure}

\begin{example}\label{ex:runningexample}
Consider the program depicted in Figure~\ref{fig:runningexample}, where $n$ is a program variable and $r$ is a sampling variable.
The semantics of $r$ here is a sampling from the two-point distribution $q$ such that $q(1)=\frac{1}{4}$ and $q(-1)=\frac{3}{4}$
(cf. the semantics to be demonstrated in Section~\ref{sect:semantics}).
Basically, the program executes around the value held by $n$.
In the function body for $\mathsf{f}$, it is nondeterministic whether two function calls at lines 3--4
or a single function call at line 5 are executed.
In the function body of $\mathsf{g}$, a sampling w.r.t $q$ for the value held by $r$ is executed at line 2 and then a function call is executed at line 3.
Both the function bodies of $\mathsf{f}$ and $\mathsf{g}$ terminate when the logical formula $n\le 0$ is satisfied.\hfill\IEEEQEDclosed
\end{example}

\subsection{The Semantics}\label{sect:semantics}

We use control-flow graphs (CFGs) and discrete-time Markov decision processes (MDPs) to specify the semantics of recursive programs.
We first illustrate the notion of CFGs.

\begin{definition}[Control-Flow Graphs]\label{def:cfg}
A \emph{control-flow graph} (CFG) is a triple which takes the form
\begin{equation}\label{eq:cfg}
\left(\fnames,\rvars,\left\{\left(\locs{f}, \clocs{f}, \alocs{f}, \flocs{f},\dlocs{f},\lin{\fn{f}},\lout{\fn{f}},\pvars{f},\transitions{\fn{f}}\right)\right\}_{\fn{f}\in\fnames}\right)
\end{equation}
where:
\begin{compactitem}
\item $\fnames$ is a finite set of \emph{function names};
\item $\rvars$ is a finite set of \emph{sampling variables};
\item each $\locs{f}$ is a finite set of \emph{labels} attached to the function name $\fn{f}$, which is partitioned into (i) the set $\clocs{f}$ of \emph{branching} labels, (ii) the set $\alocs{f}$ of \emph{assignment} labels, (iii) the set $\flocs{f}$ of \emph{call} labels and (iv) the set  $\dlocs{f}$ of \emph{nondeterministic} labels;
\item each $\pvars{f}$ is the set of \emph{program variables} attached to $\fn{f}$;
\item each $\lin{\fn{f}}$ (resp. $\lout{\fn{f}}$) is the \emph{initial label} (resp.  \emph{terminal label}) in $\locs{f}$;
\item each $\transitions{\fn{f}}$ is a relation whose every member is a triple of the form $(\loc,\alpha,\loc')$ for which $\loc$ (resp. $\loc'$) is the source label (resp. target label) of the triple such that $\loc\in\locs{f}$ (resp. $\loc'\in\locs{f}$), and $\alpha$ is either a propositional arithmetic predicate $\phi$ over $\pvars{f}$ if $\loc\in\clocs{f}$, or an \emph{update function} $u:\val{\pvars{f}}\times\val{\rvars}\rightarrow \val{\pvars{f}}$
    if $\loc\in\alocs{f}$, or a pair $(\fn{g}, v)$ with $\fn{g}\in\fnames$ and $v:\val{\pvars{f}}\rightarrow\val{\pvars{g}}$ being a \emph{value-passing function} if $\loc\in\flocs{f}$, or $\star$ if $\loc\in\dlocs{f}$.
\hfill\IEEEQEDclosed
\end{compactitem}
\end{definition}
W.l.o.g, we assume that all labels and function names are encoded by natural numbers.
For the sake of convenience, we abbreviate $\val{\pvars{f}}$ as $\val{\fn{f}}$ ($\fn{f}\in\fnames$).

Informally, a control-flow graph specifies how values for program variables and the program counter change in a program.
We refer to the status of the program counter as a \emph{label}, and assign an initial label and a terminal label to the function body of each function entity.
Moreover, we have four types of labels, namely \emph{branching}, \emph{assignment}, \emph{call} and \emph{nondeterministic} labels.
A branching label corresponds to a conditional-branching statement indicated by the keyword `\textbf{if}' or `\textbf{while}' together with some propositional arithmetic predicate $\phi$,
and leads to the next label in the current function body determined by $\phi$ without change on values.
An assignment label corresponds to an assignment statement indicated by `$:=$' or $\mathbf{skip}$,
and leads to the next label right after the statement in the current function body with change of values specified by the update function determined at the right-hand-side of `$:=$', for which an update function gives the next valuation over program variables based on the current valuation and the sampled values;
($\mathbf{skip}$ is deemed as an assignment statement that does not change values).
A call label corresponds to a function call with some function name $\fn{g}$ and initial values determined by the value-passing function specified by the call, leads to the label right after the call in the current function body,
and does not change  values in the original function body.
Finally, a nondeterministic label corresponds to a demonic nondeterministic statement indicated by `\textbf{if}' and `$\star$', and leads to two labels specified by the `\textbf{then}' and the `\textbf{else}' branches.

It is intuitively clear that every nondeterministic probabilistic recursive program can be equivalently transformed into a CFG. Due to page limit, we put the detailed transformation in Appendix~\ref{app:cfg}.
An example CFG is given in Example~\ref{ex:cfg}.

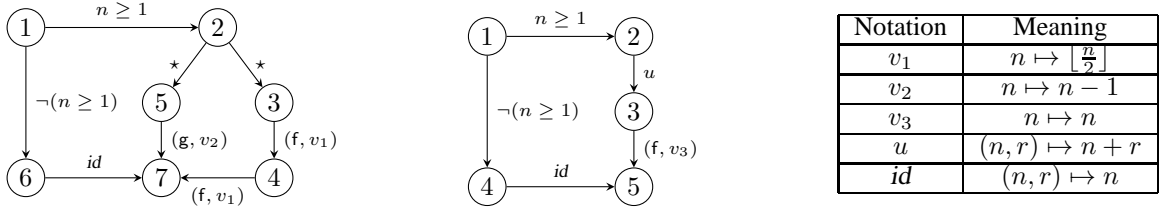
\begin{figure*}
\begin{minipage}{0.3\textwidth}
\centering
\begin{tikzpicture}[x = 1.5cm]
\node[proglabel] (outerif)         at (-1.2,0)      {$1$};
\node[proglabel] (innerif)         at (0.5,0)      {$2$};
\node[proglabel] (lfloorcall)      at (1,-1)        {$3$};
\node[proglabel] (rfloorcall)      at (1,-2)        {$4$};
\node[proglabel] (gcall)           at (0, -1)       {$5$};
\node[proglabel] (skip)            at (-1.2,-2)     {$6$};
\node[proglabel] (end)             at (0,-2)        {$7$};

\draw[tran] (outerif)     to node[auto, font=\scriptsize] {$n\ge 1$}           (innerif);
\draw[tran] (outerif)     to node[right, font=\scriptsize] {$\neg(n\ge 1)$}    (skip);
\draw[tran] (innerif)     to node[right,font=\scriptsize] {$\star$}              (lfloorcall);
\draw[tran] (lfloorcall)  to node[auto, font=\scriptsize] {$(\fn{f}, v_1)$}     (rfloorcall);
\draw[tran] (rfloorcall)  to node[auto, font=\scriptsize] {$(\fn{f}, v_1)$}     (end);
\draw[tran] (innerif)     to node[left, font=\scriptsize] {$\star$}        (gcall);
\draw[tran] (gcall)       to node[auto, font=\scriptsize] {$(\mathsf{g}, v_2)$} (end);
\draw[tran] (skip)        to node[auto, font=\scriptsize] {$\id$}     (end);
\end{tikzpicture}
\end{minipage}
\begin{minipage}{0.3\textwidth}
\centering
\begin{tikzpicture}[x = 1.5cm]

\node[proglabel] (outerif)         at (-0.8,0)         {$1$};
\node[proglabel] (sample)          at (0.5,0)      {$2$};
\node[proglabel] (innerif)         at (0.5,-1)      {$3$};
\node[proglabel] (skip)            at (-0.8,-2)     {$4$};
\node[proglabel] (end)             at (0.5,-2)        {$5$};

\draw[tran] (outerif) to node[auto, font=\scriptsize] {$n\ge 1 $}         (sample);
\draw[tran] (outerif) to node[right, font=\scriptsize] {$\neg(n\ge 1)$}     (skip);
\draw[tran] (sample)  to node[auto, font=\scriptsize] {$u$}          (innerif);
\draw[tran] (innerif) to node[right, font=\scriptsize] {$(\fn{f}, v_3)$} (end);
\draw[tran] (skip) to node[auto, font=\scriptsize] {$\id$} (end);
\end{tikzpicture}
\end{minipage}
\begin{minipage}{0.3\textwidth}
\centering
\begin{tabular}{|c|c|}
\hline
Notation & Meaning \\
\hline
$v_1$ & $n\mapsto \lfloor\frac{n}{2}\rfloor$ \\
\hline
$v_2$ & $n\mapsto n-1$   \\
\hline
$v_3$ & $n \mapsto n$ \\
\hline
$u$ & $(n,r) \mapsto n+r$\\
\hline
$\id$ & $(n,r)\mapsto n$ \\
\hline
\end{tabular}
\end{minipage}
\caption{The CFG for the Program in Figure~\ref{fig:runningexample} with Notations Given in the Rightmost Table}
\label{fig:cfgrunningexample}
\end{figure*}

\begin{example}\label{ex:cfg}
Figure~\ref{fig:cfgrunningexample} shows the CFG for Example~\ref{ex:runningexample}.
The left (resp. middle) part  of the figure is for $\fn{f}$ (resp. $\fn{g}$). \hfill\IEEEQEDclosed
\end{example}

Based on CFGs, we illustrate the semantics of nondeterministic recursive probabilistic programs as follows.
The semantics is defined through (i) samplings for sampling variables right before execution of every statement
and (ii) the standard notion of call stack.

Below we fix a nondeterministic recursive probabilistic program $W$ with its CFG taking the form (\ref{eq:cfg}).
We first define the notion of \emph{stack elements} which captures all information within a function call.

\begin{definition}[Stack Elements $\stackelems$]
A \emph{stack element} $\mathfrak{c}$ (of $W$) is a
triple $(\fn{f},\loc,\nu)$ where $\fn{f}\in\fnames$, $\loc\in\locs{f}$ and
$\nu\in\val{\fn{f}}$.
The stack element $\mathfrak{c}$ is \emph{terminal} if $\loc=\lout{\fn{f}}$; $\mathfrak{c}$ is \emph{nondeterministic} if $\loc\in\dlocs{f}$.
The set of stack elements is denoted by $\stackelems$.\hfill\IEEEQEDclosed
\end{definition}
Informally, a stack element $(\fn{f},\loc,\nu)$ specifies that the current function name is $\fn{f}$, the next statement to be executed is the one labelled with $\loc$ and the current valuation is $\nu$.

Then we define the notion of \emph{configurations} which captures all information needed to describe the current status of $W$, i.e., a configuration records the whole trace of the call stack.

\begin{definition}[Configurations $\configs$]
A \emph{configuration} (of $W$) is a finite word $w$ of nonterminal stack elements (including the empty word $\varepsilon$).
A configuration $w$ is \emph{nondeterministic} if (i)~$w\ne\varepsilon$
and (ii)~the first letter of $w$ (which is a stack element) is nondeterministic.
The set of configurations is denoted by $\configs$.\hfill\IEEEQEDclosed
\end{definition}

\begin{remark}
A configuration $w=\mathfrak{c}_0\dots \mathfrak{c}_n$ is organized such that $\mathfrak{c}_0$ reflects the current function call (i.e, top of the call stack) and $\mathfrak{c}_n$ reflects the last (bottom of the call stack).\hfill\IEEEQEDclosed
\end{remark}

We also need to assign meanings to sampling variables through discrete probability distributions.

\begin{definition}
A \emph{sampling function} $\Upsilon$ is a function assigning to every sampling variable $r\in\rvars$ a discrete probability distribution over $\Zset$.
The discrete probability distribution $\sampdpd$ over $\val{\rvars}$ is defined by:
$\sampdpd(\mu):=\prod_{r\in\rvars}\left(\Upsilon(r)\right)(\mu(r))$~.\hfill\IEEEQEDclosed
\end{definition}

Below we fix a sampling function $\Upsilon$.
Now the semantics of $W$ is described by a Markov decision process (MDP) (cf.~\cite[Chapter 10]{BaierBook}) $\MDP{W}$ as follows.
Informally, $\MDP{W}$ describes the execution of $W$, where states of $\MDP{W}$ are pairs of the form $(w,\mu)$
where $w$ reflects the current configuration while $\mu$ reflects the sampling of the \emph{previous} step.

\begin{definition}[The Semantics ($\MDP{W}$)]\label{def:semantics}
The Markov decision process (MDP) $\MDP{W}=(\MDPstates_W,\MDPactions, \MDPkernel_W)$ is defined as follows.
\begin{compactitem}
\item The \emph{state space} $\MDPstates_W$ is $\mathcal{C}\times \val{\rvars}$.
\item The \emph{action set} $\MDPactions$ is $\left\{\tau, \mathbf{th}, \mathbf{el}\right\}$.
Intuitively, $\tau$ refers to absence of nondeterminism and \textbf{th} (resp. \textbf{el}) refers to the \textbf{then}- (resp. \textbf{else}-) branch of a nondeterministic label.
\item The \emph{probability transition function}
\[
\MDPkernel_W:\MDPstates_W\times\MDPactions\times\MDPstates_W\rightarrow [0,1]
\]
is given as follows. First, for all $a\in \MDPactions$ and $\mu\in\val{\rvars}$,
\[
\MDPkernel_W((\varepsilon,\mu),a,(w,\mu'))=
\begin{cases}
\sampdpd(\mu') & \mbox{if }w=\varepsilon\\
0 & \mbox{otherwise}
\end{cases}
\]
which clarifies the case for termination.
Second, for all nonterminal stack elements $(\fn{f}, \loc, \nu)$, configurations $w$ and $\mu\in\val{\rvars}$ grouped as a state $s=\left((\fn{f},\loc,\nu)\cdot w,\mu\right)$:
\begin{compactenum}
\item {\em Assignment:} If $\loc\in\alocs{f}$ and $(\loc,u,\loc')$ is the only triple in $\transitions{\fn{f}}$ with source label $\loc$ and update function $u$, then
\begin{align*}
&\MDPkernel_W\left(s, \tau, \left(w',\mu'\right)\right):=\\
&~\begin{cases}
\sampdpd(\mu') & \mbox{ if } \loc'\ne\lout{\fn{f}} \mbox{ and }  w'=(\fn{f},\loc',u(\nu,\mu'))\cdot w\\
\sampdpd(\mu') & \mbox{ if }  \loc'=\lout{\fn{f}}  \mbox{ and } w'=w \\
0 & \mbox{ otherwise }\\
\end{cases}
\end{align*}
and $\MDPkernel_W\left(s, \mbox{\textbf{th}}, \centerdot\right), \MDPkernel_W\left(s, \mbox{\textbf{el}}, \centerdot\right)$
are identically zero;
\item {\em Call:} If $\loc\in\flocs{f}$ and $(\loc,(\fn{g},v),\loc')$ is the only triple in $\transitions{\fn{f}}$ with source label $\loc$ and value-passing function $v$, then
\begin{align*}
&\MDPkernel_W\left(s, \tau, \left(w',\mu'\right)\right):=\\
&\begin{cases}
\sampdpd(\mu')  & \mbox{ if } \loc'\ne\lout{\fn{f}}, w'=(\fn{g}, \lin{\fn{g}}, v(\nu))\cdot(\fn{f},\loc',\nu)\cdot w \\
\sampdpd(\mu') & \mbox{ if } \loc'=\lout{\fn{f}}, w'=(\fn{g}, \lin{\fn{g}}, v(\nu))\cdot w  \\
0 & \mbox{ otherwise }\\
\end{cases}
\end{align*}
and
$\MDPkernel_W\left(s, \mbox{\textbf{th}}, \centerdot\right),
\MDPkernel_W\left(s, \mbox{\textbf{el}}, \centerdot\right)$
are identically zero;
\item {\em Branching:} If $\loc\in\clocs{f}$ and $(\loc, \phi, \loc_1),(\loc, \neg\phi, \loc_2)$ are namely two triples in $\transitions{\fn{f}}$ with source label $\loc$ and propositional arithmetic predicate $\phi$, then
\begin{align*}
&\MDPkernel_W\left(s, \tau, (w',\mu')\right):=\\
&~~\begin{cases}
\sampdpd(\mu')  & \mbox{ if } \loc^*\ne\lout{\fn{f}}\mbox{ and }w'=(\fn{f},\loc^*,\nu)\cdot w  \\
\sampdpd(\mu') & \mbox{ if } \loc^*=\lout{\fn{f}}\mbox{ and }w'=w   \\
0 & \mbox{ otherwise }\\
\end{cases}
\end{align*}
where (i) $\loc^*$ is defined as $\loc_1$ if $\nu\models\phi$, and as $\loc_2$ otherwise,
and (ii) $\MDPkernel_W\left(s, \mbox{\textbf{th}}, \centerdot\right),  \MDPkernel_W\left(s, \mbox{\textbf{el}}, \centerdot\right)$ are identically zero;
\item {\em Nondeterminism:} If $\loc\in\dlocs{f}$ and $(\loc, \star, \loc_1),(\loc, \star, \loc_2)$ are namely two triples in $\transitions{\fn{f}}$ with source label $\loc$ such that $\loc_1$ (resp. $\loc_2$) refers to the \textbf{then}-(resp. \textbf{else}-)branch, then
\begin{align*}
&\MDPkernel_W\left(s, \mbox{\textbf{th}}, (w',\mu')\right):=\\
&~~\begin{cases}
\sampdpd(\mu')  & \mbox{ if } \loc_1\ne\lout{\fn{f}}\mbox{ and }w'=(\fn{f},\loc_1,\nu)\cdot w  \\
\sampdpd(\mu') & \mbox{ if } \loc_1=\lout{\fn{f}} \mbox{ and }w'=w \\
0 & \mbox{ otherwise }\\
\end{cases}
\end{align*}
and
\begin{align*}
&\MDPkernel_W\left(s, \mbox{\textbf{el}}, (w',\mu')\right):=\\
&~~\begin{cases}
\sampdpd(\mu')  & \mbox{ if } \loc_2\ne\lout{\fn{f}}\mbox{ and }w'=(\fn{f},\loc_2,\nu)\cdot w \\
\sampdpd(\mu') & \mbox{ if } \loc_2=\lout{\fn{f}}\mbox{ and }w'=w  \\
0 & \mbox{ otherwise }\\
\end{cases}
\end{align*}
while $\MDPkernel_W\left(s, \tau, \centerdot\right)$ is identically zero.
\end{compactenum}
\end{compactitem}
We say that an action $a\in\MDPactions$ is \emph{enabled} at a state $s$ if $\sum_{s'\in S_W}\MDPkernel_W(s, a, s')=1$.
The set of enabled actions at a state $s$ is denoted by $\enabled{s}$.\hfill\IEEEQEDclosed
\end{definition}

By definition, $\tau$ is the only enabled action at states with configurations that are not nondeterministic,
and $\mbox{\textbf{th}},\mbox{\textbf{el}}$ are namely two enabled actions at states with nondeterministic configurations.
Nondeterminism in MDPs are resolved by schedulers. Below we illustrate the notion of schedulers over $\MDP{W}$.
First we describe the notion of histories upon which schedulers make decisions.

\begin{definition}[Histories]\label{def:histories}
A \emph{history} is a finite word $\rho=(w_0,\mu_0)\dots (w_n,\mu_n)$ ($n\ge 0$) of states in $\MDPstates_W$
such that for all $k$, we have $\MDPkernel_W((w_k,\mu_k),a,(w_{k+1},\mu_{k+1}))>0$ for some $a\in\MDPactions$; 
the last state $(w_n,\mu_n)$ of $\rho$ is denoted by $\last{\rho}$. \hfill\IEEEQEDclosed
\end{definition}


Now the notion of schedulers is as follows.
Informally, a scheduler resolves nondeterminism at nondeterministic labels by discrete probability distributions over actions specifying the probabilities based on which actions are taken.

\begin{definition}[Schedulers]
A \emph{scheduler} $\sigma$ for $W$ is a function which maps every history $\rho$ to a discrete probability distribution $\sigma(\rho)$ over $\enabled{\last{\rho}}$.\hfill\IEEEQEDclosed
\end{definition}

By the standard construction (cf.~\cite[Chapter 10]{BaierBook}), applying a scheduler $\sigma$ to $\MDP{W}$ yields an infinite-state discrete-time Markov chain $\MDP{W,\sigma}$ whose state space is the set of all histories.
We put the detailed construction of $\MDP{W,\sigma}$ in Appendix~\ref{app:markovchains}.

By standard definition, the probability space  for $\MDP{W,\sigma}$ is define over runs, as follows.

\begin{definition}[Runs]
A \emph{finite run} is a finite sequence $\rho_0\dots\rho_n$ ($n\ge 0$) of histories  such that
\begin{compactitem}
\item $\rho_0=(\mathfrak{c},\mu)$ for some non-terminal stack element $\mathfrak{c}$ (viewed as a one-letter configuration) and valuation $\mu\in\val{\rvars}$, and
\item for all $0\le k<n$, there is some state $s\in\MDPstates_W$ such that $\rho_{k+1}=\rho_k\cdot s$.
\end{compactitem}
An \emph{infinite run} is an infinite sequence $\{\rho_n\}_{n\in\Nset_0}$ of histories such that
$\rho_0\dots\rho_n$ is a finite run for all $n\in\Nset_0$. \hfill\IEEEQEDclosed
\end{definition}

Finally, the probability space $\left(\infruns^W, \mathcal{H}^W, \probm_{\mathfrak{c}}^{W,\sigma}\right)$ with initial stack element $\mathfrak{c}$ and scheduler $\sigma$
for $\MDP{W,\sigma}$ is defined through standard \emph{cylinder} construction (cf. Appendix~\ref{app:markovchains}). We use $\expv^{W,\sigma}_{\mathfrak{c}}(\centerdot)$ to denote expectation for random variables over infinite runs (w.r.t the probability measure $\probm_{\mathfrak{c}}^{W,\sigma}$).
We omit `$W$' whenever $W$ is clear from the context.

\begin{remark}\label{rmk:runs}
A finite run $\rho_0\dots\rho_n$ can be deemed equivalently as the single history $\rho_n$ as $\rho_0,\dots,\rho_{n-1}$ are the all $n$ prefixes of $\rho_n$ (excluding $\rho_n$ itself).
Similarly, an infinite run $\{\rho_n\}_{n\in\Nset_0}$ can be deemed equivalently as an infinite sequence $\{(w_n,\mu_n)\}_{n\in\Nset_0}$ of states such that every $\rho_n$ equals $(w_0,\mu_0)\dots (w_n,\mu_n)$.
For the sake of convenience, we deem each finite run $\rho_0\dots\rho_n$ (resp. infinite run $\{\rho_n\}_{n\in\Nset_0}$) equivalently as $\rho_n$ (resp. its corresponding infinite sequence $\{(w_n,\mu_n)\}_{n\in\Nset_0}$ of states)
in the sequel.
No ambiguity will arise from the underlying context.
\hfill\IEEEQEDclosed
\end{remark}

\begin{remark}\label{rmk:semantics}
Our semantics is different from~\cite{HolgerPOPL}:
we design our probability space over infinite runs, while in~\cite{HolgerPOPL} the probability space is defined directly
over sampled values of sampling variables.
Given the infinite-state MDP, we follow the standard MDP semantics, where each scheduler defines a probability measure.
Thus in our setting, we have only one termination time random variable $\tertime$, but each scheduler $\sigma$ defines
a probability measure $\probm^\sigma_{\mathfrak{c}}$.
In contrast,  in~\cite{HolgerPOPL} there is only one probability measure $\probm$ (generated by samplings for sampling variables)
but many termination time random variables $\tertime^\sigma$ (each corresponding to a scheduler $\sigma$).
\hfill\IEEEQEDclosed
\end{remark}

\section{Termination Questions}\label{sect:termination}

In this section, we define the notions of almost-sure/bounded termination over nondeterministic recursive probabilistic programs. We also discuss some aspects on tail probabilities.

Below we fix a recursive probabilistic program $W$ with its associated CFG in the form (\ref{eq:cfg}) and a sampling function $\Upsilon$.
Recall that we adopt succinct representations for finite and infinite runs (cf. Remark~\ref{rmk:runs}).

\begin{definition}[Termination Time~\cite{HolgerPOPL,DBLP:conf/popl/ChatterjeeFNH16}]
The \emph{termination time} $\tertime$ for $W$ is a random variable on $\infruns$ defined by:
\[
\tertime(\{\left(w_n,\mu_n\right)\}_{n\in\Nset_0}):=\min\left\{n\in\Nset_0\mid w_n=\varepsilon\right\}
\]
for any infinite sequence $\{\left(w_n,\mu_n\right)\}_{n\in\Nset_0}$ of states in $\MDPstates_W$ (as an infinite run),
where $\min\emptyset:=\infty$.
The function $\expvt:\stackelems\rightarrow [0,\infty]$ is defined by
\[
\expvt(\mathfrak{c}):=
\begin{cases}
\sup_{\sigma}\expv^\sigma_{\mathfrak{c}}(\tertime) & \mbox{if }\mathfrak{c}\mbox{ is non-terminal}\\
0 & \mbox{otherwise}
\end{cases}
\]
where $\sigma$ ranges over all schedulers for $W$.
The program $W$ is \emph{almost-surely terminating} from a non-terminal stack element $\mathfrak{c}$ if $\probm^\sigma_{\mathfrak{c}}(\tertime<\infty)=1$ for all schedulers $\sigma$;
and $W$ is \emph{boundedly terminating} from $\mathfrak{c}$ if  $\expvt(\mathfrak{c})<\infty$.\hfill\IEEEQEDclosed
\end{definition}

Thus, $\tertime$ is the random variable which measures the amount of computational steps $W$ takes until termination.

\noindent{\bf Tail (Non-termination) Probabilities.} In this paper, we also focus on tail probabilities
$\sup_\sigma\probm^\sigma_\mathfrak{c}(\tertime>n)$ ($n\in\Nset_0$).
Our motivation is that these probabilities characterize quality of termination.
Algorithms that approximates $\expvt(\mathfrak{c})$ through tail probabilities have already been proposed in
~\cite{DBLP:conf/popl/ChatterjeeFNH16,DBLP:conf/sas/Monniaux01}.






\section{Bounded Termination: Expectation and Probability Bounds}\label{sect:supermartingales}

In this section, we extend the notion of ranking supermartingales to ranking measure functions, and establish the relationship of
ranking measure functions and bounded termination of nondeterministic probabilistic recursive programs.
In detail, we show the following results:
\begin{compactenum}
\item ranking measure functions provide a sound and complete approach for bounded termination;
\item conditionally difference-bounded ranking  measure functions provide lower bounds for expected termination time;
\item efficient bounds on tail probabilities can be ensured by ranking measure functions.
\end{compactenum}
We fix a nondeterministic recursive probabilistic program $W$ together with its associated CFG taking the form~(\ref{eq:cfg}) and a sampling function $\Upsilon$.
We define $\samples:=\supp{\sampdpd}$.

\subsection{Soundness and Completeness}\label{sect:btermination}

We first present the notion of ranking supermartingales.
\begin{definition}[Ranking Supermartingales~\cite{SriramCAV,HolgerPOPL, DBLP:conf/popl/ChatterjeeFNH16}]\label{def:rsupm}
A discrete-time stochastic process $\Gamma=\{X_n\}_{n\in\Nset_0}$ adapted to a filtration $\{\mathcal{F}_n\}_{n\in\Nset_0}$ is a \emph{ranking supermartingale}
if there exists an $\epsilon\in (0,\infty)$ such that for all $n\in\Nset_0$, the following conditions hold:
\begin{compactitem}
\item {\em Integrability Condition.} $\expv(|X_n|)<\infty$;
\item {\em Non-negativity Condition.} it holds a.s. that $X_n\ge 0$;
\item {\em Ranking Condition.} it holds a.s. that
\[
\condexpv{X_{n+1}}{\mathcal{F}_n}\le X_n-\epsilon\cdot\mathbf{1}_{X_n>0}\enskip. \qquad \qquad \qquad \quad \hfill\IEEEQEDclosed
\]
\end{compactitem}
\end{definition}

The following known proposition clarifies the relationship between ranking supermartingales and bounded termination
(detailed proof in Appendix~\ref{app:btermination}).

\begin{proposition}[\cite{HolgerPOPL,DBLP:conf/popl/ChatterjeeFNH16}]\label{prop:rsupm}
Let $\Gamma=\{X_n\}_{n\in\Nset_0}$ be a ranking supermartingale adapted to a filtration $\{\mathcal{F}_n\}_{n\in\Nset_0}$ with $\epsilon$ given as in Definition~\ref{def:rsupm}.
Then $\probm(\stopping{\Gamma}<\infty)=1$ and $\expv(\stopping{\Gamma})\le\frac{\expv(X_0)}{\epsilon}$.\hfill\IEEEQEDclosed
\end{proposition}

We complement Proposition~\ref{prop:rsupm} by an example showing that the Non-negativity condition cannot be dropped.

\begin{example}\label{ex:special:nonnegativity}
In Definition~\ref{def:rsupm}, the Non-negativity condition is necessary; in other words,  it is necessary having  $X_{\stopping{\Gamma}}=0$ rather than $X_{\stopping{\Gamma}}\le 0$ when $\stopping{\Gamma}<\infty$.
This can be observed as follows.
Consider the discrete-time stochastic processes $\{X_n\}_{n\in\Nset_0}$ and $\Gamma=\{Y_n\}_{n\in\Nset_0}$ given as follows:
\begin{compactitem}
\item the random variables $X_0,\dots,X_n,\dots$ are independent, $X_0$ is the random variable with constant value $\frac{1}{2}$ and each $X_n$ ($n\ge 1$) satisfies that $\probm\left(X_n=1\right)=e^{-\frac{1}{n^2}}$ and $\probm\left(X_n=-4\cdot n^2\right)=1-e^{-\frac{1}{n^2}}$;
\item $Y_n:=\sum_{j=0}^{n}X_j$ for $n\ge 0$.
\end{compactitem}
Let the filtration $\{\mathcal{F}_n\}_{n\in\Nset_0}$ be given such that each $\mathcal{F}_n$ is the $\sigma$-algebra generated by $X_0,\dots,X_n$ (i.e., the smallest $\sigma$-algebra that makes  $X_0,\dots,X_n$ measurable).
Then one can show that $\Gamma$ (adapted to $\{\mathcal{F}_n\}_{n\in\Nset_0}$) satisfies Integrability and Ranking conditions, but
$\probm\left(\stopping{\Gamma}=\infty\right)=e^{-\frac{\pi^2}{6}}>0$\enskip.
Detailed justifications are available in Appendix~\ref{app:examples}.
\hfill\IEEEQEDclosed
\end{example}

Now we present the notion of ranking measure functions, which
generalizes ranking supermartingales for nonrecursive probabilistic programs (see~\cite[Definition~8]{DBLP:conf/popl/ChatterjeeFNH16}).




\begin{definition}[Ranking Measure Functions]\label{def:mfunc}
A \emph{ranking measure function} (for $W$) is a function $h:\stackelems\rightarrow [0,\infty]$ satisfying that
there exists an $\epsilon\in (0,\infty)$ such that for all stack elements $(\fn{f},\loc,\nu)$, the following conditions hold:
\begin{compactitem}
\item[(C1)] if $\loc=\lout{\fn{f}}$, then $h(\fn{f},\loc,\nu)=0$;
\item[(C2)] if $\loc\in\alocs{f}\setminus\{\lout{\fn{f}}\}$ and $(\loc, u,\loc')$ is the only triple in $\transitions{\fn{f}}$ with source label $\loc$ and update function $u$, then
\[
\epsilon+\sum_{\mu\in \samples}\sampdpd(\mu)\cdot h\left(\fn{f},\loc',u(\nu,\mu)\right)\le h(\fn{f},\loc,\nu)\enskip;
\]
\item[(C3)] if $\loc\in\flocs{f}\setminus\{\lout{\fn{f}}\}$ and $(\loc,(\fn{g},v),\loc')$ is the only triple in $\transitions{\fn{f}}$ with source label $\loc$ and value-passing function $v$, then
\[
\epsilon+h\left(\fn{g},\lin{\fn{g}}, v(\nu)\right)+h(\fn{f},\loc',\nu)\le h(\fn{f},\loc,\nu);
\]
\item[(C4)] if $\loc\in\clocs{f}\setminus\{\lout{\fn{f}}\}$ and $(\loc, \phi,\loc_1),(\loc, \neg\phi,\loc_2)$ are namely two triples in $\transitions{\fn{f}}$ with source label $\loc$ and propositional arithmetic predicate $\phi$, then
\[
\mathbf{1}_{\nu\models\phi}\cdot h(\fn{f},\loc_1,\nu)+\mathbf{1}_{\nu\models\neg\phi}\cdot h(\fn{f},\loc_2,\nu)+\epsilon\le h(\fn{f},\loc,\nu);
\]
\item[(C5)] if $\loc\in\dlocs{f}\setminus\{\lout{\fn{f}}\}$ and $(\loc, \star,\loc_1),(\loc, \star,\loc_2)$ are namely two triples in $\transitions{\fn{f}}$ with source label $\loc$, then
\[
\max\{h(\fn{f},\loc_1,\nu),h(\fn{f},\loc_2,\nu)\}+\epsilon\le h(\fn{f},\loc,\nu).
\]
\end{compactitem}
($d\cdot\infty:=\infty$ for $d\in(0,\infty]$, $0\cdot\infty:=0$ by convention.)\hfill\IEEEQEDclosed
\end{definition}
Intuitively, a ranking measure function is a function whose (expected) values decrease by a positive stepwise amount $\epsilon$ along the execution of a recursive program.

\begin{remark}\label{rmk:rfunc}
The notion of ranking measure functions is a direct generalization of ranking supermartingales to recursion because once (C3) is omitted, then
Definition~\ref{def:mfunc} coincides with ranking supermartingales
for nonrecursive probabilistic programs
(i.e., \cite[Definition 8]{DBLP:conf/popl/ChatterjeeFNH16}).
\hfill\IEEEQEDclosed
\end{remark}

\begin{remark}\label{rmk:differencersm}
Our notion of ranking supermartingales is standard and different from~\cite{HolgerPOPL} when applied to
 programs in the following sense:
in our setting, ranking supermartingales are applied directly to the set of valuations;
in contrast, in~\cite{HolgerPOPL} ranking supermartingales are applied directly to termination time
random variables.
\hfill\IEEEQEDclosed
\end{remark}

The following lemma establishes the soundness of ranking measure functions for bounded termination (proof in Appendix~\ref{app:btermination}).


\begin{lemma}[Soundness]\label{thm:soundness}
For all ranking measure functions $h$ with $\epsilon$ given in Definition~\ref{def:mfunc} and for all stack elements $\mathfrak{c}=(\fn{f}, \loc, \nu)$, we have
$\expvt(\mathfrak{c})\le \frac{h(\mathfrak{c})}{\epsilon}$.\hfill\IEEEQEDclosed
\end{lemma}

\smallskip\noindent{\em Key Proof Ideas.}
We highlight some important aspects of the proof.
\begin{compactenum}
\item {\em Non-triviality.}
It is known that ranking supermartingales provide a sound approach~\cite{DBLP:conf/popl/ChatterjeeFNH16, HolgerPOPL}.
The key non-trivial aspect is to come up with ranking supermartingales
(from ranking measure functions) for infinite-state MDPs, where the infiniteness
is both due to recursion and countably-many valuations.
\item {\em Key intuition and technical aspects.}
The key intuition is the construction of ranking supermartingales from
ranking measure functions by summing up values taken by ranking measure
functions in an arbitrary configuration.
A key technical aspect is to consider arbitrary configurations rather than
arbitrary valuations.
Moreover, we carefully handle the integrability condition in the proof
(using Dominated and Monotone Convergence Theorem for Lesbegue Integrals)
so that the soundness statement does not require integrability restrictions.
\end{compactenum}

We illustrate our soundness result on the running example.
\begin{example}
Consider again our running example (cf. Example~\ref{ex:runningexample}).
A ranking measure function $h$ for this example is given in Table~\ref{tab:runningexample}.
One can easily verify (through, e.g., distinguishing two cases $n=1$ and $n\ge 2$ for the step from $(\fn{g},2)$ to $(\fn{g},3)$) that $h$ is a ranking measure function with corresponding $\epsilon=1$ (cf. Definition~\ref{def:mfunc}).
It follows that
\[
\expvt\left(\fn{f},1,n\right)\le h(\fn{f},1,n)=\mathbf{1}_{n\ge 1}\cdot(12\cdot n-4)+\mathbf{1}_{n\le 0}\cdot 2
\]
for all $n\in\Zset$. \hfill\IEEEQEDclosed
\end{example}

\begin{table}
\caption{A Ranking Measure Function $h$ for Example~\ref{ex:runningexample}}
\label{tab:runningexample}
\centering
\begin{tabular}{|c|c|}
\hline
Coordinate & Representation\\
\hline
$h(\fn{f},1,n)$ & $\mathbf{1}_{n\ge 1}\cdot(12\cdot n-4)+\mathbf{1}_{n\le 0}\cdot 2$ \\
\hline
$h(\fn{f},2,n)$ & $\mathbf{1}_{n\ge 1}\cdot(12\cdot n-5)+\mathbf{1}_{n\le 0}\cdot \infty$\\
\hline
$h(\fn{f},3,n)$ & $\mathbf{1}_{n\ge 1}\cdot(12\cdot n-6)+\mathbf{1}_{n\le 0}\cdot \infty$\\
\hline
$h(\fn{f},4,n)$ & $\mathbf{1}_{n\ge 2}\cdot(12\cdot \lfloor\frac{n}{2}\rfloor-3)+\mathbf{1}_{n\le 1}\cdot 3$\\
\hline
$h(\fn{f},5,n)$ & $\mathbf{1}_{n\ge 2}\cdot(12\cdot n-13.5)+\mathbf{1}_{n\le 1}\cdot 3$\\
\hline
$h(\fn{f},6,n)$ & $1$ \\
\hline
$h(\fn{f},7,n)$ & $0$ \\
\hline
\hline
Coordinate & Representation\\
\hline
$h(\fn{g},1,n)$ & $\mathbf{1}_{n\ge 1}\cdot(12\cdot n-2.5)+\mathbf{1}_{n\le 0}\cdot 2$ \\
\hline
$h(\fn{g},2,n)$ & $\mathbf{1}_{n\ge 1}\cdot (12\cdot n-3.5)+\mathbf{1}_{n\le 0}\cdot\infty$ \\
\hline
$h(\fn{g},3,n)$ & $\mathbf{1}_{n\ge 1} \cdot(12\cdot n -3)+\mathbf{1}_{n\le 0}\cdot 3$ \\
\hline
$h(\fn{g},4,n)$ & $1$  \\
\hline
$h(\fn{g},5,n)$ & $0$  \\
\hline
\end{tabular}
\end{table}

Below we show the completeness which states that $\expvt$ is a ranking measure function for $W$.

\begin{lemma}[Completeness]\label{thm:completeness}
$\expvt$ is a ranking measure function with corresponding $\epsilon=1$ (cf. Definition~\ref{def:mfunc}).\hfill\IEEEQEDclosed
\end{lemma}

\smallskip\noindent{\em Key Proof Ideas.}
We highlight some important aspects of the proof.
\begin{compactenum}
\item {\em Non-triviality.}
The completeness is quite non-trivial and previous works claimed that it
is not complete for bounded termination even for nonrecursive programs
(see Example~\ref{ex:refutation}).
The non-triviality arises from the handling of schedulers as well as recursion.

\item {\em Key intuition and technical aspects.}
The key intuition is as follows: the function $\expvt$ preserves one-step
properties wrt to arbitrary schedulers as well as recursion.
The key technical aspect to establish the one-step property  involves
representation of termination probabilities and expected termination time
through cylinders.
\end{compactenum}

By combining Lemma~\ref{thm:soundness} and Lemma~\ref{thm:completeness}, we obtain the following theorem.

\begin{theorem}
Ranking measure functions provide a sound and complete approach for bounded termination over
recursive probabilistic programs with nondeterminism.\hfill\IEEEQEDclosed
\end{theorem}

We note that our soundness and completeness results not only apply to bounded termination, but
also provide an upper bound on expected termination time (Lemma~\ref{thm:soundness}).
Below we compare our results with~\cite{HolgerPOPL}.

\begin{example}\label{ex:refutation}
Consider the nonrecursive program (that appear in the second paragraph on Page 2,
right column of~\cite{HolgerPOPL}) depicted in Figure~\ref{fig:refutation}.
In~\cite[Theorem 5.7]{HolgerPOPL}, this program is used as the only counterexample to show that generally
no ranking supermartingale exists even for nonrecursive nondeterministic probabilistic programs for bounded
termination.
Here we present a ranking supermartingale (as a ranking measure function) with $\epsilon=1$
(cf. Definition~\ref{def:mfunc}) for this program in Table~\ref{tab:refutation}.
In the table, the column ``Invariants'' specifies logical formulae at labels that reachable valuations
(from the initial label) satisfy, while ``Coordinate Functions'' presents the part of the ranking measure
functions at corresponding labels restricted to reachable valuations satisfying logical formulae
under ``Invariants'';  valuations not satisfying logical formulae under ``Invariants'' are irrelevant here
(e.g., they can be assigned $\infty$).
Note that although we replace uniform distribution (in the original program) by Bernoulli distribution to fit
our integer setting, the ranking supermartingale given in Table~\ref{tab:refutation} remains to be effective
for the original program as it preserves probability value for the guard $c<0.5$.
Note that the semantics of~\cite{HolgerPOPL} is different from our setting (Remark~\ref{rmk:semantics}),
and the notion of ranking supermartingales is also different when applied to programs (Remark~\ref{rmk:differencersm}).
The counterexample of~\cite{HolgerPOPL} might be valid with additional restrictions,
but as shown in this example the program admits a ranking supermartingale.
\hfill\IEEEQEDclosed
\end{example}

\begin{table}
\caption{The Ranking Supermartingale for Example~\ref{ex:refutation}}
\label{tab:refutation}
\centering
\begin{tabular}{|c|c|c|}
\hline
Label & Invariant & Coordinate Function \\
\hline
$1$  & $\mathsf{true}$ & $19$ \\
\hline
$2$ & $n=0$ & $18$\\
\hline
$3$ & $i=0\wedge n=0$ & $17$ \\
\hline
\multirow{2}{*}{$4$} & \multirow{2}{*}{$i\ge 0\wedge n\ge 0$} & $\mathbf{1}_{c\ge 0.5}\cdot (2\cdot i+2)+$ \\
 & & $\mathbf{1}_{c< 0.5}\cdot (2^{n+1}+2\cdot n+14) $ \\
\hline
$5$ & $i\ge 0\wedge n\ge 0\wedge c< 0.5$ & $2^{n+1}+2\cdot n+13$ \\
\hline
$6$ & $i\ge 0\wedge n\ge 0\wedge c< 0.5$ & $2^{n+1}+2\cdot n+12$ \\
\hline
\multirow{2}{*}{$7$} & \multirow{2}{*}{$i\ge 0\wedge n\ge 0$}  & $\mathbf{1}_{c< 0.5}\cdot(2^{n+2}+2\cdot n+18)+$\\
 & & $\mathbf{1}_{c\ge 0.5}\cdot(2\cdot n+4)$ \\
\hline
$8$ & $i\ge 0\wedge n\ge 0\wedge c< 0.5$ & $2^{n+2}+2\cdot n+17$ \\
\hline
$9$ & $i\ge 0\wedge n\ge 0\wedge c\ge 0.5$ & $2\cdot n+3$ \\
\hline
$10$ & $i\ge 0\wedge n\ge 0 \wedge c< 0.5$ & $2^n+4$ \\
\hline
$11$ & $i\ge 0\wedge n\ge 0 \wedge c< 0.5$ & $2\cdot i+3$\\
\hline
$12$ & $i\ge 0\wedge n\ge 0$ & $2\cdot i+1$ \\
\hline
$13$ & $i\ge 1\wedge n\ge 0$ & $2\cdot i$ \\
\hline
$14$ & $i=0\wedge n\ge 0$ & $0$ \\
\hline
\end{tabular}
\end{table}

\lstset{language=prog}
\lstset{tabsize=3}
\newsavebox{\progrefutation}
\begin{lrbox}{\progrefutation}
\begin{lstlisting}[mathescape]
$1$: $n:= 0$; $2$: $i:=0$; $3$: $c:=0$;
$4$: while $c=0$ do
$5$:   if $\star$ then
$6$:    $c:=\mathrm{Bernoulli}\left(0.5\right)$;
$7$:    if $c=0$ then
$8$:      $n:=n+1$
     else
$9$:      $i := n$
     fi
    else
$10$:    $i:=2^n$; $11$: $c:=1$
    fi
   od;
$12$: while $i > 0$ do $13$: $i := i - 1$ od
$14$:
\end{lstlisting}
\end{lrbox}
\begin{figure}
\centering
\usebox{\progrefutation}
\caption{The Example in~\cite{HolgerPOPL}}
\label{fig:refutation}
\end{figure}

\subsection{Lower Bound for Expected Termination Time}\label{sect:lowerbound}
In this section we show that a subclass of ranking measure functions can be used to derive a lower bound for $\expvt$.
We first show the relationship between conditionally difference-bounded ranking supermartingales (recall definition
from Section~\ref{sect:basicnotations})
and lower bound for expected termination time.


\begin{proposition}\label{prop:cbrsupm}
Consider any conditionally difference-bounded ranking supermartingale $\Gamma=\{X_n\}_{n\in\Nset_0}$ adapted to a filtration $\{\mathcal{F}_n\}_{n\in\Nset_0}$ with $\epsilon$ given as in Definition~\ref{def:rsupm}.
If (i) for every $n\in\Nset_0$, it holds for all $\omega$ that $X_n(\omega)=0$ implies  $X_{n+1}(\omega)=0$, and (ii)
there exists $\delta\in (0,\infty)$ such that for all $n\in\Nset_0$, it holds a.s. that
$\condexpv{X_{n+1}}{\mathcal{F}_n}\ge X_n-\delta\cdot\mathbf{1}_{X_n>0}$~,
then $\expv(\stopping{\Gamma})\ge\frac{\expv(X_0)}{\delta}$.\hfill\IEEEQEDclosed
\end{proposition}

\noindent{\em Key Proof Ideas.}
The key idea is to construct a conditionally difference-bounded submartingale
from $\Gamma$ and apply Optional Stopping Theorem (cf. Theorem~\ref{thm:optstopping} in the appendix).

\begin{example}\label{ex:special:cboundedness}
The conditionally difference-bounded condition in Proposition~\ref{prop:cbrsupm} cannot be dropped.
Consider the family $\{Y_n\}_{n\in\Nset_0}$ of independent random variables defined by:
$Y_0:=3$ and each $Y_n$ ($n\ge 1$) satisfies that
\[
\probm\left(Y_n=2^{n-1}\right)=\frac{1}{2} \text{ and } \probm\left(Y_n=-2^{n-1}-2\right)=\frac{1}{2}.
\]
Let the stochastic process $\Gamma=\{X_n\}_{n\in\Nset_0}$ be inductively defined by: $X_0:=Y_0$ and for all $n\in\Nset_0$,
we have
$X_{n+1}:=\mathbf{1}_{X_n>0}\cdot\left(X_n+Y_{n+1}\right)$.
Let $\{\mathcal{F}_n\}_{n\in\Nset_0}$ be the filtration such that each $\mathcal{F}_n$ is the smallest $\sigma$-algebra that makes all $Y_0,\dots,Y_n$ measurable, so that $\Gamma$ is adapted to $\{\mathcal{F}_n\}_{n\in\Nset_0}$.
Then we obtain that for all $n\in\Nset_0$, we have
$\condexpv{X_{n+1}}{\mathcal{F}_n}-X_n= -\mathbf{1}_{X_n>0}$\enskip.
Moreover, for all $n$ and $\omega$, we have $X_n(\omega)=0\Rightarrow X_{n+1}(\omega)=0$.
However, $\expv(\stopping{\Gamma})=2<\frac{\expv(X_0)}{1}$.
Details are available in Appendix~\ref{app:examples}.
\hfill\IEEEQEDclosed
\end{example}

\noindent{\bf Embedding of Proposition~\ref{prop:cbrsupm} to Recursion.}
Now we embed ranking measure functions into prerequisites
of Proposition~\ref{prop:cbrsupm} so as to obtain the notion of \emph{conditionally difference-bounded ranking measure functions}.
Intuitively, a ranking measure function is conditionally difference-bounded if there exist $\delta,\zeta\in(0,\infty)$ such that the stepwise conditional decrease is between $[\epsilon,\delta]$
($\epsilon$ being
given in Definition~\ref{def:mfunc})  and the stepwise conditional absolute difference is no greater than $\zeta$.
Details are given by Definition~\ref{def:bmfunc} in Appendix~\ref{app:lowerbound}. \hfill\IEEEQEDclosed


The main theorem for this section is as follows.

\begin{theorem}\label{thm:lowerbound}
For any conditionally difference-bounded ranking measure function $h$ with $\delta,\zeta$ given, we have
$\expvt(\mathfrak{c})\ge \frac{h(\mathfrak{c})}{\delta}$ for all stack elements $\mathfrak{c}$ such that $h(\mathfrak{c})<\infty$.\hfill\IEEEQEDclosed
\end{theorem}

\noindent{\em Key Proof Ideas.}
The key proof idea is to use Proposition~\ref{prop:cbrsupm} and embed conditional
difference-boundedness into the proof of Lemma~\ref{thm:soundness}.
The details are in Appendix~\ref{app:lowerbound}.

We now illustrate on our running example.

\begin{example}
Consider our running example (cf. Example~\ref{ex:runningexample}) and the ranking measure function $h$ given in Table~\ref{tab:runningexample}. One can verify easily that $h$ is conditionally
difference-bounded with $\delta=\zeta=13$.
Hence $\expvt\left(\fn{f},1,n\right)\ge \frac{h(\fn{f},1,n)}{13}$.\hfill\IEEEQEDclosed
\end{example}

\begin{remark}
A proof-rule based approach for lower bound on expected termination time has been considered in~\cite{DBLP:conf/lics/OlmedoKKM16},
whereas our approach is completely different and based on ranking supermartingales.\hfill\IEEEQEDclosed
\end{remark}

\subsection{Tail Probabilities}\label{sect:concentration}
In this section we establish the following:
(a)~for difference-bounded ranking measure functions we establish exponential decrease in tail
probabilities;
(b)~we then show if the difference-bounded condition is dropped then the optimal bound for
tail probabilities is obtained by Markov's inequality.
The following result is known~\cite{DBLP:conf/popl/ChatterjeeFNH16} and the extension from ranking supermartingales
to ranking measure functions is as follows: by an embedding similar to conditional difference-boundedness in Section~\ref{sect:lowerbound},
one can restrict ranking measure functions to be difference-bounded and apply Theorem~\ref{thm:concentration};
since this embedding is technical, we put the details (cf. Definition~\ref{def:dbmfunc} and Theorem~\ref{thm:dbmfunc})
in Appendix~\ref{app:concentration}.

\begin{theorem}[\cite{DBLP:conf/popl/ChatterjeeFNH16}]\label{thm:concentration}
Consider any difference-bounded ranking supermartingale $\Gamma=\{X_n\}_{n\in\Nset_0}$ adapted to a filtration $\{\mathcal{F}_n\}_{n\in\Nset_0}$ with $\epsilon$ given in Definition~\ref{def:rsupm}. If (i) $X_0$ is a constant random variable and (ii) for all $n\in\Nset_0$ and $\omega$, $X_n(\omega)=0$ implies $X_{n+1}(\omega)=0$.
Then for all natural numbers $n>\frac{\expv(X_0)}{\epsilon}$,
\[
\probm(\stopping{\Gamma}>n)\le e^{-\frac{(\epsilon\cdot n-\expv(X_0))^2}{2\cdot n\cdot (c+\epsilon)^2}}\le e^{\frac{\epsilon\cdot \expv(X_0)}{(c+\epsilon)^2}}\cdot e^{-\frac{\epsilon^2}{2\cdot{(c+\epsilon)}^2}\cdot n},
\]
where $c\in (0,\infty)$ is any number satisfying  that $|X_{n+1}-X_n|\le c$ a.s. for all $n\in\Nset_0$.\hfill\IEEEQEDclosed
\end{theorem}

We now present an example to show the importance of the difference-boundedness condition.

\begin{example}\label{ex:special:noconcentration}
In general, the difference-boundedness condition cannot be dropped in Theorem~\ref{thm:concentration}.
Fix any $\alpha\in (1,\infty)$ and consider the family $\{Y_n\}_{n\in\Nset_0}$ of independent random variables defined as follows:
$Y_0:=3$ and each $Y_n$ ($n\ge 1$) satisfies that
\[
\probm\left(Y_n\!=\!2\right)=\frac{n^\alpha}{(n+1)^\alpha}; \ \ \probm\left(Y_n\!=\!-2\cdot n\!-1\right)=1-\frac{n^\alpha}{(n+1)^\alpha}.
\]
Let the stochastic process $\Gamma=\{X_n\}_{n\in\Nset_0}$ be inductively defined by: $X_0:=Y_0$ and for all $n\in\Nset_0$, we have
$X_{n+1}:=\mathbf{1}_{X_n>0}\cdot\left(X_n+Y_{n+1}\right)$.
Let $\{\mathcal{F}_n\}_{n\in\Nset_0}$ be the filtration such that each $\mathcal{F}_n$ is the smallest $\sigma$-algebra that makes all $Y_0,\dots,Y_n$ measurable.
Then we obtain that for all $n\in\Nset_0$, we have
\[
\condexpv{X_{n+1}}{\mathcal{F}_n}-X_n \le -(2\cdot\alpha - 2)\cdot {\left(\frac{1}{2}\right)}^\alpha\cdot \mathbf{1}_{X_n>0} \enskip.
\]
Hence $\{X_n\}_{n\in\Nset_0}$ is a ranking supermartingale and for all $n$ and $\omega$, $X_n(\omega)=0$ implies $X_{n+1}(\omega)=0$.
However, for $n\ge 1$, we have
$\probm(\stopping{\Gamma}> n)=\frac{1}{{(n+1)}^\alpha}$\enskip.
Thus $\{X_n\}_{n\in\Nset_0}$ does not admit exponential decrease of tail probabilities.
See Appendix~\ref{app:examples} for more details.
\hfill\IEEEQEDclosed
\end{example}

Below we show that when a ranking supermartingale is not difference bounded, a direct application of Markov's Inequality and Proposition~\ref{prop:rsupm} gives optimal bounds on tail probabilities.

\begin{theorem}\label{thm:nonbrsupm}
For any ranking supermartingale $\Gamma=\{X_n\}_{n\in\Nset_0}$ with $\epsilon$ given in Definition~\ref{def:rsupm},
\[
\probm(\stopping{\Gamma}\ge k)\le \frac{\expv(\stopping{\Gamma})}{k}\le \frac{\expv(X_0)}{\epsilon\cdot k}
\]
for all $k\in\Nset$. \hfill\IEEEQEDclosed
\end{theorem}

By Example~\ref{ex:special:noconcentration}, the bound in Theorem~\ref{thm:nonbrsupm} is asymptotically optimal and
can be directly applied to ranking measure functions.


\section{Almost-Sure Termination}\label{sect:astermination}

In this section, we present a sound approach for almost-sure termination (i.e., not necessarily bounded termination)
of nondeterministic recursive probabilistic programs along with efficient bounds on tail probabilities.
We first demonstrate the relationships between supermartingales and almost-sure termination/tail probabilities.
We start with difference-bounded supermartingales, then general supermartingales.

\begin{theorem}\label{thm:supm}
Consider any difference-bounded supermartingale $\Gamma=\{X_n\}_{n\in\Nset_0}$ adapted to a filtration $\{\mathcal{F}_n\}_{n\in\Nset_0}$ satisfying the following conditions:
\begin{compactenum}
\item $X_0$ is a constant random variable;
\item for all $n\in\Nset_0$, it holds for all $\omega$ that (i) $X_n(\omega)\ge 0$ and (ii) $X_n(\omega)=0$ implies  $X_{n+1}(\omega)=0$;
\item {\em Lower Bound on Conditional Absolute Difference.} there exists $\delta\in(0,\infty)$ such that for all $n\in\Nset_0$, it holds a.s. that $X_n>0$ implies $\condexpv{|X_{n+1}-X_n|}{\mathcal{F}_n}\ge \delta$.
\end{compactenum}
Then $\probm(\stopping{\Gamma}<\infty)=1$ and $k\mapsto\probm\left(\stopping{\Gamma}\ge k\right)\in \mathcal{O}\left(\frac{1}{\sqrt{k}}\right)$.\hfill\IEEEQEDclosed
\end{theorem}

\noindent{\em Key Proof Ideas.} The main idea is a thorough analysis of the
martingale
\[
Y_n:=\frac{e^{-t\cdot X_n}}{\prod_{j=0}^{n-1} \condexpv{e^{-t\cdot \left(X_{j+1}-X_{j}\right)}}{\mathcal{F}_j}} ~~(n\in\Nset_0)
\]
for some sufficiently small $t>0$ and its limit through Optional Stopping Theorem (cf. Theorem~\ref{thm:optstopping} in the appendix).
The details are in Appendix~\ref{app:astermination}.

We now show an application on symmetric random walk.

\begin{example}\label{ex:special:supmoptimal}
Consider the family $\{Y_n\}_{n\in\Nset_0}$ of independent random variables defined as follows:
$Y_0:=1$ and each $Y_n$ ($n\ge 1$) satisfies 
\[
\probm\left(Y_n=1\right)=\frac{1}{2} \text { and } \probm\left(Y_n=-1\right)=\frac{1}{2}.
\]
Let the stochastic process $\Gamma=\{X_n\}_{n\in\Nset_0}$ be inductively defined by: $X_0:=Y_0$ and for all $n\in\Nset_0$, we have
$X_{n+1}:=\mathbf{1}_{X_n>0}\cdot\left(X_n+Y_{n+1}\right)$.
Choose the filtration $\{\mathcal{F}_n\}_{n\in\Nset_0}$ such that every $\mathcal{F}_n$ is the smallest $\sigma$-algebra that makes $Y_0,\dots, Y_n$ measurable.
Then $\Gamma$ models the classical symmetric random walk and $X_n>0$ implies $\condexpv{|X_{n+1}-X_n|}{\mathcal{F}_n}=1$ a.s.
From Theorem~\ref{thm:supm}, we obtain that $\probm(\stopping{\Gamma}<\infty)=1$ and $k\mapsto\probm\left(\stopping{\Gamma}\ge k\right)\in \mathcal{O}\left(\frac{1}{\sqrt{k}}\right)$.
In~\cite[Theorem 4.1]{DBLP:journals/jcss/BrazdilKKV15}, it is shown (in the context of pBPA)
that $k\mapsto\probm\left(\stopping{\Gamma}\ge k\right)\in \Omega\left(\frac{1}{\sqrt{k}}\right)$.
Hence the tail bound in Theorem~\ref{thm:supm} is optimal. See Appendix~\ref{app:examples} for details. \hfill\IEEEQEDclosed
\end{example}

In general, the third item in Theorem~\ref{thm:supm} cannot be relaxed.
To clarify this fact, we first show the following result which states that having $\condexpv{|X_{n+1}-X_n|}{\mathcal{F}_n}$ to be zero leads to almost-sure non-termination over martingales (proof in Appendix~\ref{app:astermination}).

\begin{proposition}\label{prop:varcezero}
Let $\Gamma=\{X_n\}_{n\in\Nset_0}$ be a martingale adapted to a filtration $\{\mathcal{F}_n\}_{n\in\Nset_0}$ such that $X_0>0$ and $\condexpv{|X_{n+1}-X_n|}{\mathcal{F}_n}=0$ a.s. for all $n\in\Nset_0$.
Then $\probm\left(\stopping{\Gamma}=\infty\right)=1$. \hfill\IEEEQEDclosed
\end{proposition}

Furthermore, the third item in Theorem~\ref{thm:supm} cannot be relaxed to positivity (i.e., not necessarily bounded from below). The following example illustrates this point.

\begin{example}\label{ex:special:positivity}
Consider the discrete-time stochastic process $\{X_n\}_{n\in\Nset_0}$ such that all $X_n$'s are independent, $X_0=1$ and every $X_n$ ($n\ge 1$) observes the two-point distribution such that
\[
\probm\left(X_n=2^{-n+1}\right)=\probm\left(X_n=-2^{-n+1}\right)=\frac{1}{2}.
\]
Choose the filtration $\{\mathcal{F}_n\}_{n\in\Nset_0}$ such that every $\mathcal{F}_n$ is the smallest $\sigma$-algebra that makes $X_0,\dots, X_n$ measurable.
Let the stochastic process $\Gamma=\{Y_n\}_{n\in\Nset_0}$ be inductively defined by: $Y_0:=X_0$ and for all $n\in\Nset_0$,
$Y_{n+1}:=\mathbf{1}_{Y_n>0}\cdot\left(Y_n+X_{n+1}\right)$.
Then $\condexpv{\left|Y_{n+1}-Y_n\right|}{\mathcal{F}_n} = 2^{-n}\cdot \mathbf{1}_{Y_n>0}$ a.s.
However, $\probm\left(\stopping{\Gamma}=\infty\right)=\frac{1}{2}$ as whether $\stopping{\Gamma}=\infty$ or not relies only on $X_1$.
(cf. Appendix~\ref{app:examples} for more details)\hfill\IEEEQEDclosed
\end{example}

Now we extend Theorem~\ref{thm:supm} to general supermartingales with a weaker bound over tail probabilities.

\begin{theorem}\label{thm:supmextended}
Consider any supermartingale $\Gamma=\{X_n\}_{n\in\Nset_0}$ adapted to a filtration $\{\mathcal{F}_n\}_{n\in\Nset_0}$ satisfying the following conditions:
\begin{compactenum}
\item $X_0$ is a constant random variable;
\item for all $n\in\Nset_0$, it holds for all $\omega$ that (i) $X_n(\omega)\ge 0$ and (ii) $X_n(\omega)=0$ implies  $X_{n+1}(\omega)=0$;
\item there exists $\delta\in(0,\infty)$ such that for all $n\in\Nset_0$, it holds that a.s. $\condexpv{|X_{n+1}-X_n|}{\mathcal{F}_n}\ge \delta\cdot\mathbf{1}_{X_n>0}$.
\end{compactenum}
Then $\probm(\stopping{\Gamma}<\infty)=1$ and $k\mapsto\probm\left(\stopping{\Gamma}\ge k\right)\in \mathcal{O}\left(k^{-\frac{1}{6}}\right)$.\hfill\IEEEQEDclosed
\end{theorem}

\noindent{\em Key Proof Ideas.} The key idea is to extend the proof of Theorem~\ref{thm:supm} with the stopping times $R_M$'s ($M\in (0,\infty)$) defined by
\[
R_M(\omega):=  \min\{n\mid X_{n}(\omega)\le 0\mbox{ or } X_{n}(\omega)\ge M\}\enskip.
\]
We first show that each $R_M$ is almost-surely finite.
Then we choose appropriate $M$ for each $k$ such that the tail bounds can be derived.

\begin{remark}
We note that Theorem~\ref{thm:supm} and~\ref{thm:supmextended} are for supermartingales, not necessarily ranking supermartingales. \hfill\IEEEQEDclosed
\end{remark}

\smallskip\noindent{\em Towards Application to Probabilistic Programs.}
Below we apply Theorem~\ref{thm:supm} and Theorem~\ref{thm:supmextended} to almost-sure termination of probabilistic recursive programs
with nondeterminism.
A major obstacle is that statements without sampling in programs lead typically to zero change over valuations of program variables,
which leads to zero conditional difference.
To tackle this problem, we extend Theorem~\ref{thm:supm} as follows (proof in Appendix~\ref{app:astermination}).
Informally, Lemma~\ref{thm:supm2} allows zero change over valuations by imposing non-increasing condition and repeated visit to positive-change situations.

\begin{lemma}\label{thm:supm2}
Consider any difference-bounded supermartingale $\Gamma=\{X_n\}_{n\in\Nset_0}$ adapted to a filtration $\{\mathcal{F}_n\}_{n\in\Nset_0}$ satisfying that there exist $\delta\in(0,\infty)$ and $K\in\Nset$ such that the following conditions hold:
\begin{compactenum}
\item $X_0$ is a constant random variable;
\item for every $n\in\Nset_0$,  it holds for all $\omega$ that (i) $X_n(\omega)\ge 0$ and (ii) $X_n(\omega)=0$ implies  $X_{n+1}(\omega)=0$;
\item for every $n\in\Nset_0$, it holds a.s. that either $X_{n+1}\le X_n$ or $\condexpv{|X_{n+1}-X_n|}{\mathcal{F}_n}\ge \delta$;
\item for every $n\in\Nset_0$ it holds a.s. that there exists an $k$ such that (i) $n\le k< n+K$ and (ii) either $X_k=0$ or $\condexpv{|X_{k+1}-X_k|}{\mathcal{F}_k}\ge \delta$.
\end{compactenum}
Then $\probm(\stopping{\Gamma}<\infty)=1$ and $k\mapsto\probm\left(\stopping{\Gamma}\ge k\right)\in \mathcal{O}\left(\frac{1}{\sqrt{k}}\right)$. \hfill\IEEEQEDclosed
\end{lemma}

Similarly, Theorem~\ref{thm:supmextended} can be extended to handle zero change over valuations (cf. Lemma~\ref{thm:supm3} in Appendix~\ref{app:astermination}).

Below we fix a 
program $W$ together with its CFG taking the form~(\ref{eq:cfg}) and a sampling function $\Upsilon$.

\smallskip\noindent{\bf Super-measure Functions.}
To apply Lemma~\ref{thm:supm2}, we introduce the notion of super-measure functions which is similar to ranking measure
functions with constraints for difference-bounded supermartingales and lower bounds on conditional absolute difference.
In extra, a super-measure function fulfills Item 3) of Lemma~\ref{thm:supm2} in the way that ``$X_{n+1}\le X_n$'' is fulfilled by non-assignment statements
and ``$\condexpv{|X_{n+1}-X_n|}{\mathcal{F}_n}\ge \delta$'' by assignment statements.
For details see Definition~\ref{def:supmfunc} in Appendix~\ref{app:astermination}.



To apply Lemma~\ref{thm:supm2}, we also need to fulfill Item 4) of Lemma~\ref{thm:supm2}.
This is done by a set $\Theta_{m^*}$ of pairs $(\fn{f},\loc)$ that can reach some assignment statement within a bounded number of steps (cf. Appendix~\ref{app:astermination} for details).
Now we apply Lemma~\ref{thm:supm2} and obtain the following corollary (proof in Appendix~\ref{app:astermination}).
Informally, Corollary~\ref{thm:supmfunc} says that if (i) every label in $W$ can lead to some assignment label or termination within a bounded number of steps and (ii) there is a super-measure function for $W$, then $W$ terminates a.s.

\begin{corollary}\label{thm:supmfunc}
If it holds that (i) $(\fn{f},\loc)\in \Theta_{m^*}$ for all $\fn{f}\in\fnames,\loc\in\locs{f}$ and (ii) there exists a super-measure function $h$ (for $W$),
then $\probm_\mathfrak{c}^\sigma (T<\infty)=1$ and $k\mapsto\probm_\mathfrak{c}^\sigma\left(T\ge k\right)\in \mathcal{O}\left(\frac{1}{\sqrt{k}}\right)$
for all schedulers $\sigma$ and non-terminal stack elements $\mathfrak{c}$ such that $h(\mathfrak{c})\in (0,\infty)$.\hfill\IEEEQEDclosed
\end{corollary}

\noindent{\em Key Proof Ideas.}
The proof is similar to the one for Lemma~\ref{thm:soundness}, however with supermartingales rather than ranking supermartingales,
and uses Lemma~\ref{thm:supm2} for almost-sure termination and bounds on tail probabilities.

\lstset{language=prog}
\lstset{tabsize=3}
\newsavebox{\progasterminationa}
\begin{lrbox}{\progasterminationa}
\begin{lstlisting}[mathescape]
$\mathsf{f}(n)~\{$
$1$: if $n\ge 1$ then
$2$:   $n:=n+r$;
$3$:   $\fn{f}(n)$
   else
$4$:   skip
   fi
$5$: $\}$
\end{lstlisting}
\end{lrbox}

\lstset{language=prog}
\lstset{tabsize=3}
\newsavebox{\progasterminationb}
\begin{lrbox}{\progasterminationb}
\begin{lstlisting}[mathescape]
|
|$\mathsf{g}(n)~\{$
|$1$: while $n\ge 1$ do
|$2$:   $n:=n+r$;
|   od
|$3$: $\}$
|
|
\end{lstlisting}
\end{lrbox}

\begin{figure}
\centering
\usebox{\progasterminationa}
\usebox{\progasterminationb}
\caption{Classical Random Walk}
\label{fig:astermination}
\end{figure}

\begin{example}
Consider the classical symmetric random walk in~\cite[Chapter~10.12]{probabilitycambridge} implemented by the programs in Figure~\ref{fig:astermination}, one with recursion and the other without recursion, where the sampling variable $r$ samples values from the probability distribution $q$ such that $q(-1)=q(1)=\frac{1}{2}$.
A super-measure function is illustrated in Table~\ref{tab:supmfunc}.
By Corollary~\ref{thm:supmfunc}, it holds that both the programs in Figure~\ref{fig:astermination} terminates almost-surely under any initial stack element with tail probabilities bounded by reciprocal of square root of the thresholds.\hfill\IEEEQEDclosed
\end{example}

\begin{table}
\caption{A Super-Measure Function $h$ for Figure~\ref{fig:astermination}}
\label{tab:supmfunc}
\centering
\begin{tabular}{|c|c|}
\hline
Coordinate & Representation\\
\hline
$h(\fn{f},1,n)$ & $\mathbf{1}_{n\ge 1}\cdot(1+n)+\mathbf{1}_{n\le 0}\cdot 1$ \\
\hline
$h(\fn{f},2,n)$ & $\mathbf{1}_{n\ge 1}\cdot(1+n)+\mathbf{1}_{n\le 0}\cdot \infty$\\
\hline
$h(\fn{f},3,n)$ & $\mathbf{1}_{n\ge 1}\cdot(1+n)+\mathbf{1}_{n\le 0}\cdot 1$\\
\hline
$h(\fn{f},4,n)$ & $1$\\
\hline
$h(\fn{f},5,n)$ & $0$\\
\hline
\hline
Coordinate & Representation\\
\hline
$h(\fn{g},1,n)$ & $\mathbf{1}_{n\ge 1}\cdot(1+n)+\mathbf{1}_{n\le 0}\cdot 1$ \\
\hline
$h(\fn{g},2,n)$ & $\mathbf{1}_{n\ge 1}\cdot(1+n)+\mathbf{1}_{n\le 0}\cdot \infty$ \\
\hline
$h(\fn{g},3,n)$ & $0$ \\
\hline
\end{tabular}
\end{table}

\begin{remark}
Theorem~\ref{thm:supmextended} can be embedded to recursive programs in a way similar to Theorem~\ref{thm:supm} and Corollary~\ref{thm:supmfunc}, where the only differences are that (i) difference-boundedness is not required and (ii) tail bound is $\mathcal{O}(k^{-\frac{1}{6}})$. \hfill\IEEEQEDclosed
\end{remark}

\begin{remark}
A recent independent work~\cite{MclverMorgan2016} has also considered supermartingale approach for almost-sure
termination, but our work present explicit bounds on tail probabilities along with almost-sure termination.\hfill\IEEEQEDclosed
\end{remark}

\section{Conclusion and Future Work}

In this work we studied termination of nondeterministic probabilistic recursive programs with integer-valued variables.
We first presented a sound and complete approach for bounded termination through ranking supermartingales, and 
 with additional restrictions a sound approach for lower bound on
expected termination time.
Then, we demonstrated a sound approach for almost-sure termination through supermartingales. 
Finally, we proved efficient bounds on tail probabilities and 
provided several illuminating counterexamples to establish the necessity of some important
prerequisites.
For simplicity, we focussed on integer-valued variables that leads to
countable-state-space MDPs.
A future work for the fuller version 
would be to extend our results to
real-valued variables. 
A second direction is more practical.
For ranking supermartingales algorithmic approaches exist for subclasses~\cite{DBLP:conf/popl/ChatterjeeFNH16,DBLP:conf/cav/ChatterjeeFG16}.
The algorithmic study of subclasses of ranking measure functions is another interesting
direction. 

{\scriptsize
\bibliographystyle{IEEEtran}
\bibliography{PL}

\begin{thebibliography}{10}
\providecommand{\url}[1]{#1}
\csname url@samestyle\endcsname
\providecommand{\newblock}{\relax}
\providecommand{\bibinfo}[2]{#2}
\providecommand{\BIBentrySTDinterwordspacing}{\spaceskip=0pt\relax}
\providecommand{\BIBentryALTinterwordstretchfactor}{4}
\providecommand{\BIBentryALTinterwordspacing}{\spaceskip=\fontdimen2\font plus
\BIBentryALTinterwordstretchfactor\fontdimen3\font minus
  \fontdimen4\font\relax}
\providecommand{\BIBforeignlanguage}[2]{{%
\expandafter\ifx\csname l@#1\endcsname\relax
\typeout{** WARNING: IEEEtran.bst: No hyphenation pattern has been}%
\typeout{** loaded for the language `#1'. Using the pattern for}%
\typeout{** the default language instead.}%
\else
\language=\csname l@#1\endcsname
\fi
#2}}
\providecommand{\BIBdecl}{\relax}
\BIBdecl

\bibitem{BaierBook}
C.~Baier and J.-P. Katoen, \emph{Principles of model checking}.\hskip 1em plus
  0.5em minus 0.4em\relax MIT Press, 2008.

\bibitem{prism}
M.~Z. Kwiatkowska, G.~Norman, and D.~Parker, ``Prism 4.0: Verification of
  probabilistic real-time systems,'' in \emph{CAV}, ser. LNCS 6806, 2011, pp.
  585--591.

\bibitem{kaelbling1998planning}
L.~P. Kaelbling, M.~L. Littman, and A.~R. Cassandra, ``Planning and acting in
  partially observable stochastic domains,'' \emph{Artificial intelligence},
  vol. 101, no.~1, pp. 99--134, 1998.

\bibitem{Durrett}
R.~Durrett, \emph{Probability: Theory and Examples (Second Edition)}.\hskip 1em
  plus 0.5em minus 0.4em\relax Duxbury Press, 1996.

\bibitem{Howard}
H.~Howard, \emph{Dynamic Programming and {Markov} Processes}.\hskip 1em plus
  0.5em minus 0.4em\relax MIT Press, 1960.

\bibitem{Kemeny}
J.~Kemeny, J.~Snell, and A.~Knapp, \emph{Denumerable {Markov} Chains}.\hskip
  1em plus 0.5em minus 0.4em\relax D. Van Nostrand Company, 1966.

\bibitem{Rabin63}
M.~Rabin, ``Probabilistic automata,'' \emph{Information and Control}, vol.~6,
  pp. 230--245, 1963.

\bibitem{PazBook}
A.~Paz, \emph{Introduction to probabilistic automata (Computer science and
  applied mathematics)}.\hskip 1em plus 0.5em minus 0.4em\relax Academic Press,
  1971.

\bibitem{LearningSurvey}
L.~P. Kaelbling, M.~L. Littman, and A.~W. Moore, ``Reinforcement learning: A
  survey,'' \emph{JAIR}, vol.~4, pp. 237--285, 1996.

\bibitem{SriramCAV}
A.~Chakarov and S.~Sankaranarayanan, ``Probabilistic program analysis with
  martingales,'' in \emph{CAV}, 2013, pp. 511--526.

\bibitem{HolgerPOPL}
L.~M.~F. Fioriti and H.~Hermanns, ``Probabilistic termination: Soundness,
  completeness, and compositionality,'' in \emph{POPL}, 2015, pp. 489--501.

\bibitem{SumitPLDI}
S.~Sankaranarayanan, A.~Chakarov, and S.~Gulwani, ``Static analysis for
  probabilistic programs: inferring whole program properties from finitely many
  paths,'' in \emph{PLDI}, 2013, pp. 447--458.

\bibitem{EGK12}
J.~Esparza, A.~Gaiser, and S.~Kiefer, ``Proving termination of probabilistic
  programs using patterns,'' in \emph{CAV}, 2012, pp. 123--138.

\bibitem{DBLP:conf/popl/ChatterjeeFNH16}
K.~Chatterjee, H.~Fu, P.~Novotn{\'{y}}, and R.~Hasheminezhad, ``Algorithmic
  analysis of qualitative and quantitative termination problems for affine
  probabilistic programs,'' in \emph{POPL}, 2016, pp. 327--342.

\bibitem{rwfloyd1967programs}
R.~W. Floyd, ``Assigning meanings to programs,'' \emph{Mathematical Aspects of
  Computer Science}, vol.~19, pp. 19--33, 1967.

\bibitem{DBLP:conf/cav/BradleyMS05}
A.~R. Bradley, Z.~Manna, and H.~B. Sipma, ``Linear ranking with reachability,''
  in \emph{CAV}, 2005, pp. 491--504.

\bibitem{DBLP:conf/tacas/ColonS01}
M.~Col{\'{o}}n and H.~Sipma, ``Synthesis of linear ranking functions,'' in
  \emph{TACAS}, 2001, pp. 67--81.

\bibitem{DBLP:conf/vmcai/PodelskiR04}
A.~Podelski and A.~Rybalchenko, ``A complete method for the synthesis of linear
  ranking functions,'' in \emph{VMCAI}, 2004, pp. 239--251.

\bibitem{DBLP:conf/pods/SohnG91}
K.~Sohn and A.~V. Gelder, ``Termination detection in logic programs using
  argument sizes,'' in \emph{PODS}, 1991, pp. 216--226.

\bibitem{MM04}
A.~McIver and C.~Morgan, ``Developing and reasoning about probabilistic
  programs in \emph{pGCL},'' in \emph{PSSE}, 2004, pp. 123--155.

\bibitem{MM05}
------, \emph{Abstraction, Refinement and Proof for Probabilistic Systems},
  ser. Monographs in Computer Science.\hskip 1em plus 0.5em minus 0.4em\relax
  Springer, 2005.

\bibitem{BG05}
O.~Bournez and F.~Garnier, ``Proving positive almost-sure termination,'' in
  \emph{RTA}, 2005, pp. 323--337.

\bibitem{Foster53}
F.~G. Foster, ``On the stochastic matrices associated with certain queuing
  processes,'' \emph{The Annals of Mathematical Statistics}, vol.~24, no.~3,
  pp. 355--360, 1953.

\bibitem{ChatterjeeNZ2017}
K.~Chatterjee, P.~Novotn\'{y}, and {\DJ}.~\v{Z}ikeli\'{c}, ``Stochastic
  invariants for probabilistic termination,'' in \emph{POPL}, 2017, to appear.

\bibitem{DBLP:journals/jacm/EtessamiY15}
K.~Etessami and M.~Yannakakis, ``Recursive {M}arkov decision processes and
  recursive stochastic games,'' \emph{J. {ACM}}, vol.~62, no.~2, pp.
  11:1--11:69, 2015.

\bibitem{DBLP:journals/iandc/BrazdilBFK08}
T.~Br{\'{a}}zdil, V.~Brozek, V.~Forejt, and A.~Kucera, ``Reachability in
  recursive {M}arkov decision processes,'' \emph{Inf. Comput.}, vol. 206,
  no.~5, pp. 520--537, 2008.

\bibitem{DBLP:journals/jcss/BrazdilKKV15}
T.~Br{\'{a}}zdil, S.~Kiefer, A.~Kucera, and I.~H. Varekov{\'{a}}, ``Runtime
  analysis of probabilistic programs with unbounded recursion,'' \emph{J.
  Comput. Syst. Sci.}, vol.~81, no.~1, pp. 288--310, 2015.

\bibitem{DBLP:conf/lics/OlmedoKKM16}
F.~Olmedo, B.~L. Kaminski, J.-P. Katoen, and C.~Matheja, ``Reasoning about
  recursive probabilistic programs,'' in \emph{LICS}, 2016, pp. 672--681.

\bibitem{probabilitycambridge}
D.~Williams, \emph{{P}robability with {M}artingales}.\hskip 1em plus 0.5em
  minus 0.4em\relax Cambridge University Press, 1991.

\bibitem{DBLP:conf/sas/Monniaux01}
D.~Monniaux, ``An abstract analysis of the probabilistic termination of
  programs,'' in \emph{SAS}, 2001, pp. 111--126.

\bibitem{MclverMorgan2016}
\BIBentryALTinterwordspacing
A.~McIver and C.~Morgan, ``A new rule for almost-certain termination of
  probabilistic and demonic programs,'' \emph{arXiv}, 2016. [Online].
  Available: \url{https://arxiv.org/abs/1612.01091}
\BIBentrySTDinterwordspacing

\bibitem{DBLP:conf/cav/ChatterjeeFG16}
K.~Chatterjee, H.~Fu, and A.~K. Goharshady, ``Termination analysis of
  probabilistic programs through {P}ositivstellensatz's,'' in \emph{CAV}, 2016,
  pp. 3--22.

\bibitem{Azuma1967inequality}
K.~Azuma, ``Weighted sums of certain dependent random variables,'' \emph{Tohoku
  Mathematical Journal}, vol.~19, no.~3, pp. 357--367, 1967.

\end{thebibliography}
}

\clearpage
\appendices

\section{Properties for Conditional Expectation}\label{app:condexpv}

Conditional expectation has the following properties for any random variables $X,Y$ and $\{X_n\}_{n\in\Nset_0}$ (from a same probability space) satisfying $\expv(|X|)<\infty,\expv(|Y|)<\infty, \expv(|X_n|)<\infty$ ($n\ge 0$) and any suitable sub-$\sigma$-algebras $\mathcal{G},\mathcal{H}$:
\begin{compactitem}
\item[(E4)] $\expv\left(\condexpv{X}{\mathcal{G}}\right)=\expv(X)$ ;
\item[(E5)] if $X$ is $\mathcal{G}$-measurable, then $\condexpv{X}{\mathcal{G}}=X$ a.s.;
\item[(E6)] for any real constants $b,d$,
\[
\condexpv{b\cdot X+d\cdot Y}{G}=b\cdot\condexpv{X}{G}+d\cdot \condexpv{Y}{G}\mbox{ a.s.;}
\]
\item[(E7)] if $\mathcal{H}\subseteq\mathcal{G}$, then $\condexpv{\condexpv{X}{\mathcal{G}}}{\mathcal{H}}=\condexpv{X}{\mathcal{H}}$ a.s.;
\item[(E8)] if $Y$ is $\mathcal{G}$-measurable and $\expv(|Y|)<\infty$, $\expv(|Y\cdot X|)<\infty$,
then
\[
\condexpv{Y\cdot X}{\mathcal{G}}=Y\cdot\condexpv{X}{\mathcal{G}}\mbox{ a.s.;}
\]
\item[(E9)] if $X$ is independent of $\mathcal{H}$, then $\condexpv{X}{\mathcal{H}}=\expv(X)$ a.s., where $\expv(X)$ here is deemed as the random variable with constant value $\expv(X)$;
\item[(E10)] if it holds a.s that $X\ge 0$, then $\condexpv{X}{\mathcal{G}}\ge 0$ a.s.;
\item[(E11)] if it holds a.s. that (i) $X_n\ge 0$ and $X_n\le X_{n+1}$ for all $n$ and (ii) $\lim\limits_{n\rightarrow\infty}X_n=X$, then
\[
\lim\limits_{n\rightarrow\infty}\condexpv{X_n}{\mathcal{G}}=\condexpv{X}{\mathcal{G}}\mbox{ a.s.}
\]
\item[(E12)] if (i) $|X_n|\le Y$ for all $n$ and (ii) $\lim\limits_{n\rightarrow\infty} X_n=X$, then
\[
\lim\limits_{n\rightarrow\infty}\condexpv{X_n}{\mathcal{G}}=\condexpv{X}{\mathcal{G}}\mbox{ a.s.}
\]
\item[(E13)] if $g:\Rset\rightarrow\Rset$ is a convex function and $\expv(|g(X)|)<\infty$, then $g(\condexpv{X}{\mathcal{G}})\le \condexpv{g(X)}{\mathcal{G}}$ a.s.
\end{compactitem}
We refer to~\cite[Chapter~9]{probabilitycambridge} for more details.

\section{Detailed Syntax}\label{app:syntax}

In the sequel, we fix two countable sets of \emph{program variables} and \emph{sampling variables}.
We also fix a countable set of \emph{function names}.
W.l.o.g, these three sets are pairwise disjoint.

Informally, program variables are variables that are directly related to the control-flow of a program, while sampling variables reflect randomized inputs to the program.
Every program variable holds an integer upon instantiation,
while every sampling variable is bound to a discrete probability distribution (cf. Section~\ref{sect:semantics}).

\noindent{\bf The Syntax.} The syntax of our recursive programs is illustrated by the grammar in Figure~\ref{fig:syntax}.
Below we explain the grammar.
\begin{compactitem}
\item \emph{Variables.} Expressions $\langle\mathit{pvar}\rangle$ (resp. $\langle\mathit{rvar}\rangle$) range over program (resp. sampling) variables.
\item \emph{Function Names.} Expressions $\langle\mathit{fname}\rangle$ range over function names.
\item \emph{Constants.} Expressions $\langle\mathit{const}\rangle$ range over decimal integers.
\item \emph{Arithmetic Expressions.} Expressions $\langle\mathit{expr}\rangle$ (resp. $\langle\mathit{pexpr}\rangle$) range over arithmetic expressions over both program and sampling variables (resp. program variables). As a theoretical paper, we do not fix the syntax for $\langle\mathit{expr}\rangle$ and $\langle\mathit{pexpr}\rangle$.
\item \emph{Parameters.} Expressions $\langle\mathit{plist}\rangle$ range over finite lists of program variables, and expressions $\langle\mathit{vlist}\rangle$ range over finite lists of arithmetic expressions over program variables.
\item \emph{Boolean Expressions.} Expressions $\langle\mathit{bexpr}\rangle$ range over propositional arithmetic predicates over program variables.
\item \emph{Demonic Nondeterminism.} The symbol `$\star$' indicates a nondeterministic choice to be resolved in a demonic way.
\item \emph{Statements $\langle \mathit{stmt}\rangle$.} Assignment statements are indicated by `$:=$';
`\textbf{skip}' is the statement that does nothing;
conditional branches and demonic nondeterminism are both indicated by the keyword `\textbf{if}';
while-loops are indicated by the keyword `\textbf{while}';
sequential compositions are indicated by semicolon;
finally, function calls are indicated by $\mathit{fname}\left(\langle\mathit{vlist}\rangle\right)$.
\item \emph{Programs.} Each recursive program $\langle\mathit{prog}\rangle$ is a sequence of function entities, for which each function entity $\langle\mathit{func}\rangle$ consists of a function name followed by a list of parameters (composing a function declaration) and a curly-braced statement (serving as the function body).
\end{compactitem}

\begin{figure}
\begin{align*}
\langle \mathit{prog}\rangle &::= \,\langle\mathit{func}\rangle\langle\mathit{prog}\rangle \mid \langle\mathit{func}\rangle
\\
\vspace{\baselineskip}
\\
\langle \mathit{func}\rangle &::= \,\langle\mathit{fname}\rangle\mbox{`$($'}\langle plist\rangle\mbox{`$)$'}\mbox{`$\{$'}\langle stmt\rangle\mbox{`$\}$'}
\\
\langle plist\rangle &::= \,\langle \mathit{pvar}\rangle\mid \langle \mathit{pvar}\rangle\mbox{`$,$'}\langle \mathit{plist}\rangle
\\
\vspace{\baselineskip}
\\
\langle \mathit{stmt}\rangle &::= \mbox{`\textbf{skip}'}\\
&\mid \langle\mathit{pvar}\rangle \,\mbox{`$:=$'}\, \langle\mathit{expr} \rangle\\
& \mid \mbox{`\textbf{if}'} \, \langle\mathit{bexpr}\rangle\,\mbox{`\textbf{then}'} \, \langle \mathit{stmt}\rangle \, \mbox{`\textbf{else}'} \, \langle \mathit{stmt}\rangle \,\mbox{`\textbf{fi}'}
\\
& \mid \mbox{`\textbf{if}'} \, \mbox{`$\star$'}\,\mbox{`\textbf{then}'} \, \langle \mathit{stmt}\rangle \, \mbox{`\textbf{else}'} \, \langle \mathit{stmt}\rangle \,\mbox{`\textbf{fi}'}\\
&\mid  \mbox{`\textbf{while}'}\, \langle\mathit{bexpr}\rangle \, \text{`\textbf{do}'} \, \langle \mathit{stmt}\rangle \, \text{`\textbf{od}'}
\\
& \mid \langle\mathit{fname}\rangle\mbox{`$($'}\langle\mathit{vlist}\rangle\mbox{`$)$'} \mid \langle\mathit{stmt}\rangle \, \text{`;'} \, \langle \mathit{stmt}\rangle
\\
\vspace{\baselineskip}
\langle\mathit{vlist}\rangle &::= \langle\mathit{pexpr}\rangle \mid \langle\mathit{pexpr}\rangle\mbox{`,'} \langle\mathit{vlist}\rangle
\\
\vspace{\baselineskip}
\\
\langle\mathit{literal} \rangle &::= \langle\mathit{pexpr} \rangle\, \mbox{`$\leq$'} \,\langle\mathit{pexpr} \rangle \mid \langle\mathit{pexpr} \rangle\, \mbox{`$\geq$'} \,\langle\mathit{pexpr} \rangle
\\
\langle \mathit{bexpr}\rangle &::=  \langle \mathit{literal} \rangle \mid \neg \langle\mathit{bexpr}\rangle\\
&\mid \langle \mathit{bexpr} \rangle \, \mbox{`\textbf{or}'} \, \langle\mathit{bexpr}\rangle
\mid \langle \mathit{bexpr} \rangle \, \mbox{`\textbf{and}'} \, \langle\mathit{bexpr}\rangle
\end{align*}
\caption{Syntax of Recursive Probabilistic Programs}
\label{fig:syntax}
\end{figure}

\smallskip\noindent{\bf Assumptions.}
W.l.o.g, we adopt further syntactical restrictions for simplicity:
\begin{compactitem}
\item \emph{Distinct Sampling Variables.} every sampling variable appears exactly once in any program.
\item \emph{Function Declarations.} every parameter list $\langle\mathit{plist}\rangle$ contains no duplicate program variables, and function names are distinct.
\item \emph{Function Calls.} no function call involves some function name without function entity (i.e., undeclared function names).
\end{compactitem}

\section{Control-Flow Graphs for Nondeterministic Recursive Probabilistic Programs}\label{app:cfg}

In this part, we demonstrate inductively how the control-flow graph of a recursive program can be constructed.

Recall that given an arithmetic expression $\mathfrak{e}$ over $V$, we define the \emph{evaluation} $\mathfrak{e}(\nu)\in\Zset$ under a valuation $\nu\in\val{V}$ as the result by substituting $\nu(x)$ for $x$, for every $x\in V$ appearing in $\mathfrak{e}$.
Then given an arithmetic expression $\mathfrak{e}$ over $V_1\cup V_2$ where $V_1,V_2$ are two disjoint sets of variables,
the evaluation $\mathfrak{e}(\nu\cup\mu)\in\Zset$ under $\nu\in\val{V_1}$ and $\mu\in\val{V_2}$ is the result by substituting (i) $\nu(x)$ for $x$ and (ii) $\mu(y)$ for $y$ (in $\mathfrak{e}$), for every $x\in V_1$ and $y\in V_2$ appearing in $\mathfrak{e}$.

Below we fix a recursive program $W$ and denote by $\fnames$ the set of function names appearing in $W$.
For each function name $\fn{f}\in\fnames$, we define $P_\fn{f}$ to be the function body of
$\fn{f}$, and define $\pvars{f}$ to be the set of program variables appearing in $P_\fn{f}$ and the parameter list of $\fn{f}$.
We let $\rvars$ to be the set of sampling variables appearing in $W$.

The control-flow graph of $W$ is constructed by first constructing the counterparts $\{\transitions{\fn{f}}\}_{\fn{f}\in\fnames}$ for each of its function bodies and then grouping them together.
To construct each $\transitions{\fn{f}}$, we first construct the partial relation $\transitions{P,\fn{f}}$ inductively on the structure of $P$ for each statement $P$ which involves programs variables solely from $\pvars{f}$, then define $\transitions{\fn{f}}$ as $\transitions{P_\fn{f},\fn{f}}$.

Let $\fn{f}\in\fnames$. Given an arithmetic expression ${\mathfrak{e}}$ over $\pvars{f}\cup\rvars$, a program variable $x\in\pvars{f}$ and two valuations $\nu\in\val{\fn{f}},\mu\in\samples$, we denote by ${{\nu}\assgn{\mathfrak{e(\nu\cup\mu)}}{x}}$ the valuation over $\pvars{f}$ such that
\[
\left({\nu}{\assgn{\mathfrak{e(\nu\cup\mu)}}{x}}\right)(y)=
\begin{cases} \nu(y) & \mbox{ if } y\in\pvars{f}\backslash\{x\} \\
\mathfrak{e}(\nu\cup\mu) & \mbox{ if }y=x
\end{cases}\enskip.
\]

Given a function call $\fn{g}(\mathfrak{e}_1,\dots,\mathfrak{e}_k)$ with variables solely from $\pvars{f}$ and its declaration being $\fn{g}(y_1,\dots,y_k)$, and
a valuation $\nu\in\val{\fn{f}}$, we define ${\nu}{[\fn{g},\{\mathfrak{e}_j\}_{1\le j\le k}]}$ to be a valuation over $\pvars{g}$ by:
\[
{\nu}{[\fn{g},\{\mathfrak{e}_j\}_{1\le j\le k}]}(y):=\begin{cases}
\mathfrak{e}_j(\nu) & \mbox{ if }y=y_j\mbox{ for some }j \\
0 & \mbox{ if }y\in \pvars{g}\backslash\{y_1,\dots,y_k\}
\end{cases}\enskip.
\]

Now the inductive construction for each $\transitions{P,\fn{f}}$ is demonstrated as follows.
For each statement $P$ which involves program variables solely from $\pvars{f}$, the relation $\transitions{P,\fn{f}}$ involves two distinguished labels, namely $\lin{P,\fn{f}}$ and $\lout{P,\fn{f}}$, that intuitively represent the label assigned to the first instruction to be executed in $P$ and the terminal program counter of $P$, respectively.
After the inductive construction, $\lin{\fn{f}},\lout{\fn{f}}$ are defined as $\lin{P_{\fn{f}},\fn{f}},\lout{P_{\fn{f}},\fn{f}}$, respectively.

\begin{compactenum}
\item {\em Assignments.}
For $P$ of the form ${x}{:=}{\mathfrak{e}}$ or resp. $\mbox{\textbf{skip}}$ where $\mathfrak{e}$ is an arithmetic expression over program and sampling variables, $\transitions{P,\fn{f}}$ involves a new assignment label $\lin{P,\fn{f}}$ (as the initial label) and a new branching label $\lout{P,\fn{f}}$ (as the terminal label), and contains a sole triple
\[
\left(\lin{P,\fn{f}},(\nu,\mu)\mapsto {\nu}{\assgn{\mathfrak{e(\nu\cup\mu)}}{x}},\lout{P,\fn{f}}\right)
\]
or resp.
\[
\left(\lin{P,\fn{f}},(\nu,\mu)\mapsto\nu,\lout{P,\fn{f}}\right),
\]
respectively.
\item {\em Function Calls.}
For $P$ of the form $\fn{g}(\mathfrak{e}_1,\dots,\mathfrak{e}_k)$ involving solely program variables,
$\transitions{P,\fn{f}}$ involves a new call label $\lin{P,\fn{f}}$ and a new branching label $\lout{P,\fn{f}}$, and contains a sole triple
\[
\left(\lin{P,\fn{f}},\left(\fn{g},\nu\mapsto\nu[\fn{g},\{\mathfrak{e}_j\}_{1\le j\le k}]\right),\lout{P,\fn{f}}\right)\enskip.
\]

\item {\em Sequential Statements.}
For ${P}{=}{Q_1;Q_2}$, we take the disjoint union of $\transitions{Q_1,\fn{f}}$ and  $\transitions{Q_2,\fn{f}}$, while redefining $\lout{Q_1,\fn{f}}$ to be $\lin{Q_2,\fn{f}}$ and putting $\lin{P,\fn{f}}:=\lin{Q_1,\fn{f}}$ and $\lout{P,\fn{f}}:=\lout{Q_2,\fn{f}}$.
\item {\em If-Branches.}
For ${P}{=}{\textbf{if $\phi$ then }Q_1 \textbf{ else } Q_2 \textbf{ fi}}$ with $\phi$ being a propositional arithmetic predicate, we first add two new branching labels $\lin{P},\lout{P}$, then take the disjoint union of $\transitions{Q_1,\fn{f}}$ and $\transitions{Q_2,\fn{f}}$ while simultaneously identifying both $\lout{Q_1,\fn{f}}$ and $\lout{Q_2,\fn{f}}$ with $\lout{P,\fn{f}}$, and finally obtain $\transitions{P,\fn{f}}$ by adding two triples $(\lin{P,\fn{f}},\phi,\lin{Q_1,\fn{f}})$ and $(\lin{P,\fn{f}},\neg\phi,\lin{Q_2,\fn{f}})$ into the disjoint union of $\transitions{Q_1,\fn{f}}$ and $\transitions{Q_2,\fn{f}}$.
\item {\em While-Loops.}
For ${P}{=}{\mbox{ \textbf{while} } \phi \mbox{ \textbf{do} }Q \mbox{ \textbf{od}}}$,
we add a new branching label $\lout{P,\fn{f}}$ as a terminal label and obtain $\transitions{P,\fn{f}}$ by adding triples $(\lout{Q,\fn{f}},\phi,\lin{Q,\fn{f}})$ and $(\lout{Q,\fn{f}},\neg\phi,\lout{P,\fn{f}})$ into $\transitions{Q,\fn{f}}$, and define $\lin{P,\fn{f}}:=\lout{Q,\fn{f}}$.
\item {\em Nondeterminism.} For ${P}{=}{\textbf{if $\star$ then }Q_1 \textbf{ else } Q_2 \textbf{ fi}}$,
we first add a new nondeterministic label $\lin{P}$ and a new branching labels $\lout{P}$, then take the disjoint union of $\transitions{Q_1,\fn{f}}$ and $\transitions{Q_2,\fn{f}}$ while simultaneously identifying both $\lout{Q_1,\fn{f}}$ and $\lout{Q_2,\fn{f}}$ with $\lout{P,\fn{f}}$, and finally obtain $\transitions{P,\fn{f}}$ by adding two triples $(\lin{P,\fn{f}},\star,\lin{Q_1,\fn{f}})$ and $(\lin{P,\fn{f}},\star,\lin{Q_2,\fn{f}})$ into the disjoint union of $\transitions{Q_1,\fn{f}}$ and $\transitions{Q_2,\fn{f}}$.
\end{compactenum}

\section{Details for $\MDP{W,\sigma}$ and $\left(\infruns^W, \mathcal{H}^W, \probm_{\mathfrak{c}}^{W,\sigma}\right)$}\label{app:markovchains}

\begin{definition}[$\MDP{W,\sigma}$ from~{\cite[Chapter 10]{BaierBook}}]
Let $\sigma$ be any scheduler for $W$. The Markov chain $\MDP{W,\sigma}=(\MDPstates_W^*,\MDPkernel_{W,\sigma}^*)$ is given as follows.
\begin{compactitem}
\item The \emph{state space} $S^*_W$ is the set of all histories.
\item The \emph{probability transition function}
\[
\MDPkernel^*_{W,\sigma}:\MDPstates^*_{W}\times \MDPstates^*_{W}\rightarrow[0,1]
\]
is defined by: (i) if $\rho'=\rho\cdot s$ for $s\in \MDPstates_W$ then
\[
\MDPkernel^*_{W,\sigma}(\rho,\rho'):=\sum_{a\in\enabled{\last{\rho}}}
\sigma(\rho)(a)\cdot \MDPkernel_W(\last{\rho}, a, s)
\]
and (ii) $\MDPkernel^*_{W,\sigma}(\rho,\rho'):=0$ otherwise.
\end{compactitem}
\hfill\IEEEQEDclosed
\end{definition}


\noindent{\bf The Probability Space $\left(\infruns^W, \mathcal{H}^W, \probm_{\mathfrak{c}}^{W,\sigma}\right)$.}
Consider any scheduler for $W$ and any non-terminal stack element $\mathfrak{c}$.
The initial state of $\MDP{W,\sigma}$ is set to $(\mathfrak{c},\mathbf{0}_\mathrm{r})$ where
$\mathfrak{c}$ is viewed as a single-letter configuration and
$\mathbf{0}_\mathrm{r}$ is the valuation over $\rvars$ assigning to every sampling variable $0$.
Then the probability space $\left(\infruns^W, \mathcal{H}^W, \probm_{\mathfrak{c}}^{W,\sigma}\right)$ is given as follows:
\begin{compactitem}
\item the sample space $\infruns^W$ is the set of infinite runs;
\item $\mathcal{H}^W$ is the smallest $\sigma$-algebra generated by the set
\[
\{\cyl_\varpi\mid \varpi\mbox{ is a finite run}\}
\]
of all cylinder sets where each $\cyl_\varpi\subseteq\infruns^W$ is defined as the set
$\left\{\omega\mid \varpi \mbox{ is a finite prefix of }\omega\right\}$~;
\item $\probm_{\mathfrak{c}}^{W,\sigma}$ is the unique probability measure on $\mathcal{H}^W$ such that for any cylinder set $\cyl_\varpi$ with $\varpi=\rho_0\dots\rho_n$,
\[
\probm_{\mathfrak{c}}^{W,\sigma}\left(\cyl_\varpi\right)=\begin{cases}
{\displaystyle\prod_{k=0}^{n-1}\MDPkernel_{W,\sigma}^*(\rho_k,\rho_{k+1})} & \mbox{if }\rho_0=(\mathfrak{c},\mathbf{0}_\mathrm{r})\\
0 & \mbox{otherwise }\\
\end{cases}.
\]
\end{compactitem}
For more details, we refer to~\cite[Chapter 10]{BaierBook}.\hfill\IEEEQEDclosed

\section{Proofs for Section~\ref{sect:btermination}}\label{app:btermination}

In this part, we fix a nondeterministic recursive probabilistic program $W$ together with its associated CFG taking the form~(\ref{eq:cfg}) and a sampling function $\Upsilon$.

\subsection{Soundness}

\noindent{\bf Proposition~\ref{prop:rsupm}.}
Let $\Gamma=\{X_n\}_{n\in\Nset_0}$ be a ranking supermartingale adapted to a filtration $\{\mathcal{F}_n\}_{n\in\Nset_0}$ with $\epsilon$ given as in Definition~\ref{def:rsupm}.
Then $\probm(\stopping{\Gamma}<\infty)=1$ and $\expv(\stopping{\Gamma})\le\frac{\expv(X_0)}{\epsilon}$.
\begin{IEEEproof}
We first prove by induction on $n\ge 0$ that
\[
\expv(X_n)\le \expv(X_0)-\epsilon\cdot\sum_{k=0}^{n-1}\probm(X_k>0)\enskip.
\]
The base step $n=0$ is clear.
The inductive step can be carried out as follows:
\begin{eqnarray*}
\expv(X_{n+1})&= & \mbox{\S~By E4~\S} \\
& & \expv\left(\expv(X_{n+1}\mid \mathcal{F}_n)\right) \\
&\le & \mbox{\S~By Definition~\ref{def:rsupm}~\S} \\
& &\expv(X_n)-\epsilon\cdot\expv(\mathbf{1}_{X_n> 0}) \\
&=& \S~\expv(\mathbf{1}_{X_n>0})=\probm(X_n>0)~\S\\
& & \expv(X_n)-\epsilon\cdot\probm(X_n> 0) \\
&\le & \mbox{\S~By Induction Hypothesis~\S} \\
& & \expv(X_0)-\epsilon\cdot\sum_{k=0}^{n-1}\probm(X_k> 0)-\epsilon\cdot\probm(X_n> 0) \\
&= & \mbox{\S~By Direct Algebraic Computation~\S} \\
& &\expv(X_0)-\epsilon\cdot\sum_{k=0}^{n}\probm(X_k> 0)\enskip.
\end{eqnarray*}
From the Non-negativity Condition in Definition~\ref{def:rsupm}, one has that
$\expv(X_{n}) \geq 0$, for all $n$.
Hence it holds for all $n$ that
\[
\sum_{k=0}^{n}\probm(X_k> 0) \leq \frac{\expv(X_0) - \expv(X_{n+1})}{\epsilon} \leq \frac{\expv(X_0)}{\epsilon}~~.
\]
Hence, the series $\sum_{k=0}^{\infty}\probm(X_k> 0)$ converges and
\[
\sum_{k=0}^{\infty}\probm(X_k> 0)\le \frac{\expv(X_0)}{\epsilon}\enskip.
\]
It follows from $\stopping{\Gamma}(\omega)> k\Rightarrow X_k(\omega)> 0$ (for all $k,\omega$) that
\[
\probm(\stopping{\Gamma}=\infty)=\lim_{k\rightarrow\infty}\probm(\stopping{\Gamma}> k)=0
\]
and
\begin{eqnarray*}
\expv(\stopping{\Gamma})&=& \sum_{k=0}^{\infty}\probm(k<\stopping{\Gamma}<\infty)\\
&\le&\sum_{k=0}^{\infty}\probm(X_k> 0)\le \frac{\expv(X_0)}{\epsilon}\enskip.
\end{eqnarray*}
The desired result follows.
\end{IEEEproof}

To prove Lemma~\ref{thm:soundness}, we need the following proposition.

\begin{proposition}\label{prop:independence}
For any two independent random variables $X,Y$ (from some probability space), if it holds a.s. that $X\ge 0$ and $Y\ge 0$, then
$\expv(X\cdot Y)=\expv(X)\cdot \expv(Y)$ (where $d\cdot\infty:=\infty$ for $d\in(0,\infty]$ and $0\cdot\infty:=0$ by convention).
\end{proposition}
\begin{IEEEproof}
Define random variables $X_n, Y_n$'s ($n\in\Nset_0$) by $X_n:=\min\{X, n\}$ and $Y_n:=\min\{Y,n\}$.
Then it holds for all $n$ that
\begin{compactitem}
\item $X_n$ and $Y_n$ are independent, and
\item $\expv(|X_n|)<\infty$ and $\expv(|Y_n|)<\infty$.
\end{compactitem}
Hence, from~\cite[Chapter 7]{probabilitycambridge}, one obtains that for all $n$,
\[
\expv(X_n\cdot Y_n)=\expv(X_n)\cdot \expv(Y_n)\enskip.
\]
By (i) $X,Y\ge 0$ a.s., (ii) taking the limit $n\rightarrow\infty$ and (iii) Monotone Convergence Theorem (cf.~\cite[Chapter 6]{probabilitycambridge}), one obtains directly that
$\expv(X\cdot Y)=\expv(X)\cdot \expv(Y)$\enskip.
\end{IEEEproof}

In order to prove Lemma~\ref{thm:soundness}, we further need to define the following random variables.

\noindent\textbf{Random Variables w.r.t $W$.} We define random variables
\[
\rvlen_{n},~\rvsam_{n},~\rvfn_{n,k},~\rvlb_{n,k},~\rvval^{\fn{f},x}_{n,k}
\]
for $n,k\in\Nset_0$, $\fn{f}\in\fnames$, $x\in\pvars{f}$ over infinite runs as infinite sequences $\{(w_n,\mu_n)\}_{n\in\Nset_0}$ of states in $\MDPstates_W$ (cf. Remark~\ref{rmk:runs}) as follows.
For any infinite run
$\omega=\{(w_n,\mu_n)\}_{n\in\Nset_0}$
for which
$w_n=(\fn{f}_{n,0}, \loc_{n,0}, \nu_{n,0})\dots(\fn{f}_{n,l_n}, \loc_{n,l_n}, \nu_{n,l_n})$
(where $w_n=\varepsilon$ when $l_n=-1$),
\[
\rvlen_n(\omega):=l_n+1,~\rvsam_{n}(\omega):=\mu_{n}
\]
and
\begin{align*}
& (\rvfn_{n,k}\left(\omega\right),~\rvlb_{n,k}\left(\omega\right),~\rvval^{\fn{f},x}_{n,k}\left(\omega\right)):=\\
& ~~\begin{cases}
(\fn{f}_{n,k},~\loc_{n,k},~\nu_{n,k}(x)) & \mbox{if }k\le l_n\mbox{ and } \fn{f}=\fn{f}_{n,k} \\
(\fn{f}_{n,k},~\loc_{n,k},~0) & \mbox{if }k\le l_n\mbox{ and }\fn{f}\ne\fn{f}_{n,k}\\
(\fn{f}_\bot,~\loc_\bot,~0) & \mbox{otherwise}
\end{cases}
\end{align*}
where $\fn{f}_\bot,\loc_\bot$ are arbitrarily fixed elements only to handle the invalid case $k>l_n$.
We denote by $\rvval^{\fn{f}}_{n,k}$ the function $\omega\mapsto \rvval^{\fn{f}}_{n,k}(\omega)$ over infinite runs where $\rvval^{\fn{f}}_{n,k}(\omega)$ is the valuation $x\mapsto \rvval^{\fn{f},x}_{n,k}(\omega)$ on  $\pvars{f}$.
Note that $\rvval^{\rvfn_{n,k}(\omega)}_{n,k}(\omega)=\nu_{n,k}$ when $k\le l_n$.
Moreover, we define $\mathcal{H}_n\subseteq\mathcal{H}$ ($n\in\Nset_0$) to be the smallest $\sigma$-algebra such that all random variables
\[
\rvlen_{m},~\rvsam_{m-1},~\rvfn_{m,k},~\rvlb_{m,k},~\rvval^{\fn{f},x}_{m,k}~
\]
($0\le m\le n, k\ge 0, \fn{f}\in\fnames, x\in\pvars{f}$) are $\mathcal{H}_n$-measurable.

Informally, $\rvlen_{n}$ (resp. $\rvsam_{n}$) is the length of the configuration (resp. the valuation sampled for sampling variables) right before the $n$-th (resp. $(n-1)$-th) execution of the program;
$\rvfn_{n,k}$ (resp. $\rvlb_{n,k}, \rvval^{\fn{f}}_{n,k}$) is the $k$-th function name (resp. label, valuation) in the configuration right before the $n$-th execution. \hfill\IEEEQEDclosed

Now we prove Lemma~\ref{thm:soundness}.

\noindent{\bf Lemma~\ref{thm:soundness}.}
For all ranking measure functions $h$ with $\epsilon$ given in Definition~\ref{def:mfunc} and for all stack elements $\mathfrak{c}=(\fn{f}, \loc, \nu)$, we have
$\expvt(\mathfrak{c})\le \frac{h(\mathfrak{c})}{\epsilon}$.
\begin{IEEEproof}
Let $h$ be any ranking measure function with $\epsilon$ given in Definition~\ref{def:mfunc} and $\mathfrak{c}=(\fn{f}, \loc, \nu)$ be any stack element.
The case when either $\mathfrak{c}$ is terminal or $h(\mathfrak{c})=\infty$ is straightforward.
Below we consider that $\mathfrak{c}$ is non-terminal and $h(\mathfrak{c})<\infty$.

Let $\sigma$ be any scheduler.
Define the stochastic process $\Gamma=\{X_n\}_{n\in\Nset_0}$ adapted to $\{\mathcal{H}_n\}_{n\in\Nset_0}$ by:
\begin{equation}\label{eq:proof:xnprocess}
X_n(\omega):=\sum_{k=0}^{\rvlen_n(\omega)-1} h\left(\rvfn_{n,k}(\omega),~\rvlb_{n,k}(\omega), ~\rvval^{\rvfn_{n,k}(\omega)}_{n,k}(\omega)\right)
\end{equation}
for all $n$ and all infinite runs $\omega$.
Note that all $X_n$'s are $\mathcal{H}_n$-measurable since (i) every $X_n$ depends only on histories formed by first $n+1$ configurations and samplings and (ii) there are countably many histories.
Moreover, by Definition~\ref{def:mfunc} and the fact that we do not allow terminal stack elements to appear in a configuration,
$X_n(\omega)>0$ iff $\rvlen_n(\omega)\ge 1$ for all $\omega$; it follows that $\tertime=\stopping{\Gamma}$.
We show that $\{X_n\}_{n\in\Nset_0}$ is a ranking supermartingale (under $\probm^\sigma_\mathfrak{c}$).
Below for the sake of simplicity, we abbreviate `$\expv^\sigma_\mathfrak{c}$' simply as `$\expv$'.

By definition, the Non-negativity Condition for ranking supermartingales holds naturally for $\{X_n\}_{n\in\Nset_0}$.
We then prove the Integrability Condition.
Fix any $n\in\Nset_0$.
From our semantics and (C1), it holds a.s. that
\begin{equation}\label{eq:proof:xnplusone}
X_{n+1}= \mathbf{1}_{\rvlen_n\ge 1}\cdot\left[D+Y^{n,\mathrm{a}}+Y^{n,\mathrm{c}}+Y^{n,\mathrm{b}}+Y^{n,\mathrm{d}}\right]
\end{equation}
(cf. below for the definitions of $D, Y^{n,\mathrm{a}}, Y^{n,\mathrm{c}}, Y^{n,\mathrm{b}}, Y^{n,\mathrm{d}}$).
\[
D:=\sum_{k=1}^{\rvlen_n-1} h\left(\rvfn_{n,k},~\rvlb_{n,k}, ~\rvval^{\rvfn_{n,k}}_{n,k}\right)
\]
reflects the stack after removing the top stack element.
\[
Y^{n,\mathrm{a}}:=\sum_{\mu\in\samples}\mathbf{1}_{\rvsam_{n+1}=\mu}\cdot Y^{n,\mathrm{a}}_{\mu}
\]
reflects assignment labels at the top of the stack, where
\[
Y^{n,\mathrm{a}}_{\mu}:=\sum_{\fn{f}\in\fnames}\sum_{\loc\in\alocs{f}\setminus\{\lout{\fn{f}}\}}\mathbf{1}_{\left(\rvfn_{n,0}, \rvlb_{n,0}\right)=
(\fn{f},\loc)}\cdot Y^{n,\mathrm{a}}_{\fn{f},\loc,\mu}~
\]
and
\[
Y^{n,\mathrm{a}}_{\fn{f},\loc,\mu} :=   h\left(\fn{f},\loc',u(\rvval^{\fn{f}}_{n,0},\mu)\right)
\]
with $(\loc,u,\loc')\in\transitions{\fn{f}}$ being the sole triple in $\transitions{\fn{f}}$ with source label $\loc$ and update function $u$.
\[
Y^{n,\mathrm{c}}:=\sum_{\fn{f}\in\fnames}\sum_{\loc\in\flocs{f}\setminus\{\lout{\fn{f}}\}}\mathbf{1}_{\left(\rvfn_{n,0}, \rvlb_{n,0}\right)=
(\fn{f},\loc)}\cdot Y^{n,\mathrm{c}}_{\fn{f},\loc}
\]
reflects call labels at the top of the stack, where
\[
Y^{n,\mathrm{c}}_{\fn{f},\loc} := h\left(\fn{g},\lin{\fn{g}},v\left(\rvval^{\fn{f}}_{n,0}\right)\right)+h\left(\fn{f},\loc',\rvval^{\fn{f}}_{n,0}\right)
\]
with $(\loc, (\fn{g},v), \loc')$ being the sole triple in $\transitions{\fn{f}}$ with source label $\loc$ and value-passing function $v$.
\[
Y^{n,\mathrm{b}}:=\sum_{\fn{f}\in\fnames}\sum_{\loc\in\clocs{f}\setminus\{\lout{\fn{f}}\}}\mathbf{1}_{\left(\rvfn_{n,0}, \rvlb_{n,0}\right)=(\fn{f},\loc)}\cdot Y^{n,\mathrm{b}}_{\fn{f},\loc}
\]
reflects branching labels at the top of the stack, where
\[
Y^{n,\mathrm{b}}_{\fn{f},\loc}:=\sum_{i\in\{1,2\}}\mathbf{1}_{\rvval^{\fn{f}}_{n,0}\models\phi_i}\cdot h\left(\fn{f},\loc_i,\rvval^{\fn{f}}_{n,0}\right)
\]
with $(\loc, \phi, \loc_1)$, $(\loc, \neg\phi, \loc_2)$ being namely the two triples in $\transitions{\fn{f}}$ with source label $\loc$ and propositional arithmetic predicate $\phi$, $\phi_1:=\phi$ and $\phi_2:=\neg\phi$.
\[
Y^{n,\mathrm{d}}:=\sum_{\fn{f}\in\fnames}\sum_{\loc\in\dlocs{f}\setminus\{\lout{\fn{f}}\}}\mathbf{1}_{\left(\rvfn_{n,0}, \rvlb_{n,0}\right)=(\fn{f},\loc)}\cdot Y^{n,\mathrm{d}}_{\fn{f},\loc}
\]
reflects nondeterministic labels at the top of the stack, where
\[
Y^{n,\mathrm{d}}_{\fn{f},\loc}:=\sum_{i\in\{1,2\}}\mathbf{1}_{\rvlb_{n+1}=\loc_i}\cdot h\left(\fn{f},\loc_i,\rvval^{\fn{f}}_{n,0}\right)
\]
with $(\loc, \star, \loc_1)$, $(\loc, \star, \loc_2)$ being namely the two triples in $\transitions{\fn{f}}$ with source label $\loc$.

For each $n\in\Nset_0$, we define the random variable
\[
\widehat{Y}^{n,\mathrm{d}}:=\sum_{\fn{f}\in\fnames}\sum_{\loc\in\dlocs{f}\setminus\{\lout{\fn{f}}\}}\mathbf{1}_{\left(\rvfn_{n,0}, \rvlb_{n,0}\right)=(\fn{f},\loc)}\cdot \widehat{Y}^{n,\mathrm{d}}_{\fn{f},\loc}
\]
where
\[
\widehat{Y}^{n,\mathrm{d}}_{\fn{f},\loc}:=\max\left\{h\left(\fn{f},\loc_1,\rvval^{\fn{f}}_{n,0}\right), h\left(\fn{f},\loc_2,\rvval^{\fn{f}}_{n,0}\right)\right\}
\]
with $(\loc, \star, \loc_1)$, $(\loc, \star, \loc_2)$ being namely the two triples in $\transitions{\fn{f}}$ with source label $\loc$.

By definition, it holds that $Y^{n,\mathrm{d}}_{\fn{f},\loc}\le \widehat{Y}^{n,\mathrm{d}}_{\fn{f},\loc}$ for any suitable $n, \fn{f},\loc$.
Hence,
\begin{equation}\label{eq:proof:xxp}
X_{n+1}\le X'_{n+1}
\end{equation}
from (\ref{eq:proof:xnplusone}) where
\begin{equation}\label{eq:proof:xprimenplusone}
X'_{n+1}:= \mathbf{1}_{\rvlen_n\ge 1}\cdot\left[D+Y^{n,\mathrm{a}}+Y^{n,\mathrm{c}}+Y^{n,\mathrm{b}}+\widehat{Y}^{n,\mathrm{d}}\right].
\end{equation}
By the fact that random variables $\mathbf{1}_{\rvsam_{n+1}=\mu}$'s ($\mu\in\samples$) are independent of $\mathcal{H}_n$ (as they depend only on the sampling at the $n$-step), we obtain from Monotone Convergence Theorem~\cite[Chapter 6]{probabilitycambridge} (for possibly infinite sum) and Proposition~\ref{prop:independence} that
\begin{equation}\label{eq:proof:xpxpp}
\expv(X'_{n+1})= \expv(X''_{n+1})
\end{equation}
where
\begin{equation}\label{eq:proof:xppnplusone}
X''_{n+1}:= \mathbf{1}_{\rvlen_n\ge 1}\cdot\left[D+\widehat{Y}^{n,\mathrm{a}}+Y^{n,\mathrm{c}}+Y^{n,\mathrm{b}}+\widehat{Y}^{n,\mathrm{d}}\right]
\end{equation}
and
\[
\widehat{Y}^{n,\mathrm{a}}:=\sum_{\mu\in\samples}\sampdpd(\mu)\cdot Y^{n,\mathrm{a}}_{\mu}\enskip.
\]
Furthermore, by our semantics and conditions (C1)--(C5) for ranking measure functions, one has that
\begin{eqnarray}\label{eq:proof:xppxn}
X''_{n+1} &\le & \mathbf{1}_{\rvlen_n\ge 1}\cdot \Bigg[D+h\left(\rvfn_{n,0},~\rvlb_{n,0}, ~\rvval^{\rvfn_{n,0}}_{n,0}\right)-\epsilon\Bigg]\nonumber\\
&=& \mathbf{1}_{\rvlen_n\ge 1}\cdot (X_n-\epsilon)\nonumber\\
&=& \mathbf{1}_{\rvlen_n\ge 1}\cdot (X_n-\epsilon)+\mathbf{1}_{\rvlen_n=0}\cdot X_n\nonumber\\
&=& \mathbf{1}_{X_n>0}\cdot (X_n-\epsilon)+\mathbf{1}_{X_n=0}\cdot X_n\nonumber\\
&=& X_n- \mathbf{1}_{X_n>0}\cdot \epsilon\enskip.
\end{eqnarray}
Hence, we obtain that
\begin{equation}\label{eq:proof:xxpxpp}
\expv(X_{n+1})\le\expv(X'_{n+1})=\expv(X''_{n+1})\le \expv(X_n)\enskip.
\end{equation}
The Integrability Condition now follows from an easy induction on $n$ with the base step $\expv^\sigma_\mathfrak{c}(X_0)=h(\mathfrak{c})<\infty$.

Now we prove the Ranking Condition for $\{X_n\}_{n\in\Nset_0}$.
Since $\expv(X'_{n+1})=\expv(X''_{n+1})<\infty$, one has the following:
\begin{compactitem}
\item $\expv\left(\mathbf{1}_{\rvlen_n\ge 1}\cdot Y^{n,\mathrm{a}}\right)<\infty$, which further implies that
$\expv\left(\mathbf{1}_{\rvlen_n\ge 1}\cdot Y^{n,\mathrm{a}}_{\mu}\right)<\infty$ as long as $\sampdpd(\mu)>0$;
\item $\expv\left(\mathbf{1}_{\rvlen_n\ge 1}\cdot \left(D+Y^{n,\mathrm{c}}+Y^{n,\mathrm{b}}+\widehat{Y}^{n,\mathrm{d}}\right)\right)<\infty$.
\end{compactitem}
Then the followings hold a.s.:
\begin{eqnarray}\label{eq:proof:xnplusonexpxpp}
& &\condexpv{X_{n+1}}{\mathcal{H}_n}\nonumber\\
 &\le &\condexpv{X'_{n+1}}{\mathcal{H}_n}\quad\mbox{(by (E6), (E10))} \nonumber\\
&=& \condexpv{\mathbf{1}_{\rvlen_n\ge 1}\cdot Y^{n,\mathrm{a}}}{\mathcal{H}_n} \quad\mbox{(by (E6))}\nonumber\\
& &\quad{}+ \condexpv{\mathbf{1}_{\rvlen_n\ge 1}\cdot \left(D+Y^{n,\mathrm{c}} + Y^{n,\mathrm{b}} + \widehat{Y}^{n,\mathrm{d}}\right)}{\mathcal{H}_n} \nonumber\\
&=& \condexpv{\mathbf{1}_{\rvlen_n\ge 1}\cdot Y^{n,\mathrm{a}}}{\mathcal{H}_n} \quad\mbox{(by (E5))}\nonumber\\
& &\quad{}+ \mathbf{1}_{\rvlen_n\ge 1}\cdot \left(D+Y^{n,\mathrm{c}} + Y^{n,\mathrm{b}} + \widehat{Y}^{n,\mathrm{d}}\right) \nonumber\\
&=& \condexpv{\sum_{\mu\in\samples} \mathbf{1}_{\rvsam_{n+1}=\mu}\cdot\mathbf{1}_{\rvlen_n\ge 1}\cdot Y^{n,\mathrm{a}}_\mu}{\mathcal{H}_n}\nonumber\\
& &\quad{}+ \mathbf{1}_{\rvlen_n\ge 1}\cdot \left(D+Y^{n,\mathrm{c}} + Y^{n,\mathrm{b}} + \widehat{Y}^{n,\mathrm{d}}\right) \nonumber\\
&=& ~\mbox{\S~by (E6), (E11) \S} \nonumber\\
& &\sum_{\mu\in\samples}\condexpv{ \mathbf{1}_{\rvsam_{n+1}=\mu}\cdot\mathbf{1}_{\rvlen_n\ge 1}\cdot Y^{n,\mathrm{a}}_\mu}{\mathcal{H}_n}\nonumber\\
& &\quad{}+ \mathbf{1}_{\rvlen_n\ge 1}\cdot \left(D+Y^{n,\mathrm{c}} + Y^{n,\mathrm{b}} + \widehat{Y}^{n,\mathrm{d}}\right) \nonumber\\
&=& \sum_{\mu\in\samples} \condexpv{ \mathbf{1}_{\rvsam_{n+1}=\mu}}{\mathcal{H}_n}\cdot \mathbf{1}_{\rvlen_n\ge 1}\cdot Y^{n,\mathrm{a}}_\mu~\mbox{(by (E8))}\nonumber\\
& &\quad{}+ \mathbf{1}_{\rvlen_n\ge 1}\cdot \left(D+Y^{n,\mathrm{c}} + Y^{n,\mathrm{b}} + \widehat{Y}^{n,\mathrm{d}}\right) \nonumber\\
&=& \sum_{\mu\in\samples}\sampdpd(\mu)\cdot \left(\mathbf{1}_{\rvlen_n\ge 1}\cdot Y^{n,\mathrm{a}}_\mu\right) \quad\mbox{(by (E9))}\nonumber\\
& &\quad{}+ \mathbf{1}_{\rvlen_n\ge 1}\cdot \left(D +Y^{n,\mathrm{c}} + Y^{n,\mathrm{b}} + \widehat{Y}^{n,\mathrm{d}}\right) \nonumber\\
&=& \mathbf{1}_{\rvlen_n\ge 1}\cdot \left(D+\widehat{Y}^{n,\mathrm{a}} +Y^{n,\mathrm{c}} + Y^{n,\mathrm{b}} + \widehat{Y}^{n,\mathrm{d}}\right)\nonumber\\
&=& X''_{n+1} \\
&\le& X_n- \mathbf{1}_{X_n>0}\cdot \epsilon  ~~\mbox{(by (\ref{eq:proof:xppxn}))}\enskip.\nonumber
\end{eqnarray}
It follows that $\{X_n\}_{n\in\Nset_0}$ is a ranking supermartingale.
Hence by Proposition~\ref{prop:rsupm},
\[
\expv(\tertime) =\expv(\stopping{\Gamma}) \le \frac{\expv\left(X_0\right)}{\epsilon} =\frac{h(\mathfrak{c})}{\epsilon}\enskip.
\]
Thus,  $\expvt\left(\mathfrak{c}\right)\le \frac{h(\mathfrak{c})}{\epsilon}$ by the arbitrary choice of $\sigma$.
\end{IEEEproof}

\subsection{Completeness}

In this part, we prove Lemma~\ref{thm:completeness}.
We fix a nondeterministic recursive probabilistic program $W$ together with its associated CFG taking the form~(\ref{eq:cfg}) and a sampling function $\Upsilon$.
To prove the lemma, we introduce more definitions as follows.

Consider a history $\rho=(w_0,\mu_0)\dots (w_n,\mu_n)$ ($n\ge 0$):
\begin{compactitem}
\item we say that
$\rho$ is \emph{strictly terminating} if $w_n=\varepsilon$ and $w_k\ne\varepsilon$ for all $0\le k< n$;
\item
we denote by $\rho[k]$ the state $(w_k,\mu_k)$ for $0\le k\le n$;
\item
The \emph{length} of $\rho$, denoted by $|\rho|$, is defined as $n+1$.
\end{compactitem}

We say that a scheduler $\sigma$ for $W$ is \emph{well-behaved} if for any history
$\rho=(w_0,\mu_0)\dots (w_n,\mu_n)$ ($n\ge 0$),
$\sigma(\rho)= \sigma\left((w_0,\mathbf{0}_\mathrm{r})\dots (w_n,\mu_n)\right)$; in other words,
$\sigma$ ignores the initial sampling $\mu_0$.
W.l.o.g, we can safely assume that a scheduler is well-behaved, as we only consider histories starting from $(\mathfrak{c},\mathbf{0}_\mathrm{r})$ for non-terminal stack element $\mathfrak{c}$.

We present a succinct representation for probability values of cylinder sets.
Given a history $\rho$ and a scheduler $\sigma$ for $\MDP{\sigma}$, we define
\begin{align*}
&\probht{\rho;\sigma}:= \\
&\prod_{k=0}^{|\rho|-2}\left[\sum_{a\in\enabled{\rho[k]}}\sigma(\rho[0]\dots \rho[k])(a)\cdot
\MDPkernel_W\left(\rho[k], a, \rho[k+1]\right)\right]
\end{align*}
Note that $\probm_{\mathfrak{c}}^{\sigma}\left(\cyl_\varpi\right)=\probht{\rho_n;\sigma}$ upon $\varpi=\rho_0\dots \rho_n$ and $\rho_0=(\mathfrak{c},\mathbf{0}_\mathrm{r})$.
Moreover, if $\sigma$ is well-behaved, then the valuation for sampling variables $\mu_0$ in $\rho_0=(w_0,\mu_0)$ plays a dummy role in the definition of $\probht{\rho;\sigma}$ as $\probht{\rho;\sigma}$ remains the same no matter how $\mu_0$ varies.

For each non-terminal stack element $\mathfrak{c}$, valuation $\mu\in\samples$ and positive integer $n$,
we define $H_{n,\mathfrak{c},\mu}$ to be the set of histories $\rho$ such that
(i) $\rho$ is strictly terminating, (ii) $|\rho|=n$ and (iii) $\rho[0]=(\mathfrak{c},\mu)$.
Furthermore, we let $H_{\mathfrak{c},\mu}:=\bigcup_{n\in\Nset}H_{n,\mathfrak{c},\mu}$ and abbreviate $H_{n,\mathfrak{c},\mathbf{0}_\mathrm{r}}, H_{\mathfrak{c},\mathbf{0}_\mathrm{r}}$ as $H_{n,\mathfrak{c}},H_{\mathfrak{c}}$, respectively.

By definition, $\probm^\sigma_\mathfrak{c}\left(\tertime<\infty\right)$ is the reachability probability to strictly-terminating histories under initial stack element $\mathfrak{c}$ and scheduler $\sigma$.
In the following proposition, we present direct representations of $\probm^\sigma_\mathfrak{c}\left(\tertime<\infty\right),\expv^\sigma_\mathfrak{c}\left(\tertime\right)$ under cylinder sets.

\begin{proposition}\label{prop:reachability}
For any non-terminal stack element $\mathfrak{c}$ and scheduler $\sigma$, one has that
\begin{compactenum}
\item
$\probm^\sigma_\mathfrak{c}\left(\tertime<\infty\right)=\sum_{\rho\in H_{\mathfrak{c}}} \probht{\rho;\sigma}$, and
\item
$\expv^\sigma_\mathfrak{c}\left(\tertime\right)=\sum_{\rho\in H_{\mathfrak{c}}} (|\rho|-1)\cdot \probht{\rho;\sigma}$ provided that $\probm^\sigma_\mathfrak{c}\left(\tertime<\infty\right)=1$\enskip.
\end{compactenum}
\end{proposition}
\begin{IEEEproof}
From the semantics (cf.~Section~\ref{sect:preliminaries} or~\cite[Chapter 10]{BaierBook}), we have
\[
\probm^\sigma_\mathfrak{c}\left(\tertime<\infty\right)=\sum_{\rho\in H_{\mathfrak{c}}} \probm^\sigma_\mathfrak{c}\left(\cyl_\rho\right) = \sum_{\rho\in H_{\mathfrak{c}}} \probht{\rho;\sigma}
\]
and $\probm^\sigma_\mathfrak{c}\left(\tertime<\infty\right)=1$ implies
\begin{eqnarray*}
\expv^\sigma_\mathfrak{c}\left(\tertime\right)&=&\sum_{\rho\in H_{\mathfrak{c}}} (|\rho|-1)\cdot\probm^\sigma_\mathfrak{c}\left(\cyl_\rho\right) \\
&=& \sum_{\rho\in H_{\mathfrak{c}}} (|\rho|-1)\cdot\probht{\rho;\sigma}\enskip.
\end{eqnarray*}
\end{IEEEproof}

Below we prove Lemma~\ref{thm:completeness}.
For every action $a\in\MDPactions$, let $\Dirac{a}$ be the Dirac distribution at $a$, i.e., $\Dirac{a}(a)=1$ and $\Dirac{a}(b)=0$ for $b\ne a$.
Given any history $\rho=(w_0,\mu_0)\dots(w_n,\mu_n)$ and any non-terminal stack element $\mathfrak{c}$, we define the histories $\stksch{\rho}{\mathfrak{c}}:=(w_0\cdot\mathfrak{c},\mu_0)\dots(w_n\cdot\mathfrak{c},\mu_n)$
and $\wbsch{\rho}:=(w_0,\mathbf{0}_\mathrm{r})\dots(w_n,\mu_n)$.

\noindent{\bf Lemma~\ref{thm:completeness}.}
$\expvt$ is a ranking measure function with corresponding $\epsilon=1$ (cf. Definition~\ref{def:mfunc}).
\begin{IEEEproof}
Let $\mathfrak{c}=(\fn{f},\loc,\nu)$ be any stack element.
We consider that $\mathfrak{c}$ is non-terminal and $\expvt(\mathfrak{c})<\infty$, since otherwise the situations are straightforward to prove.
Note that from  $\expvt(\mathfrak{c})<\infty$, $\probm_\mathfrak{c}^\sigma(T<\infty)=1$ for any scheduler $\sigma$ for $W$.
We clarify four cases below.

\noindent{\em Case 1: Assignment.} $\loc\in\alocs{f}$ and $(\loc,u,\loc')$ is the only triple in $\transitions{\fn{f}}$ with source label $\loc$ and update function $u$. Let $\sigma'$ be any well-behaved scheduler. Furthermore, let $\sigma$ be any scheduler such that
    $\sigma((\mathfrak{c},\mathbf{0}_\mathrm{r}))=\Dirac{\tau}$ and
    $\sigma((\mathfrak{c},\mathbf{0}_\mathrm{r})\cdot\rho)=\sigma'(\rho)$ whenever $\rho\ne\varepsilon$.
    Denote $\mathfrak{c}'_\mu:=(\fn{f},\loc',u(\nu,\mu))$ for $\mu\in\samples$.
    Since the case $\loc'=\lout{\fn{f}}$ is straightforward, we only consider the case that $\loc'\ne\lout{\fn{f}}$.
    From Proposition~\ref{prop:reachability} and the semantics at assignment labels, one has that
\begin{eqnarray*}
1&=&\probm^{\sigma}_{\mathfrak{c}}\left(T<\infty\right)\\
&=&\sum_{\rho\in H_{\mathfrak{c}}} \probht{\rho;\sigma}\\
&=& \sum_{\mu\in\samples}\sum_{\rho'\in H_{\mathfrak{c}'_\mu,\mu}} \probht{(\mathfrak{c},\mathbf{0}_\mathrm{r})\cdot\rho';\sigma} \\
&=& \sum_{\mu\in\samples}\sum_{\rho'\in H_{\mathfrak{c}'_\mu,\mu}}\sampdpd(\mu)\cdot \probht{\rho';\sigma'} \\
&=& \sum_{\mu\in\samples}\sampdpd(\mu)\cdot\left[\sum_{\rho'\in H_{\mathfrak{c}'_\mu,\mu}} \probht{\rho';\sigma'}\right] \\
&=& \sum_{\mu\in\samples}\sampdpd(\mu)\cdot\left[\sum_{\rho'\in H_{\mathfrak{c}'_\mu,\mu}} \probht{\wbsch{\rho'};\sigma'}\right] \\
&=& \sum_{\mu\in\samples}\sampdpd(\mu)\cdot\left[\sum_{\rho'\in H_{\mathfrak{c}'_\mu}} \probht{\rho';\sigma'}\right]\\
&=& \sum_{\mu\in\samples}\sampdpd(\mu)\cdot\probm_{\mathfrak{c}'_\mu}^{\sigma'}(T<\infty)\enskip,\\
\end{eqnarray*}
which implies that $\probm_{\mathfrak{c}'_\mu}^{\sigma'}(T<\infty)=1$ for $\mu\in\samples$. Then,
\begin{eqnarray*}
& &\expv^{\sigma}_{\mathfrak{c}}\left(T\right)\\
&=&\sum_{\rho\in H_{\mathfrak{c}}} (|\rho|-1)\cdot\probht{\rho;\sigma}\\
&=&\sum_{\rho\in H_{\mathfrak{c}}} (|\rho|-2)\cdot\probht{\rho;\sigma}+ \sum_{\rho\in H_{\mathfrak{c}}}\probht{\rho;\sigma}\\
&=&\probm_{\mathfrak{c}}^\sigma (T<\infty)+\sum_{\rho\in H_{\mathfrak{c}}} (|\rho|-2)\cdot\probht{\rho;\sigma} \\
&=& 1+\sum_{\mu\in\samples}\sum_{\rho'\in H_{\mathfrak{c}'_\mu,\mu}}(|(\mathfrak{c},\mathbf{0}_\mathrm{r})\cdot\rho'|-2)\cdot \probht{(\mathfrak{c},\mathbf{0}_\mathrm{r})\cdot\rho';\sigma} \\
&=& 1+\sum_{\mu\in\samples}\sum_{\rho'\in H_{\mathfrak{c}'_\mu,\mu}}(|\rho'|-1)\cdot \left(\sampdpd(\mu)\cdot \probht{\rho';\sigma'}\right) \\
&=& 1+\sum_{\mu\in\samples}\sampdpd(\mu)\cdot\left[\sum_{\rho'\in H_{\mathfrak{c}'_\mu,\mu}}(|\rho'|-1)\cdot \probht{\rho';\sigma'}\right] \\
&=& 1+\sum_{\mu\in\samples}\sampdpd(\mu)\cdot\left[\sum_{\rho'\in H_{\mathfrak{c}'_\mu,\mu}}(|\wbsch{\rho'}|-1)\cdot \probht{\wbsch{\rho'};\sigma'}\right] \\
&=& 1+\sum_{\mu\in\samples}\sampdpd(\mu)\cdot\left[\sum_{\rho'\in H_{\mathfrak{c}'_\mu}}(|\rho'|-1)\cdot \probht{\rho';\sigma'}\right]\\
&=& 1+\sum_{\mu\in\samples}\sampdpd(\mu)\cdot\expv_{\mathfrak{c}'_\mu}^{\sigma'}(T)\\
\end{eqnarray*}
Then by taking the supremum at the both sides of the equality
\[
\expv^{\sigma}_{\mathfrak{c}}\left(T\right)=1+\sum_{\mu\in\samples}\sampdpd(\mu)\cdot\expv_{\mathfrak{c}'_\mu}^{\sigma'}(T)\enskip,
\]
one obtains that
\[
1+\sum_{\mu\in\samples}\sampdpd(\mu)\cdot \expvt(\mathfrak{c}'_\mu)\le \expvt(\mathfrak{c})\enskip.
\]

\noindent{\em Case 2: Call.} $\loc\in\flocs{f}$ and $(\loc,(\fn{g},v),\loc')$ is the only triple in $\transitions{\fn{f}}$ with source label $\loc$ and value-passing function $v$. Let $\mathfrak{c}_1=(\fn{g},\lin{\fn{g}},v(\nu))$ and $\mathfrak{c}_2=(\fn{f},\loc',\nu)$.
    Let $\sigma_1,\sigma_2$ be any two well-behaved schedulers.

    We first consider the case when $\loc'\ne\lout{\fn{f}}$.
    Let $\sigma$ be any scheduler such that for any history $\rho$ ($\rho',\rho_1,\rho_2$ can be $\varepsilon$ below):
\begin{compactitem}
\item if $\rho=(\mathfrak{c},\mathbf{0}_{\mathrm{r}})$ then $\sigma(\rho):=\Dirac{\tau}$;
\item if $\rho=(\mathfrak{c},\mathbf{0}_{\mathrm{r}})\cdot\stksch{\left((\mathfrak{c}_1,\mu')\cdot \rho'\right)}{\mathfrak{c}_2}$ for some history $\rho'$ and valuation $\mu'\in\samples$, then
$\sigma(\rho):=\sigma_1\left((\mathfrak{c}_1,\mu')\cdot\rho'\right)$;
\item if
$\rho=(\mathfrak{c},\mathbf{0}_{\mathrm{r}})\cdot\stksch{\left((\mathfrak{c}_1,\mu_1)\cdot \rho_1\right)}{\mathfrak{c}_2}
\cdot (\mathfrak{c}_2,\mu_2)\cdot\rho_2$ for some histories $\rho_1,\rho_2$ and valuations $\mu_1,\mu_2\in\samples$, then $\sigma(\rho):=\sigma_2\left((\mathfrak{c}_2,\mu_2)\cdot\rho_2\right)$.
\end{compactitem}
For each $\mu\in\samples$, define $H'_\mu$ to be the set of all histories $\rho'$ such that $\rho'\cdot(\varepsilon,\mu_\bot)\in H_{\mathfrak{c}_1,\mu}$, where $\mu_\bot$ is any element in $\samples$.
Note that from our semantics, if $\rho'\cdot(\varepsilon,\mu_\bot)\in H_{\mathfrak{c}_1,\mu}$ (for $\mu\in\samples$) then $\last{\rho'}=((\fn{f}',\loc',\nu'),\mu')$ for some stack element $(\fn{f}',\loc',\nu')$ and $\mu'\in\samples$ such that (i) $\loc'\in\alocs{f'}\cup\clocs{f'}$ and (ii) $(\fn{f}',\loc',\nu')$ leads to $\lout{\fn{f}}$ in the next step regardless of samplings.
Then we have that
\begin{eqnarray*}
\sum_{\rho'\in H'_{\mathbf{0}_\mathrm{r}}}\probht{\rho';\sigma_1} &=& \sum_{\mu\in\samples}\sum_{\rho'\in H'_{\mathbf{0}_\mathrm{r}}}\sampdpd(\mu)\cdot\probht{\rho';\sigma_1} \\
&=&\sum_{\mu\in\samples}\sum_{\rho'\in H'_{\mathbf{0}_\mathrm{r}}}\probht{\rho'\cdot (\varepsilon,\mu);\sigma_1} \\
&=&\sum_{\rho\in H_{\mathfrak{c}_1}}\probht{\rho;\sigma_1} \\
&=& \probm_{\mathfrak{c}_1}^{\sigma_1}(T<\infty) \\
&\le& 1\enskip.
\end{eqnarray*}
Furthermore, we have that (below we use $\sum_{\mu_{1},\mu_2,\rho_{1},\rho_2}$ as a short hand for $\sum_{\mu_1\in\samples}\sum_{\mu_2\in\samples}\sum_{\rho_1\in H'_{\mu_1}}\sum_{\rho_2\in H_{\mathfrak{c}_2,\mu_2}}$):
\begin{eqnarray*}
1 &=& \probm_{\mathfrak{c}}^\sigma(T<\infty) \\
&=& \sum_{\rho\in H_{\mathfrak{c}}} \probht{\rho;\sigma}\\
&=& \sum_{\mu_{1},\mu_2,\rho_{1},\rho_2} \left[ \probht{(\mathfrak{c},\mathbf{0}_\mathrm{r})\cdot\stksch{\rho_1}{\mathfrak{c}_2}\cdot \rho_2;\sigma}\right]\\
&=&\sum_{\mu_{1},\mu_2,\rho_{1},\rho_2}\left[\sampdpd(\mu_1)\cdot\probht{\rho_1\cdot(\varepsilon,\mu_2);\sigma_1}\cdot \probht{\rho_2;\sigma_2}\right]\\
&=& \sum_{\mu_{1},\mu_2,\rho_{1},\rho_2}\left[\prod_{i=1}^2\sampdpd(\mu_i)\cdot\probht{\rho_i;\sigma_i}\right]\\
&=& \sum_{\mu_{1},\mu_2,\rho_{1},\rho_2}\left[\prod_{i=1}^2\sampdpd(\mu_i)\cdot\probht{\wbsch{\rho_i};\sigma_i}
\right]\\
&=& \sum_{\mu_1\in\samples}\sum_{\mu_2\in\samples}\sum_{\rho_1\in H'_{\mathbf{0}_\mathrm{r}}}\sum_{\rho_2\in H_{\mathfrak{c}_2}}\left[\prod_{i=1}^2\sampdpd(\mu_i)\cdot\probht{\rho_i;\sigma_i}
\right]\\
&=& \sum_{\rho_1\in H'_{\mathbf{0}_\mathrm{r}}}\sum_{\rho_2\in H_{\mathfrak{c}_2}}\probht{\rho_1;\sigma_1}\cdot \probht{\rho_2;\sigma_2} \\
&=& \left(\sum_{\rho_1\in H'_{\mathbf{0}_\mathrm{r}}}\probht{\rho_1;\sigma_1}\right)\cdot \left(\sum_{\rho_2\in H_{\mathfrak{c}_2}}\probht{\rho_2;\sigma_2}\right) \\
&=& \left(\sum_{\rho_1\in H'_{\mathbf{0}_\mathrm{r}}}\probht{\rho_1;\sigma_1}\right)\cdot \probm_{\mathfrak{c}_2}^{\sigma_2} \left(T<\infty\right)\enskip.
\end{eqnarray*}
Thus,
\begin{equation}\label{eq:proof:callaux}
\sum_{\rho_1\in H'_{\mathbf{0}_\mathrm{r}}}\probht{\rho_1;\sigma_1}=\sum_{\rho_2\in H_{\mathfrak{c}_2}}\probht{\rho_2;\sigma_2}=\probm_{\mathfrak{c}_2}^{\sigma_2} \left(T<\infty\right)=1.
\end{equation}

Then we have (we continue using $\sum_{\mu_{1},\mu_2,\rho_{1},\rho_2}$ as a short hand for $\sum_{\mu_1\in\samples}\sum_{\mu_2\in\samples}\sum_{\rho_1\in H'_{\mu_1}}\sum_{\rho_2\in H_{\mathfrak{c}_2,\mu_2}}$)
\begin{eqnarray*}
& &\expv^{\sigma}_{\mathfrak{c}}\left(T\right)\\
&=&\sum_{\rho\in H_{\mathfrak{c}}} (|\rho|-1)\cdot\probht{\rho;\sigma} \\
&{=}& \sum_{\mu_{1},\mu_2,\rho_{1},\rho_2}\Big[(|(\mathfrak{c},\mathbf{0}_\mathrm{r})\cdot \stksch{\rho_1}{\mathfrak{c}_2}\cdot \rho_2|-1)\cdot{} \\
& &\qquad\probht{{(\mathfrak{c},\mathbf{0}_\mathrm{r})}{\cdot}{\stksch{\rho_1}{\mathfrak{c}_2}}{\cdot}{\rho_2};\sigma}\Big]\\
&=&\sum_{\mu_{1},\mu_2,\rho_{1},\rho_2}\Big[\sampdpd(\mu_1)\cdot\left(\sum_{i=1}^2|\rho_i|\right)\cdot{} \\
& & \qquad\probht{\rho_1\cdot(\varepsilon,\mu_2);\sigma_1}\cdot \probht{\rho_2;\sigma_2}\Big] \\
&=&\sum_{\mu_{1},\mu_2,\rho_{1},\rho_2}\left[\prod_{i=1}^2\left(\sampdpd(\mu_i)\cdot \probht{\rho_i;\sigma_i}\right)\right]\cdot
\left(\sum_{i=1}^2|\rho_i|\right) \\
&=&\sum_{\mu_{1},\mu_2,\rho_{1},\rho_2}\left[\prod_{i=1}^2\left(\sampdpd(\mu_i)\cdot \probht{\wbsch{\rho_i};\sigma_i}\right)\right]\cdot
\left(\sum_{i=1}^2|\wbsch{\rho_i}|\right) \\
&=& \sum_{\mu_1\in\samples}\sum_{\mu_2\in\samples} \sum_{\rho_1\in H'_{\mathbf{0}_\mathrm{r}}}\sum_{\rho_2\in H_{\mathfrak{c}_2}}\\
& & \quad\left[\prod_{i=1}^2\left(\sampdpd(\mu_i)\cdot \probht{\rho_i;\sigma_i}\right)\right]\cdot\left(\sum_{i=1}^2|\rho_i|\right)\\
&=& \sum_{\rho_1\in H'_{\mathbf{0}_\mathrm{r}}}\sum_{\rho_2\in H_{\mathfrak{c}_2}}
\left[\prod_{i=1}^2\probht{\rho_i;\sigma_i}\right]\cdot\left(\sum_{i=1}^2|\rho_i|\right)
\\
&=& \sum_{\rho_1\in H'_{\mathbf{0}_\mathrm{r}}}\sum_{\rho_2\in H_{\mathfrak{c}_2}} \left[|\rho_1|\cdot\prod_{i=1}^2\probht{\rho_i;\sigma_i}\right]  \\
& &{}+\sum_{\rho_1\in H'_{\mathbf{0}_\mathrm{r}}}\sum_{\rho_2\in H_{\mathfrak{c}_2}} \left[|\rho_2|\cdot\prod_{i=1}^2\probht{\rho_i;\sigma_i}\right] \\
&=& \left(\sum_{\rho_1\in H'_{\mathbf{0}_\mathrm{r}}}|\rho_1|\cdot\probht{\rho_1;\sigma_1}\right)\cdot\left(
\sum_{\rho_2\in H_{\mathfrak{c}_2}} \probht{\rho_2;\sigma_2}\right)  \\
& &{}+\left(\sum_{\rho_1\in H'_{\mathbf{0}_\mathrm{r}}}\probht{\rho_1;\sigma_1}\right)\cdot\left(
\sum_{\rho_2\in H_{\mathfrak{c}_2}} |\rho_2|\cdot\probht{\rho_2;\sigma_2}\right) \\
\end{eqnarray*}
Hence from (\ref{eq:proof:callaux}),
\begin{eqnarray*}
&=& \sum_{\rho_1\in H'_{\mathbf{0}_\mathrm{r}}}|\rho_1|\cdot\probht{\rho_1;\sigma_1}+\sum_{\rho_2\in H_{\mathfrak{c}_2}} |\rho_2|\cdot \probht{\rho_2;\sigma_2} \\
&=& \sum_{\rho_1\in H'_{\mathbf{0}_\mathrm{r}}}\sum_{\mu\in\samples}(|\rho_1\cdot (\varepsilon,\mu)|-1)\cdot\sampdpd(\mu)\cdot\probht{\rho_1;\sigma_1}\\
& &\quad{}+\sum_{\rho_2\in H_{\mathfrak{c}_2}} (|\rho_2|-1)\cdot \probht{\rho_2;\sigma_2} \\
& &\quad{}+\sum_{\rho_2\in H_{\mathfrak{c}_2}} \probht{\rho_2;\sigma_2} \\
&=& \sum_{\rho_1\in H'_{\mathbf{0}_\mathrm{r}}}\sum_{\mu\in\samples}(|\rho_1\cdot (\varepsilon,\mu)|-1)\cdot\probht{\rho_1\cdot(\varepsilon,\mu);\sigma_1} \\
& & \quad{}+1+\expv_{\mathfrak{c}_2}^{\sigma_2}(T)\\
&=& \left(\sum_{\rho'\in H_{\mathfrak{c}_1}}(|\rho'|-1)\cdot\probht{\rho';\sigma_1} \right) +1+\expv_{\mathfrak{c}_2}^{\sigma_2}(T)\\
&=& 1+\expv_{\mathfrak{c}_1}^{\sigma_1}(T)+\expv_{\mathfrak{c}_2}^{\sigma_2}(T)
\end{eqnarray*}
By taking the supremum at the both sides, one obtains directly that
\[
\expvt(\mathfrak{c})\ge 1+\expvt(\mathfrak{c}_1)+\expvt(\mathfrak{c}_2)\enskip.
\]

Now we consider the simpler case $\loc'=\lout{\fn{f}}$.
Choose $\sigma$ to be any scheduler such that for any history $\rho$ ($\rho'$ can be $\varepsilon$ below):
\begin{compactitem}
\item if $\rho=(\mathfrak{c},\mathbf{0}_{\mathrm{r}})$ then $\sigma(\rho)=\Dirac{\tau}$;
\item if $\rho=(\mathfrak{c},\mathbf{0}_{\mathrm{r}})\cdot (\mathfrak{c}_1,\mu') \cdot\rho'$ for some history $\rho'$ and valuation $\mu'\in\samples$, then
$\sigma(\rho):=\sigma_1\left((\mathfrak{c}_1,\mu')\cdot\rho'\right)$.
\end{compactitem}
We have
\begin{eqnarray*}
1&=&\probm^{\sigma}_{\mathfrak{c}}\left(T<\infty\right)\\
&=&\sum_{\rho\in H_{\mathfrak{c}}} \probht{\rho;\sigma}\\
&=& \sum_{\mu\in\samples}\sum_{\rho'\in H_{\mathfrak{c}_1,\mu}} \probht{(\mathfrak{c},\mathbf{0}_\mathrm{r})\cdot\rho';\sigma} \\
&=& \sum_{\mu\in\samples}\sum_{\rho'\in H_{\mathfrak{c}_1,\mu}}\sampdpd(\mu)\cdot \probht{\rho';\sigma_1} \\
&=& \sum_{\mu\in\samples}\sampdpd(\mu)\cdot\left[\sum_{\rho'\in H_{\mathfrak{c}_1,\mu}} \probht{\rho';\sigma_1}\right] \\
&=& \sum_{\mu\in\samples}\sampdpd(\mu)\cdot\left[\sum_{\rho'\in H_{\mathfrak{c}_1,\mu}} \probht{\wbsch{\rho'};\sigma_1}\right] \\
&=& \sum_{\mu\in\samples}\sampdpd(\mu)\cdot\left[\sum_{\rho'\in H_{\mathfrak{c}_1}} \probht{\rho';\sigma_1}\right]\\
&=& \probm_{\mathfrak{c}_1}^{\sigma_1}(T<\infty)\enskip.\\
\end{eqnarray*}

Then we can obtain that
\begin{eqnarray*}
& &\expv^{\sigma}_{\mathfrak{c}}\left(T\right)\\
&=&\sum_{\rho\in H_{\mathfrak{c}}} (|\rho|-1)\cdot\probht{\rho;\sigma}\\
&=&\sum_{\rho\in H_{\mathfrak{c}}} (|\rho|-2)\cdot\probht{\rho;\sigma}+ \sum_{\rho\in H_{\mathfrak{c}}}\probht{\rho;\sigma}\\
&=&\probm_{\mathfrak{c}}^\sigma (T<\infty)+\sum_{\rho\in H_{\mathfrak{c}}} (|\rho|-2)\cdot\probht{\rho;\sigma} \\
&=& 1+\sum_{\mu\in\samples}\sum_{\rho'\in H_{\mathfrak{c}_1,\mu}}(|(\mathfrak{c},\mathbf{0}_\mathrm{r})\cdot\rho'|-2)\cdot \probht{(\mathfrak{c},\mathbf{0}_\mathrm{r})\cdot\rho';\sigma} \\
&=& 1+\sum_{\mu\in\samples}\sum_{\rho'\in H_{\mathfrak{c}_1,\mu}}(|\rho'|-1)\cdot \left(\sampdpd(\mu)\cdot \probht{\rho';\sigma_1}\right) \\
&=& 1+\sum_{\mu\in\samples}\sampdpd(\mu)\cdot\left[\sum_{\rho'\in H_{\mathfrak{c}_1,\mu}}(|\rho'|-1)\cdot \probht{\rho';\sigma_1}\right] \\
&=& 1+\sum_{\mu\in\samples}\sampdpd(\mu)\cdot\left[\sum_{\rho'\in H_{\mathfrak{c}_1,\mu}}(|\wbsch{\rho'}|-1)\cdot \probht{\wbsch{\rho'};\sigma_1}\right] \\
&=& 1+\sum_{\mu\in\samples}\sampdpd(\mu)\cdot\left[\sum_{\rho'\in H_{\mathfrak{c}_1}}(|\rho'|-1)\cdot \probht{\rho';\sigma_1}\right] \\
&=& 1+\sum_{\rho'\in H_{\mathfrak{c}_1}}(|\rho'|-1)\cdot \probht{\rho';\sigma_1} \\
&=& 1+\expv_{\mathfrak{c}_1}^{\sigma_1}(T)\enskip.
\end{eqnarray*}
By taking the supremum at the both sides of the equality
\[
\expv^{\sigma}_{\mathfrak{c}}\left(T\right)= 1+\expv_{\mathfrak{c}_1}^{\sigma_1}(T)
\]
and the fact that $\expvt(\mathfrak{c}_2)=0$, we have again that
\[
\expvt(\mathfrak{c})\ge 1+\expvt(\mathfrak{c}_1)+\expvt(\mathfrak{c}_2)\enskip.
\]

\noindent{\em Case 3: Branching.} $\loc\in\flocs{f}$ and $(\loc,\phi,\loc_1),(\loc,\neg\phi,\loc_2)$ are namely the two triple in $\transitions{\fn{f}}$ with source label $\loc$ and propositional arithmetic predicate $\phi$. Denote $\mathfrak{c}_1:=(\fn{f}, \loc_1,\nu)$ and $\mathfrak{c}_2:=(\fn{f}, \loc_2,\nu)$.
Let $\sigma_1,\sigma_2$ be two arbitrary well-behaved schedulers.
Define
\[
\mathfrak{c}':=
\begin{cases}
\mathfrak{c}_1 & \mbox{if }\nu\models\phi \\
\mathfrak{c}_2 & \mbox{if }\nu\models\neg\phi \\
\end{cases}
\mbox{ and }
\sigma':=
\begin{cases}
\sigma_1 & \mbox{if }\nu\models\phi \\
\sigma_2 & \mbox{if }\nu\models\neg\phi \\
\end{cases}
\enskip.
\]
In the case that $\mathfrak{c}'$ is terminal, it holds straightforwardly that
\[
\expvt\left(\mathfrak{c}\right)\ge 1+\expvt\left(\mathfrak{c}'\right)= 1+\mathbf{1}_{\nu\models\phi}\cdot\expvt\left(\mathfrak{c}_1\right)+\mathbf{1}_{\nu\models\neg\phi}\cdot\expvt\left(\mathfrak{c}_2\right)\enskip.
\]
Below we consider that $\mathfrak{c}'$ is non-terminal.
Choose $\sigma$ to be any scheduler such that (below $\rho'$ can be $\varepsilon$)
\begin{compactitem}
\item $\sigma\left((\mathfrak{c},\mathbf{0}_\mathrm{r})\right)=\Dirac{\tau}$, and
\item $\sigma\left((\mathfrak{c},\mathbf{0}_\mathrm{r})\cdot (\mathfrak{c}_1,\mu)\cdot \rho'\right)=\sigma_1\left((\mathfrak{c}_1,\mu)\cdot \rho'\right)$ for any valuation $\mu\in\samples$ and history $\rho'$, and
\item $\sigma\left((\mathfrak{c},\mathbf{0}_\mathrm{r})\cdot (\mathfrak{c}_2,\mu)\cdot \rho'\right)=\sigma_2\left((\mathfrak{c}_2,\mu)\cdot \rho'\right)$ for any valuation $\mu\in\samples$ and history $\rho'$.
\end{compactitem}
We have

\begin{eqnarray*}
1&=&\probm^{\sigma}_{\mathfrak{c}}\left(T<\infty\right)\\
&=&\sum_{\rho\in H_{\mathfrak{c}}} \probht{\rho;\sigma}\\
&=&\sum_{\mu\in\samples}\sum_{\rho'\in H_{\mathfrak{c}',\mu}} \probht{(\mathfrak{c},\mathbf{0}_\mathrm{r})\cdot\rho';\sigma}\\
&=&\sum_{\mu\in\samples}\sum_{\rho'\in H_{\mathfrak{c}',\mu}}  \sampdpd(\mu)\cdot\probht{\rho';\sigma'}\\
&=&\sum_{\mu\in\samples}\sum_{\rho'\in H_{\mathfrak{c}',\mu}}  \sampdpd(\mu)\cdot\probht{\wbsch{\rho'};\sigma'}\\
&=&
\sum_{\mu\in\samples}\sum_{\rho'\in H_{\mathfrak{c}'}} \sampdpd(\mu)\cdot\probht{\rho';\sigma'}\\
&=&
\sum_{\rho'\in H_{\mathfrak{c}'}} \probht{\rho';\sigma'}\\
&=& \probm^{\sigma'}_{\mathfrak{c}'}\left(T<\infty\right)\enskip.
\end{eqnarray*}

Then
\begin{eqnarray*}
& &\expv^{\sigma}_{\mathfrak{c}}\left(T\right)\\
&=&\sum_{\rho\in H_{\mathfrak{c}}} (|\rho|-1)\cdot \probht{\rho;\sigma}\\
&=&\sum_{\rho\in H_{\mathfrak{c}}} (|\rho|-2)\cdot \probht{\rho;\sigma}+\sum_{\rho\in H_{\mathfrak{c}}}\probht{\rho;\sigma}\\
&=&\probm_\mathfrak{c}^\sigma (T<\infty) + \sum_{\rho\in H_{\mathfrak{c}}} (|\rho|-2)\cdot \probht{\rho;\sigma}\\
&=&1+\sum_{\mu\in\samples}\sum_{\rho'\in H_{\mathfrak{c}',\mu}} (|(\mathfrak{c},\mathbf{0}_\mathrm{r})\cdot\rho'|-2)\cdot\probht{(\mathfrak{c},\mathbf{0}_\mathrm{r})\cdot\rho';\sigma}\\
&=&1+\sum_{\mu\in\samples}\sum_{\rho'\in H_{\mathfrak{c}',\mu}} (|(\mathfrak{c},\mathbf{0}_\mathrm{r})\cdot\rho'|-2)\cdot \sampdpd(\mu)\cdot\probht{\rho';\sigma'}\\
&=&1+\sum_{\mu\in\samples}\sum_{\rho'\in H_{\mathfrak{c}',\mu}} (|\wbsch{\rho'}|-1)\cdot \sampdpd(\mu)\cdot\probht{\wbsch{\rho'};\sigma'}\\
&=&1+
\sum_{\mu\in\samples}\sum_{\rho'\in H_{\mathfrak{c}'}} (|\rho'|-1)\cdot \sampdpd(\mu)\cdot\probht{\rho';\sigma'}\\
&=& 1+
\sum_{\rho'\in H_{\mathfrak{c}'}} (|\rho'|-1)\cdot \probht{\rho';\sigma'}\\
&=&1+\expv^{\sigma'}_{\mathfrak{c}'}\left(T\right)\\
&=&1+\mathbf{1}_{\nu\models\phi}\cdot\expv^{\sigma_1}_{\mathfrak{c}_1}\left(T\right)+\mathbf{1}_{\nu\models\neg\phi}\cdot\expv^{\sigma_2}_{\mathfrak{c}_2}\left(T\right)\enskip.
\end{eqnarray*}
By taking the supremum at the both sides of the equality
\[
\expv^{\sigma}_{\mathfrak{c}}\left(T\right)=1+\mathbf{1}_{\nu\models\phi}\cdot\expv^{\sigma_1}_{\mathfrak{c}_1}\left(T\right)+\mathbf{1}_{\nu\models\neg\phi}\cdot\expv^{\sigma_2}_{\mathfrak{c}_2}\left(T\right)\enskip.
\]
one obtains that
\[
\expvt\left(\mathfrak{c}\right)\ge 1+\mathbf{1}_{\nu\models\phi}\cdot\expvt\left(\mathfrak{c}_1\right)+\mathbf{1}_{\nu\models\neg\phi}\cdot\expvt\left(\mathfrak{c}_2\right)\enskip.
\]

\noindent{\em Case 4: Nondeterminism.} $\loc\in\flocs{f}$ and $(\loc,\star,\loc_1),(\loc,\star,\loc_2)$ are namely the two triple in $\transitions{\fn{f}}$ with source label $\loc$ such that $\loc_1$ (resp. $\loc_2$) refers to the \textbf{then}-(resp. \textbf{else}-)branch.
    Denote $\mathfrak{c}_1:=(\fn{f}, \loc_1,\nu)$ and $\mathfrak{c}_2:=(\fn{f}, \loc_2,\nu)$.
    Let $\sigma_1,\sigma_2$ be two arbitrary well-behaved schedulers. Define
\[
(\mathfrak{c}',\sigma'):=
\begin{cases}
(\mathfrak{c}_1,\sigma_1) & \mbox{if }\expv^{\sigma_1}_{\mathfrak{c}_1}(T)\ge \expv^{\sigma_2}_{\mathfrak{c}_2}(T)\\
(\mathfrak{c}_2,\sigma_2) & \mbox{otherwise } \\
\end{cases}.
\]
Note that $\mathfrak{c}'$ is guaranteed to be non-terminal from our setting for programs and semantics.
Choose $\sigma$ to be any scheduler such that ($\rho'$ can be $\varepsilon$ below)
\begin{compactitem}
\item
\[
\sigma\left((\mathfrak{c},\mathbf{0}_\mathrm{r})\right)=
\begin{cases}
\Dirac{\mathbf{th}} & \mbox{if }\expv^{\sigma_1}_{\mathfrak{c}_1}(T)\ge \expv^{\sigma_2}_{\mathfrak{c}_2}(T)\\
\Dirac{\mathbf{el}} & \mbox{otherwise}
\end{cases},
\]
\item $\sigma\left((\mathfrak{c},\mathbf{0}_\mathrm{r})\cdot (\mathfrak{c}_1,\mu)\cdot \rho'\right)=\sigma_1\left((\mathfrak{c}_1,\mu)\cdot \rho'\right)$ for any valuation $\mu\in\samples$ and history $\rho'$, and
\item $\sigma\left((\mathfrak{c},\mathbf{0}_\mathrm{r})\cdot (\mathfrak{c}_2,\mu)\cdot \rho'\right)=\sigma_2\left((\mathfrak{c}_2,\mu)\cdot \rho'\right)$ for any valuation $\mu\in\samples$ and history $\rho'$.
\end{compactitem}
We have
\begin{eqnarray*}
1&=&\probm^{\sigma}_{\mathfrak{c}}\left(T<\infty\right)\\
&=&\sum_{\rho\in H_{\mathfrak{c}}} \probht{\rho;\sigma}\\
&=&\sum_{\mu\in\samples}\sum_{\rho'\in H_{\mathfrak{c}',\mu}} \probht{(\mathfrak{c},\mathbf{0}_\mathrm{r})\cdot\rho';\sigma'}\\
&=&\sum_{\mu\in\samples}\sum_{\rho'\in H_{\mathfrak{c}',\mu}} \sampdpd(\mu)\cdot\probht{\rho';\sigma'}\\
&=&\sum_{\mu\in\samples}\sum_{\rho'\in H_{\mathfrak{c}'}} \sampdpd(\mu)\cdot\probht{\wbsch{\rho'};\sigma'}\\
&=&\sum_{\rho'\in H_{\mathfrak{c}'}} \probht{\rho';\sigma'}\\
&=&\probm^{\sigma'}_{\mathfrak{c}'}\left(T<\infty\right)\enskip.
\end{eqnarray*}

Then
\begin{eqnarray*}
& &\expv^{\sigma}_{\mathfrak{c}}\left(T\right)\\
&=&\sum_{\rho\in H_{\mathfrak{c}}} (|\rho|-1)\cdot \probht{\rho;\sigma}\\
&=&\sum_{\rho\in H_{\mathfrak{c}}} (|\rho|-2)\cdot \probht{\rho;\sigma}+\sum_{\rho\in H_{\mathfrak{c}}}\probht{\rho;\sigma}\\
&=&\probm_\mathfrak{c}^\sigma (T<\infty) + \sum_{\rho\in H_{\mathfrak{c}}} (|\rho|-2)\cdot \probht{\rho;\sigma}\\
&=&1+\sum_{\mu\in\samples}\sum_{\rho'\in H_{\mathfrak{c}',\mu}} (|(\mathfrak{c},\mathbf{0}_\mathrm{r})\cdot\rho'|-2)\cdot \probht{(\mathfrak{c},\mathbf{0}_\mathrm{r})\cdot\rho';\sigma'}\\
&=&1+\sum_{\mu\in\samples}\sum_{\rho'\in H_{\mathfrak{c}',\mu}} (|\rho'|-1)\cdot \sampdpd(\mu)\cdot\probht{\rho';\sigma'}\\
&=&1+\sum_{\mu\in\samples}\sum_{\rho'\in H_{\mathfrak{c}'}} (|\wbsch{\rho'}|-1)\cdot \sampdpd(\mu)\cdot\probht{\wbsch{\rho'};\sigma'}\\
&=&1+\sum_{\rho'\in H_{\mathfrak{c}'}} (|\rho'|-1)\cdot \probht{\rho';\sigma'}\\
&=&1+\expv^{\sigma'}_{\mathfrak{c}'}\left(T\right)\\
&=&1+\max\left\{\expv^{\sigma_1}_{\mathfrak{c}_1}\left(T\right),\expv^{\sigma_2}_{\mathfrak{c}_2}\left(T\right)\right\}\enskip.
\end{eqnarray*}
By taking the supremum at the both sides of the equality
\[
\expv^{\sigma}_{\mathfrak{c}}\left(T\right)=1+\max\left\{\expv^{\sigma_1}_{\mathfrak{c}_1}\left(T\right),\expv^{\sigma_2}_{\mathfrak{c}_2}\left(T\right)\right\}
\]
one obtains that
\[
\expvt\left(\mathfrak{c}\right)\ge 1+\max\left\{\expvt\left(\mathfrak{c}_1\right), \expvt\left(\mathfrak{c}_2\right)\right\}\enskip.
\]
Hence, $\expvt$ is a ranking measure function with corresponding $\epsilon=1$ (cf. Definition~\ref{def:mfunc}).
\end{IEEEproof}

\section{Details for Section~\ref{sect:lowerbound}}\label{app:lowerbound}

We first introduce two classical theorems, namely Doob's Convergence Theorem and Optional Stopping Theorem as follows.

\begin{theorem}[Doob's Convergence Theorem{~\cite[Chapter~11]{probabilitycambridge}}]\label{thm:mconv}
Consider any supermartingale $\{X_n\}_{n\in\Nset_0}$ (adapted to some filtration)  such that $\sup_n\left\{\expv\left(|X_n|\right)\right\}<\infty$.
Then there exists a random variable $Y$ such that $Y=\lim\limits_{n\rightarrow\infty} X_n$ a.s. and  $\expv\left(|Y|\right)<\infty$.\hfill\IEEEQEDclosed
\end{theorem}

The following version of Optional Stopping Theorem
is an extension of the one from \cite[Chapter 10]{probabilitycambridge}.
In the proof of the following theorem, for a stopping time $R$ and a nonnegative integer $n\in\Nset_0$, we denote by $R\wedge n$ the random variable $\min\{R,n\}$.

\begin{theorem}[Optional Stopping Theorem\footnote{
cf. \url{https://en.wikipedia.org/wiki/Optional_stopping_theorem}}]\label{thm:optstopping}
Consider any stopping time $R$ w.r.t a filtration $\{\mathcal{F}_n\}_{n\in\Nset_0}$ and any martingale (resp. supermartingale) $\{X_n\}_{n\in\Nset_0}$ adapted to $\{\mathcal{F}_n\}_{n\in\Nset_0}$.
Then $\expv\left(|Y|\right)<\infty$ and $\expv\left(Y\right)=\expv(X_0)$ (resp. $\expv\left(Y\right)\le\expv(X_0)$) if one of the following conditions hold:
\begin{compactenum}
\item there exists an $M\in (0,\infty)$ such that $\left|X_{R\wedge n}\right|<M$ a.s. for all $n\in\Nset_0$, and $Y=\lim\limits_{n\rightarrow\infty}X_{R\wedge n}$ a.s., where the existence of $Y$ follows from Doob's Convergence Theorem (Theorem~\ref{thm:mconv});
\item $\expv(R)<\infty$, $Y=X_R$ and there exists a $c\in (0,\infty)$ such that for all $n\in\Nset_0$, $\condexpv{\left|X_{n+1}-X_n\right|}{\mathcal{F}_n}\le c$ a.s.
\end{compactenum}
Moreover,
\begin{compactitem}
\item[3)] if $\probm(R<\infty)=1$ and $X_n(\omega)\ge 0$ for all $n,\omega$, then  $\expv\left(X_R\right)\le\expv(X_0)$.
\end{compactitem}
\end{theorem}
\begin{IEEEproof}
The first item of the theorem follows directly from Dominated Convergence Theorem~\cite[Chapter 6.2]{probabilitycambridge} and properties for stopped processes (cf.~\cite[Chapter 10.9]{probabilitycambridge}).
The third item is from \cite[Chapter 10.10(d)]{probabilitycambridge}.
Below we prove the second item.
We have for every $n\in\Nset_0$,
\begin{eqnarray*}
\left|X_{R\wedge n}\right|&=& \left|X_0+\sum_{k=0}^{R\wedge n-1} \left(X_{k+1}-X_k\right)\right| \\
&=& \left|X_0+\sum_{k=0}^\infty \left(X_{k+1}-X_k\right)\cdot \mathbf{1}_{R>k\wedge n>k}\right|\\
&\le& \left|X_0\right|+\sum_{k=0}^\infty \left|\left(X_{k+1}-X_k\right)\cdot \mathbf{1}_{R>k\wedge n>k}\right| \\
&\le& \left|X_0\right|+\sum_{k=0}^\infty \left|\left(X_{k+1}-X_k\right)\cdot \mathbf{1}_{R>k}\right|\enskip. \\
\end{eqnarray*}
Note that
\begin{eqnarray*}
& &   \expv\left(\left|X_0\right|+\sum_{k=0}^\infty \left|\left(X_{k+1}-X_k\right)\cdot \mathbf{1}_{R>k}\right|\right) \\
&=& \mbox{(By Monotone Convergence Theorem~\cite[Chap. 6]{probabilitycambridge})} \\
& & \expv\left(\left|X_0\right|\right)+\sum_{k=0}^\infty \expv\left(\left|\left(X_{k+1}-X_k\right)\cdot \mathbf{1}_{R>k}\right|\right) \\
&=& \expv\left(\left|X_0\right|\right)+\sum_{k=0}^\infty \expv\left(\left|X_{k+1}-X_k\right|\cdot \mathbf{1}_{R>k}\right) \\
&=& \mbox{(By (E4))}\\
& & \expv\left(\left|X_0\right|\right)+\sum_{k=0}^\infty \expv\left(\condexpv{\left|X_{k+1}-X_k\right|\cdot \mathbf{1}_{R>k}}{\mathcal{F}_k}\right) \\
&=& \mbox{(by (E8))} \\
& &
\expv\left(\left|X_0\right|\right)+\sum_{k=0}^\infty \expv\left(\condexpv{\left|X_{k+1}-X_k\right|}{\mathcal{F}_k}\cdot \mathbf{1}_{R>k}\right) \\
&\le & \expv\left(\left|X_0\right|\right)+\sum_{k=0}^\infty \expv\left(c\cdot \mathbf{1}_{R>k}\right) \\
&=& \expv\left(\left|X_0\right|\right)+\sum_{k=0}^\infty c\cdot \probm\left(R>k\right) \\
&=& \expv\left(\left|X_0\right|\right)+\sum_{k=0}^\infty c\cdot \probm\left(k<R<\infty\right) \\
&=& \expv\left(\left|X_0\right|\right)+ c\cdot \expv(R) \\
&<& \infty\enskip.
\end{eqnarray*}
Thus, by Dominated Convergence Theorem~\cite[Chapter 6.2]{probabilitycambridge} and the fact that $X_R=\lim\limits_{n\rightarrow\infty} X_{R\wedge n}$ a.s.,
\[
\expv\left(X_R\right)=\expv\left(\lim\limits_{n\rightarrow\infty} X_{R\wedge n}\right)=\lim\limits_{n\rightarrow\infty}\expv\left(X_{R\wedge n}\right)\enskip.
\]
Then the result follows from properties for the stopped process $\{X_{R\wedge n}\}_{n\in\Nset_0}$ (cf.~\cite[Chapter 10.9]{probabilitycambridge}).
\end{IEEEproof}

We use Theorem~\ref{thm:optstopping} to prove Proposition~\ref{prop:cbrsupm}.

\noindent\textbf{Proposition~\ref{prop:cbrsupm}.}
Consider any conditionally difference-bounded ranking supermartingale $\Gamma=\{X_n\}_{n\in\Nset_0}$ adapted to a filtration $\{\mathcal{F}_n\}_{n\in\Nset_0}$ with $\epsilon$ given as in Definition~\ref{def:rsupm}.
If (i) for every $n\in\Nset_0$, it holds for all $\omega$ that $X_n(\omega)=0$ implies  $X_{n+1}(\omega)=0$, and (ii)
there exists $\delta\in (0,\infty)$ such that for all $n\in\Nset_0$, it holds a.s. that
$\condexpv{X_{n+1}}{\mathcal{F}_n}\ge X_n-\delta\cdot\mathbf{1}_{X_n>0}$~,
then $\expv(\stopping{\Gamma})\ge\frac{\expv(X_0)}{\delta}$.
\begin{IEEEproof}
Let $c\in (0,\infty)$ be such that for all $n\in\Nset_0$, $\condexpv{|X_{n+1}-X_n|}{\mathcal{F}_n}\le c$ a.s.
Define the stochastic process $\{Y_n\}_{n\in\Nset_0}$ adapted to $\{\mathcal{F}_n\}_{n\in\Nset_0}$ by:
\[
Y_n=X_n+\delta\cdot\min \{n,\stopping{\Gamma}\}\enskip.
\]
We prove that $Y_n$ is a submartingale.
For each $n\in\Nset_0$, define the following random variable
\[
U_n:=\min\{\stopping{\Gamma}, n+1\}-\min\{\stopping{\Gamma},n\}(=\mathbf{1}_{\stopping{\Gamma}>n})\enskip.
\]
We have that the followings hold a.s. for all $n$:
\begin{eqnarray*}
& &\condexpv{Y_{n+1}}{\mathcal{F}_n}-Y_n\\
&=& \mbox{\S~By (E5), (E6) \S} \\
& & \condexpv{X_{n+1}}{\mathcal{F}_n}-X_n +\delta\cdot\condexpv{U_n}{\mathcal{F}_n} \\
&=& \condexpv{X_{n+1}}{\mathcal{F}_n}-X_n+\delta\cdot\expv(\mathbf{1}_{\stopping{\Gamma}>n}\mid\mathcal{F}_n)\\
&=& \mbox{\S~By (E5) \S} \\
& &\condexpv{X_{n+1}}{\mathcal{F}_n}-X_n+\delta\cdot\mathbf{1}_{\stopping{\Gamma}>n} \\
&\ge &-\delta\cdot\mathbf{1}_{X_n> 0}+\delta\cdot\mathbf{1}_{\stopping{\Gamma}>n}\\
&=& 0~\mbox{\S~By $\stopping{\Gamma}(\omega)>n$ iff $X_n(\omega)> 0$ \S}\enskip.
\end{eqnarray*}
Also it holds a.s. that
\begin{eqnarray*}
& &\condexpv{\left|Y_{n+1}-Y_n\right|}{\mathcal{F}_n} \\
&=&\condexpv{\left|X_{n+1}-X_n+\delta\cdot U_n\right|}{\mathcal{F}_n} \\
&\le & \mbox{\S~By (E6), (E10) \S} \\
& & \condexpv{\left|X_{n+1}-X_n\right|}{\mathcal{F}_n}+\delta \\
&\le& c+\delta\enskip.
\end{eqnarray*}
Hence, by applying Proposition~\ref{prop:rsupm} and Optional Stopping Theorem (cf. Item 2 of Theorem~\ref{thm:optstopping}) to the supermartingale $\{-Y_n\}_{n\in\Nset_0}$ and the stopping time $\stopping{\Gamma}$, we obtain that
\begin{eqnarray*}
-\expv\left(X_{\stopping{\Gamma}}+\delta\cdot \stopping{\Gamma}\right)&=&\expv(-Y_{\stopping{\Gamma}})\\
&\le& \expv(-Y_0)=\expv(-X_0)
\end{eqnarray*}
It follows from $X_{\stopping{\Gamma}}=0$ a.s. that $\expv(\stopping{\Gamma})\ge \frac{\expv(X_0)}{\delta}$.
\end{IEEEproof}

Now we formally define the notion of conditionally difference-bounded ranking measure functions.
We fix a nondeterministic recursive probabilistic program $W$ together with its associated CFG taking the form~(\ref{eq:cfg}) and a sampling function $\Upsilon$.

\begin{definition}[Conditionally Difference-Bounded Ranking Measure Functions]\label{def:bmfunc}
A ranking measure function $h$ is \emph{conditionally difference-bounded} if there exist $\delta,\zeta\in(0,\infty)$ such that
for all stack elements $(\fn{f},\loc,\nu)$ satisfying $h(\fn{f},\loc,\nu)<\infty$,
the following conditions hold:
\begin{compactitem}
\item[(C6)] if $\loc\in\alocs{f}\setminus\{\lout{\fn{f}}\}$ and $(\loc, u,\loc')$ is the only triple in $\transitions{\fn{f}}$ with source label $\loc$ and update function $u$, then (i)
\[
\delta+\sum_{\mu\in \samples}\sampdpd(\mu)\cdot h\left(\fn{f},\loc',u(\nu,\mu)\right)\ge h(\fn{f},\loc,\nu)
\]
and (ii)
\[
\sum_{\mu\in \samples}\sampdpd(\mu)\cdot |h\left(\fn{f},\loc',u(\nu,\mu)\right)-h(\fn{f},\loc,\nu)|\le\zeta;
\]

\item[(C7)] if $\loc\in\flocs{f}\setminus\{\lout{\fn{f}}\}$ and $(\loc,(\fn{g},v),\loc')$ is the only triple in $\transitions{\fn{f}}$ with source label $\loc$ and value-passing function $v$, then
\[
\delta+h\left(\fn{g},\lin{\fn{g}}, v(\nu)\right)+h(\fn{f},\loc',\nu)\ge h(\fn{f},\loc,\nu);
\]
\item[(C8)] if $\loc\in\clocs{f}\setminus\{\lout{\fn{f}}\}$ and $(\loc, \phi,\loc_1),(\loc, \neg\phi,\loc_2)$ are namely two triples in $\transitions{\fn{f}}$ with source label $\loc$ and propositional arithmetic predicate $\phi$, then
\[
\mathbf{1}_{\nu\models\phi}\cdot h(\fn{f},\loc_1,\nu)+\mathbf{1}_{\nu\models\neg\phi}\cdot h(\fn{f},\loc_2,\nu)+\delta\ge h(\fn{f},\loc,\nu);
\]
\item[(C9)] if $\loc\in\dlocs{f}\setminus\{\lout{\fn{f}}\}$ and $(\loc, \star,\loc_1),(\loc, \star,\loc_2)$ are namely two triples in $\transitions{\fn{f}}$ with source label $\loc$, then
\[
\max\{h(\fn{f},\loc_1,\nu),h(\fn{f},\loc_2,\nu)\}+\delta\ge h(\fn{f},\loc,\nu).
\]
\end{compactitem}
\hfill\IEEEQEDclosed
\end{definition}

Below we prove Theorem~\ref{thm:lowerbound}. In the proof, we reuse many parts in the one for Lemma~\ref{thm:soundness}.
For the sake of convenience, we temporarily define $\infty-\infty:=\infty$ in the proof for Theorem~\ref{thm:lowerbound} (this situation will always happen with probability zero).

\noindent\textbf{Theorem~\ref{thm:lowerbound}.}
For any conditionally difference-bounded ranking measure function $h$ with $\delta,\zeta$ given in Definition~\ref{def:bmfunc}, $\expvt(\mathfrak{c})\ge \frac{h(\mathfrak{c})}{\delta}$ for all stack elements $\mathfrak{c}$ such that $h(\mathfrak{c})<\infty$.
\begin{IEEEproof}
Consider any conditionally difference-bounded ranking measure function $h$ with $\epsilon,\delta,\zeta$ given in Definition~\ref{def:mfunc} and Definition~\ref{def:bmfunc}, and any non-terminal stack element $\mathfrak{c}=(\fn{f}, \loc,\nu)$  such that $h(\mathfrak{c})<\infty$.

Define the scheduler $\sigma$ such that for any history $\rho$ ending in a state $\last{\rho}=((\fn{f}',\loc',\nu')\cdot w,\mu)$ with $\loc'\in\dlocs{f'}$ and $(\loc', \star,\loc'_1),(\loc', \star,\loc'_2)$ being namely the two triples in $\transitions{\fn{f}'}$ corresponding to resp. \textbf{then}- and \textbf{else}-branch, it holds that
\[
\sigma(\rho)=
\begin{cases}
\Dirac{\mathbf{th}} & \mbox{if }h(\fn{f}',\loc'_1,\nu')\ge h(\fn{f}',\loc'_2,\nu') \\
\Dirac{\mathbf{el}} & \mbox{if }h(\fn{f}',\loc'_1,\nu')< h(\fn{f}',\loc'_2,\nu') \\
\end{cases}\enskip.
\]
Then the underlying probability measure is $\probm_\mathfrak{c}^\sigma$.
For the sake of simplicity, we abbreviate `$\expv_\mathfrak{c}^\sigma$' as `$\expv$'.

Define then the stochastic process $\Gamma=\{X_n\}_{n\in\Nset_0}$ as in (\ref{eq:proof:xnprocess}).
From the proof of Theorem~\ref{thm:soundness}, $\Gamma$ is a ranking supermartingale.
Hence, $\expv(\stopping{\Gamma})\le\frac{h(\mathfrak{c})}{\epsilon}<\infty$.
For each $n\in\Nset_0$, define the random variable $\rvtop_n$ by
\[
\rvtop_n(\omega):= h\left(\rvfn_{n,0}(\omega),~\rvlb_{n,0}(\omega), ~\rvval^{\rvfn_{n,0}(\omega)}_{n,0}(\omega)\right)
\]
for all infinite runs $\omega$. Note that $\expv(\mathbf{1}_{\rvlen_n\ge 1}\cdot\rvtop_n)<\infty$ as $\Gamma$ is a ranking supermartingale.

Then by our semantics and (C1), we have that for all $n$,
\begin{eqnarray}\label{eq:proof2:absdifference}
& &|X_{n+1}-X_n|\nonumber\\
&=& \mathbf{1}_{\rvlen_n\ge 1}\cdot |Y^{n,\mathrm{a}}+Y^{n,\mathrm{c}} + Y^{n,\mathrm{b}} + Y^{n,\mathrm{d}}-\rvtop_n| ~~\mbox{(by (\ref{eq:proof:xnplusone}))}\nonumber\\
&=& \mathbf{1}_{\rvlen_n\ge 1}\cdot \left(\widetilde{Y}^{n,\mathrm{a}}+\widetilde{Y}^{n,\mathrm{c}} + \widetilde{Y}^{n,\mathrm{b}} + \widetilde{Y}^{n,\mathrm{d}}\right)
\end{eqnarray}
where
\[
\widetilde{Y}^{n,\mathrm{a}}:=\sum_{\mu\in\samples}\mathbf{1}_{\rvsam_{n+1}=\mu}\cdot \widetilde{Y}^{n,\mathrm{a}}_{\mu}
\]
with
\[
\widetilde{Y}^{n,\mathrm{a}}_{\mu}:=\sum_{\fn{f}\in\fnames}\sum_{\loc\in\alocs{f}\setminus\{\lout{\fn{f}}\}}\mathbf{1}_{\left(\rvfn_{n,0}, \rvlb_{n,0}\right)=
(\fn{f},\loc)}\cdot |Y^{n,\mathrm{a}}_{\fn{f},\loc,\mu}-\rvtop_n|,
\]
and
\[
\widetilde{Y}^{n,\mathrm{c}}:=\sum_{\fn{f}\in\fnames}\sum_{\loc\in\flocs{f}\setminus\{\lout{\fn{f}}\}}\mathbf{1}_{\left(\rvfn_{n,0}, \rvlb_{n,0}\right)=
(\fn{f},\loc)}\cdot |Y^{n,\mathrm{c}}_{\fn{f},\loc}-\rvtop_n|,
\]
\[
\widetilde{Y}^{n,\mathrm{b}}:=\sum_{\fn{f}\in\fnames}\sum_{\loc\in\clocs{f}\setminus\{\lout{\fn{f}}\}}\mathbf{1}_{\left(\rvfn_{n,0}, \rvlb_{n,0}\right)=(\fn{f},\loc)}\cdot |Y^{n,\mathrm{b}}_{\fn{f},\loc}-\rvtop_n|,
\]
\[
\widetilde{Y}^{n,\mathrm{d}}:=\sum_{\fn{f}\in\fnames}\sum_{\loc\in\dlocs{f}\setminus\{\lout{\fn{f}}\}}\mathbf{1}_{\left(\rvfn_{n,0}, \rvlb_{n,0}\right)=(\fn{f},\loc)}\cdot |Y^{n,\mathrm{d}}_{\fn{f},\loc}-\rvtop_n|.
\]
Note that any of
\begin{eqnarray*}
& \expv\left(\mathbf{1}_{\rvlen_n\ge 1}\cdot\widetilde{Y}^{n,\mathrm{a}}\right), \expv\left(\mathbf{1}_{\rvlen_n\ge 1}\cdot\widetilde{Y}^{n,\mathrm{a}}_{\mu}\right), \expv\left(\mathbf{1}_{\rvlen_n\ge 1}\cdot\widetilde{Y}^{n,\mathrm{d}}\right), \\
& \expv\left(\mathbf{1}_{\rvlen_n\ge 1}\cdot\widetilde{Y}^{n,\mathrm{b}}\right), \expv\left(\mathbf{1}_{\rvlen_n\ge 1}\cdot\widetilde{Y}^{n,\mathrm{c}}\right)
\end{eqnarray*}
is finite since $\Gamma$ is a ranking supermartingale.

Let $c:=\max\{\zeta,\delta\}$.
It holds a.s. for all $n$ that
\begin{eqnarray}\label{eq:proof:absoluteconditional}
& &\condexpv{|X_{n+1}-X_n|}{\mathcal{H}_n} \nonumber\\
&=& \condexpv{\mathbf{1}_{\rvlen_n\ge 1}\cdot \left(\widetilde{Y}^{n,\mathrm{a}}+\widetilde{Y}^{n,\mathrm{c}} + \widetilde{Y}^{n,\mathrm{b}} + \widetilde{Y}^{n,\mathrm{d}}\right)}{\mathcal{H}_n} \nonumber\\
&=& \condexpv{\mathbf{1}_{\rvlen_n\ge 1}\cdot\widetilde{Y}^{n,\mathrm{a}}}{\mathcal{H}_n}~~\mbox{(By (E6))} \nonumber\\
& & ~~{}+\condexpv{\mathbf{1}_{\rvlen_n\ge 1}\cdot\left(\widetilde{Y}^{n,\mathrm{c}} + \widetilde{Y}^{n,\mathrm{b}} + \widetilde{Y}^{n,\mathrm{d}}\right)}{\mathcal{H}_n} \nonumber\\
&=& \condexpv{\mathbf{1}_{\rvlen_n\ge 1}\cdot\widetilde{Y}^{n,\mathrm{a}}}{\mathcal{H}_n}~~\mbox{(By (E5))} \nonumber\\
& & ~~{}+\mathbf{1}_{\rvlen_n\ge 1}\cdot\left(\widetilde{Y}^{n,\mathrm{c}} + \widetilde{Y}^{n,\mathrm{b}} + \widetilde{Y}^{n,\mathrm{d}}\right) \nonumber\\
&=& ~~\mbox{\S~By (E6), (E11) \S} \nonumber\\
& &
\sum_{\mu\in\samples}\condexpv{\mathbf{1}_{\rvlen_n\ge 1}\cdot\mathbf{1}_{\rvsam_{n+1}=\mu}\cdot\widetilde{Y}^{n,\mathrm{a}}_\mu}{\mathcal{H}_n} \nonumber\\
& & ~~{}+\mathbf{1}_{\rvlen_n\ge 1}\cdot\left(\widetilde{Y}^{n,\mathrm{c}} + \widetilde{Y}^{n,\mathrm{b}} + \widetilde{Y}^{n,\mathrm{d}}\right) \nonumber\\
&=& ~~\mbox{\S~By (E8) \S} \nonumber\\
& &\sum_{\mu\in\samples} \condexpv{\mathbf{1}_{\rvsam_{n+1}=\mu}}{\mathcal{H}_n}\cdot \mathbf{1}_{\rvlen_n\ge 1}\cdot\widetilde{Y}^{n,\mathrm{a}}_\mu \nonumber\\
& & ~~{}+\mathbf{1}_{\rvlen_n\ge 1}\cdot\left(\widetilde{Y}^{n,\mathrm{c}} + \widetilde{Y}^{n,\mathrm{b}} + \widetilde{Y}^{n,\mathrm{d}}\right)\nonumber\\
&=& ~~\mbox{\S~By (E9) \S} \nonumber\\
& &\sum_{\mu\in\samples}\sampdpd(\mu)\cdot \mathbf{1}_{\rvlen_n\ge 1}\cdot\widetilde{Y}^{n,\mathrm{a}}_\mu \nonumber\\
& & ~~{}+\mathbf{1}_{\rvlen_n\ge 1}\cdot\left(\widetilde{Y}^{n,\mathrm{c}} + \widetilde{Y}^{n,\mathrm{b}} + \widetilde{Y}^{n,\mathrm{d}}\right)\enskip.
\end{eqnarray}
Hence from (C1)--(C9), for all $n$, it holds a.s. that
\[
\condexpv{|X_{n+1}-X_n|}{\mathcal{H}_n}\le c\enskip;
\]
moreover, from the choice of $\sigma$ and (C6)--(C9), we have that the followings hold a.s.:
\begin{eqnarray*}
& & \condexpv{X_{n+1}}{\mathcal{H}_n} \\
&=& \mbox{\S~By (\ref{eq:proof:xnplusone}) \S} \\
& & \condexpv{\mathbf{1}_{\rvlen_n\ge 1}\cdot\left[D+Y^{n,\mathrm{a}}+Y^{n,\mathrm{c}}+Y^{n,\mathrm{b}}+Y^{n,\mathrm{d}}\right]}{\mathcal{H}_n}\\
&=& \S\mbox{ By the Choice of }\sigma~\S \\
& & \condexpv{\mathbf{1}_{\rvlen_n\ge 1}\cdot\left[D+Y^{n,\mathrm{a}}+Y^{n,\mathrm{c}}+Y^{n,\mathrm{b}}+\widehat{Y}^{n,\mathrm{d}}\right]}{\mathcal{H}_n} \\
&=& \mbox{\S~By~(\ref{eq:proof:xprimenplusone}) \S} \\
& & \condexpv{X'_{n+1}}{\mathcal{H}_n} \\
&=& \mbox{\S~By (\ref{eq:proof:xnplusonexpxpp}) \S} \\
& & X''_{n+1} \\
&=& \mbox{\S~By (\ref{eq:proof:xppnplusone}) \S} \\
& & \mathbf{1}_{\rvlen_n\ge 1}\cdot\left[D+\widehat{Y}^{n,\mathrm{a}}+Y^{n,\mathrm{c}}+Y^{n,\mathrm{b}}+\widehat{Y}^{n,\mathrm{d}}\right]\\
&\ge& \mbox{\S~By (C1), (C6)--(C9) \S} \\
& & \mathbf{1}_{\rvlen_n\ge 1}\cdot\left[D+\mathsf{top}_n-\delta\right] \\
&=& X_n -\mathbf{1}_{X_n>0}\cdot \delta \enskip.
\end{eqnarray*}
Then the result follows directly from $T=\stopping{\Gamma}$ and Proposition~\ref{prop:cbrsupm}.
\end{IEEEproof}

\section{Details for Section~\ref{sect:concentration}}\label{app:concentration}

To prove Theorem~\ref{thm:concentration}, we need the following well-known theorem.

\begin{theorem}[Azuma's Inequality~\cite{Azuma1967inequality}]\label{thm:azuma}
Consider any supermartingale $\{X_n\}_{n\in\Nset_0}$ adapted to some filtration  $\{\mathcal{F}_n\}_{n\in\Nset_0}$ and any sequence of positive real numbers $\{c_n\}_{n\in\Nset_0}$ such that
\begin{compactitem}
\item $|X_{n+1}-X_n|\le c_n$ a.s. for all $n\in\Nset$ and
\item $X_0$ is a constant random variable.
\end{compactitem}
Then $\probm\left(X_n-X_0\ge \lambda\right)\le e^{-\frac{\lambda^2}{2\cdot\sum_{k=0}^{n-1} c_n^2}}$ for all $n\in\Nset$ and $\lambda\in (0,\infty)$.
\end{theorem}

\noindent{\bf Theorem~\ref{thm:concentration}.}
Consider any difference-bounded ranking supermartingale $\Gamma=\{X_n\}_{n\in\Nset_0}$ adapted to a filtration $\{\mathcal{F}_n\}_{n\in\Nset_0}$ with $\epsilon$ given in Definition~\ref{def:rsupm}. If (i) $X_0$ is a constant random variable and (ii) for all $n\in\Nset_0$ and $\omega$, $X_n(\omega)=0$ implies $X_{n+1}(\omega)=0$.
Then for all natural numbers $n>\frac{\expv(X_0)}{\epsilon}$,
\[
\probm(\stopping{\Gamma}>n)\le e^{-\frac{(\epsilon\cdot n-\expv(X_0))^2}{2\cdot n\cdot (c+\epsilon)^2}}\le e^{\frac{\epsilon\cdot \expv(X_0)}{(c+\epsilon)^2}}\cdot e^{-\frac{\epsilon^2}{2\cdot{(c+\epsilon)}^2}\cdot n},
\]
where $c\in (0,\infty)$ is any number satisfying  that $|X_{n+1}-X_n|\le c$ a.s. for all $n\in\Nset_0$.
\begin{IEEEproof}
W.l.o.g, we can assume that $X_0>0$. Define the stochastic process $\{Y_n\}_{\Nset_0}$ by:
\[
Y_n=X_n+\epsilon\cdot\min\{\stopping{\Gamma},n\}\enskip.
\]
We first prove that $Y_n$ is a difference-bounded supermartingale.

For each $n\in\Nset_0$, define the following random variable:
\[
U_n:=\min\{\stopping{\Gamma}, n+1\}-\min\{\stopping{\Gamma},n\}(=\mathbf{1}_{\stopping{\Gamma}>n})\enskip.
\]
Then we have that the followings hold a.s.:
\begin{eqnarray*}
& &\condexpv{Y_{n+1}}{\mathcal{F}_n}-Y_n\\
&=& \mbox{\S~By (E5), (E6) \S} \\
& & \condexpv{X_{n+1}}{\mathcal{F}_n}-X_n +\epsilon\cdot\condexpv{U_n}{\mathcal{F}_n} \\
&=& \condexpv{X_{n+1}}{\mathcal{F}_n}-X_n+\epsilon\cdot\expv(\mathbf{1}_{\stopping{\Gamma}>n}\mid\mathcal{F}_n)\\
&=& \mbox{\S~By (E5) \S} \\
& &\condexpv{X_{n+1}}{\mathcal{F}_n}-X_n+\epsilon\cdot\mathbf{1}_{\stopping{\Gamma}>n} \\
&\le &-\epsilon\cdot\mathbf{1}_{X_n> 0}+\epsilon\cdot\mathbf{1}_{\stopping{\Gamma}>n}\\
&=& 0~\mbox{\S~By $\stopping{\Gamma}(\omega)>n$ iff $X_n(\omega)> 0$ \S}\enskip.
\end{eqnarray*}
Moreover, it is straightforward to see that $|Y_{n+1}-Y_n|\le c+\epsilon$ a.s. for all $n\in\Nset_0$.
Now define random variables $\alpha_n:=\epsilon\cdot n-X_0$ and $\widehat{\alpha}_n:=\epsilon\cdot\min\{n,\stopping{\Gamma}\}-X_0$.
We have that
\begin{eqnarray*}
\probm(\stopping{\Gamma} >n) &=& \mbox{\S~By $X_m(\omega)=0\Rightarrow X_{m+1}(\omega)=0$ \S} \\
& & \probm(X_n> 0\wedge \stopping{\Gamma} >n) \\
&=&\probm((X_n+\alpha_n \ge \alpha_n ) \wedge (\stopping{\Gamma}>n))\\
&=&\probm((X_n+\widehat{\alpha}_n \ge \alpha_n ) \wedge (\stopping{\Gamma}>n))\\
&\le &\probm((X_n+ \widehat{\alpha}_n \geq \alpha_n))\\
&=&\mathbb{P}(Y_n-Y_0\ge \epsilon\cdot n-X_0)\\
&\le &e^{-\frac{(\epsilon\cdot n-\expv(X_0))^2}{2\cdot n\cdot {(c+\epsilon)}^2}} \mbox{\S~By Theorem~\ref{thm:azuma} \S}
\end{eqnarray*}
for all $n>\frac{\expv(X_0)}{\epsilon}$.
\end{IEEEproof}


We fix a nondeterministic recursive probabilistic program $W$ together with its associated CFG taking the form~(\ref{eq:cfg}) and a sampling function $\Upsilon$.

\begin{definition}[Difference-Bounded Ranking Measure Functions]\label{def:dbmfunc}
A ranking measure function $h$ is \emph{difference bounded} if there exists a $\zeta\in(0,\infty)$ such that for all stack elements $(\fn{f},\loc,\nu)$ satisfying $h(\fn{f},\loc,\nu)<\infty$, the following conditions hold:
\begin{compactitem}
\item[(C10)] if $\loc\in\alocs{f}\setminus\{\lout{\fn{f}}\}$ and $(\loc, u,\loc')$ is the only triple in $\transitions{\fn{f}}$ with source label $\loc$ and update function $u$, then for all $\mu\in\samples$, it holds that
\[
\left|h\left(\fn{f},\loc',u(\nu,\mu)\right)-h(\fn{f},\loc,\nu)\right|\le \zeta;
\]
\item[(C11)] if $\loc\in\flocs{f}\setminus\{\lout{\fn{f}}\}$ and $(\loc,(\fn{g},v),\loc')$ is the only triple in $\transitions{\fn{f}}$ with source label $\loc$ and value-passing function $v$, then
\[
\left|h\left(\fn{g},\lin{\fn{g}}, v(\nu)\right)+h(\fn{f},\loc',\nu)- h(\fn{f},\loc,\nu)\right|\le \zeta;
\]

\item[(C12)] if $\loc\in\clocs{f}\setminus\{\lout{\fn{f}}\}$ and $(\loc, \phi,\loc_1),(\loc, \neg\phi,\loc_2)$ are namely two triples in $\transitions{\fn{f}}$ with source label $\loc$ and propositional arithmetic predicate $\phi$, then
\[
\left|\left[\sum_{i\in\{1,2\}}\mathbf{1}_{\nu\models\phi_i}\cdot h(\fn{f},\loc_i,\nu)\right]-h(\fn{f},\loc,\nu)\right|\le \zeta
\]
where $\phi_1:=\phi$ and $\phi_2:=\neg\phi$;
\item[(C13)] if $\loc\in\dlocs{f}\setminus\{\lout{\fn{f}}\}$ and $(\loc, \star,\loc_1),(\loc, \star,\loc_2)$ are namely two triples in $\transitions{\fn{f}}$ with source label $\loc$, then
\[
\max_{i\in\{1,2\}}\left| h(\fn{f},\loc_i,\nu)-h(\fn{f},\loc,\nu)\right|\le \zeta\enskip.
\]
\end{compactitem}
\hfill\IEEEQEDclosed
\end{definition}

\begin{theorem}\label{thm:dbmfunc}
For any difference-bounded ranking measure function $h$ with $\epsilon, \zeta$ given in Defintion~\ref{def:mfunc} and Definition~\ref{def:dbmfunc}, it holds that for all $n\in\Nset_0$ and non-terminal stack elements $\mathfrak{c}$ such that $h(\mathfrak{c})<\infty$ and $n>\frac{h(\mathfrak{c})}{\epsilon}$,
\[
\sup_\sigma\probm_\mathfrak{c}^\sigma(T>n)\le e^{-\frac{(\epsilon\cdot n-h(\mathfrak{c}))^2}{2\cdot n\cdot (\epsilon+\zeta)^2}}
\]
where $\sigma$ ranges over all schedulers for $W$.
\end{theorem}
\begin{IEEEproof}
Consider any difference-bounded ranking measure function $h$ with $\epsilon,\zeta$ given in Definition~\ref{def:mfunc} 
and Definition~\ref{def:dbmfunc}, and any non-terminal stack element $\mathfrak{c}=(\fn{f}, \loc, \nu)$ satisfying $h(\mathfrak{c})<\infty$.
Let $\sigma$ be any scheduler so that the underlying probability measure is $\probm_c^\sigma$.
Define the stochastic process $\Gamma=\{X_n\}_{n\in\Nset_0}$ as in (\ref{eq:proof:xnprocess}).
Then by the proof of Lemma~\ref{thm:soundness}, $\Gamma$ is a ranking supermartingale.
Moreover, from (\ref{eq:proof2:absdifference}), we have that
\[
|X_{n+1}-X_n| =\mathbf{1}_{\rvlen_n\ge 1}\cdot \left(\widetilde{Y}^{n,\mathrm{a}}+\widetilde{Y}^{n,\mathrm{c}} + \widetilde{Y}^{n,\mathrm{b}} + \widetilde{Y}^{n,\mathrm{d}}\right)
\]
where all relevant random variables are defined in the proof of Theorem~\ref{thm:lowerbound}.
Hence, by (C10)--(C13), $\Gamma$ is difference-bounded.
It follows from Theorem~\ref{thm:concentration} that
\[
\probm_\mathfrak{c}^\sigma(\tertime>n)=\probm_\mathfrak{c}^\sigma(\stopping{\Gamma}>n)\le e^{-\frac{(\epsilon\cdot n-h(\mathfrak{c}))^2}{2\cdot n\cdot (\zeta+\epsilon)^2}}
\]
for all natural numbers $n>\frac{h(\mathfrak{c})}{\epsilon}$.
Then the result follows from the arbitrary choice of $\sigma$.
\end{IEEEproof}

\section{Details for Section~\ref{sect:astermination}}\label{app:astermination}



In this section, for a random variable $R$ and a real number $M$, we denote by $R\wedge M$ the random variable $\min\{R,M\}$.

\noindent{\bf Theorem~\ref{thm:supm}.}
Consider any difference-bounded supermartingale $\Gamma=\{X_n\}_{n\in\Nset_0}$ adapted to a filtration $\{\mathcal{F}_n\}_{n\in\Nset_0}$ satisfying the following conditions:
\begin{compactenum}
\item $X_0$ is a constant random variable;
\item for all $n\in\Nset_0$, it holds for all $\omega$ that (i) $X_n(\omega)\ge 0$ and (ii) $X_n(\omega)=0$ implies  $X_{n+1}(\omega)=0$;
\item there exists a $\delta\in(0,\infty)$ such that for all $n\in\Nset_0$, it holds a.s. that $X_n>0$ implies $\condexpv{|X_{n+1}-X_n|}{\mathcal{F}_n}\ge \delta$.
\end{compactenum}
Then $\probm(\stopping{\Gamma}<\infty)=1$ and $k\mapsto\probm\left(\stopping{\Gamma}\ge k\right)\in \mathcal{O}\left(\frac{1}{\sqrt{k}}\right)$.
\begin{IEEEproof}
The proof uses ideas from both~\cite[Chapter~10.12]{probabilitycambridge} and \cite[Theorem 4.1]{DBLP:journals/jcss/BrazdilKKV15}.
Let $c\in (0,\infty)$ be such that for every $n\in\Nset_0$, it holds a.s. that $|X_{n+1}-X_n|\le c$.
Let $\delta$ be given as in the statement of the theorem.
W.l.o.g, we assume that $X_0>0$.
Note that from (E13), it holds a.s. that $X_n>0$ implies
\begin{equation}\label{eq:proof:convexity}
\condexpv{{(X_{n+1}-X_n)}^2}{\mathcal{F}_n}\ge {(\condexpv{|X_{n+1}-X_n|}{\mathcal{F}_n})}^2\ge \delta^2\enskip.
\end{equation}
Fix any sufficiently small real number $t\in (0,\infty)$ such that
\[
e^{c\cdot t}-(1+c\cdot t +\frac{1}{2}\cdot c^2\cdot t^2)\left(=\sum_{j=3}^\infty\frac{(c\cdot t)^j}{{j}{!}} \right)\le \frac{\delta^2}{4}\cdot t^2\enskip.
\]
Define the discrete-time stochastic process $\{Y_n\}_{n\in\Nset_0}$ by
\begin{equation}\label{eq:proof4:definition}
Y_n:=\frac{e^{-t\cdot X_n}}{\prod_{j=0}^{n-1} \condexpv{e^{-t\cdot \left(X_{j+1}-X_{j}\right)}}{\mathcal{F}_j}}\enskip.
\end{equation}
Note that from difference-boundedness, $0<Y_n\le e^{n\cdot c\cdot t}$ a.s. for all $n\in\Nset_0$.
Then the followings hold a.s.:
\begin{eqnarray}\label{eq:proof4:martingale}
& & \condexpv{Y_{n+1}}{\mathcal{F}_n} \nonumber\\
&=& \condexpv{\frac{e^{-t\cdot X_{n+1}}}{\prod_{j=0}^{n} \condexpv{e^{-t\cdot \left(X_{j+1}-X_{j}\right)}}{\mathcal{F}_j}}}{\mathcal{F}_n} \nonumber\\
&=& \condexpv{\frac{e^{-t\cdot X_n}\cdot e^{-t\cdot \left(X_{n+1}-X_{n}\right)}}{\prod_{j=0}^{n} \condexpv{e^{-t\cdot \left(X_{j+1}-X_{j}\right)}}{\mathcal{F}_j}}}{\mathcal{F}_n} \nonumber\\
&=& \mbox{\S~By (E8), (E1) \S} \nonumber\\
& &\frac{e^{-t\cdot X_n}\cdot \condexpv{ e^{-t\cdot \left(X_{n+1}-X_{n}\right)}}{\mathcal{F}_n}}{\prod_{j=0}^{n} \condexpv{e^{-t\cdot \left(X_{j+1}-X_{j}\right)}}{\mathcal{F}_j}} \nonumber\\
&=& \frac{e^{-t\cdot X_n}}{\prod_{j=0}^{n-1} \condexpv{e^{-t\cdot \left(X_{j+1}-X_{j}\right)}}{\mathcal{F}_j}} \nonumber\\
&=& Y_n\enskip.
\end{eqnarray}
Hence, $\{Y_n\}_{n\in\Nset_0}$ is a martingale.
For every $n\in\Nset_0$, it holds a.s. that $X_n>0$ implies
\begin{eqnarray}\label{eq:proof4:exponentiation}
& &\expv\left(e^{-t\cdot \left(X_{n+1}-X_{n}\right)}\mid\mathcal{F}_n\right) \nonumber\\
&=&\expv\left(\sum_{j=0}^{\infty} \frac{(-1)^j\cdot t^j\cdot {\left(X_{n+1}-X_{n}\right)}^j}{{j}{!}}\mid\mathcal{F}_n\right) \nonumber\\
&=& \mbox{\S~By (E12) \S}\nonumber\\
& &
\sum_{j=0}^{\infty}\expv\left(\frac{(-1)^j\cdot t^j\cdot {\left(X_{n+1}-X_{n}\right)}^j}{{j}{!}}\mid\mathcal{F}_n\right) \nonumber\\
&=& \mbox{\S~By (E6) \S} \nonumber\\
& &
1-t\cdot \condexpv{X_{n+1}-X_{n}}{\mathcal{F}_n} \nonumber\\
& &{}+\frac{t^2}{2}\cdot\condexpv{(X_{n+1}-X_{n})^2}{\mathcal{F}_n}\nonumber\\
& &{}+\sum_{j=3}^{\infty}\condexpv{\frac{(-1)^j\cdot t^j\cdot {\left(X_{n+1}-X_{n}\right)}^j}{{j}{!}}}{\mathcal{F}_n}\nonumber\\
&\ge & 1+\frac{t^2}{2}\cdot\condexpv{(X_{n+1}-X_{n})^2}{\mathcal{F}_n}- \sum_{j=3}^{\infty}\frac{{(c\cdot t)}^j}{{j}{!}}~~\\
&\ge & 1+\frac{\delta^2}{4}\cdot t^2\enskip.\nonumber
\end{eqnarray}

Thus,
\begin{compactitem}
\item $|Y_{\stopping{\Gamma}\wedge n}|\le 1$ a.s. for all $n\in\Nset_0$, and
\item it holds a.s. that
\begin{equation}\label{eq:thmmartingale}
\left(\lim\limits_{n\rightarrow\infty} Y_{n\wedge \stopping{\Gamma}}\right)(\omega)=
\begin{cases}
0 & \mbox{if }\stopping{\Gamma}(\omega)=\infty\\
Y_{\stopping{\Gamma}(\omega)}(\omega) & \mbox{if }\stopping{\Gamma}(\omega)<\infty
\end{cases}\enskip.
\end{equation}
\end{compactitem}
Then from Optional Stopping Theorem (Item 1 of Theorem~\ref{thm:optstopping}), by letting $Y_\infty:=\lim\limits_{n\rightarrow\infty} Y_{n\wedge\stopping{\Gamma}}$
one has that
\[
\expv\left(Y_\infty\right)=\expv\left(Y_0\right)=e^{-t\cdot \expv(X_0)}\enskip.
\]
Moreover, from~(\ref{eq:thmmartingale}), one can obtain that
\begin{eqnarray}\label{eq:proof4:derivation}
& &\expv\left(Y_\infty\right)\nonumber\\
&=& \mbox{\S~By Definition \S} \nonumber\\
& & \int Y_\infty\,\mathrm{d}\probm \nonumber\\
&=& \mbox{\S~By Linear Property of Lebesgue Integral \S} \nonumber\\
& &\int Y_\infty\cdot\mathbf{1}_{\stopping{\Gamma}=\infty}\,\mathrm{d}\probm+\int Y_\infty\cdot\mathbf{1}_{\stopping{\Gamma}<\infty}\,\mathrm{d}\probm\nonumber\\
&=& \mbox{\S~By Monotone Convergence Theorem~\cite[Chap. 6]{probabilitycambridge} \S} \nonumber\\
& & 0\cdot\probm\left(\stopping{\Gamma}=\infty\right)+\sum_{n=0}^{\infty}\int Y_\infty\cdot\mathbf{1}_{\stopping{\Gamma}=n}\,\mathrm{d}\probm\nonumber\\
&=& \sum_{n=0}^{\infty}\int Y_n\cdot\mathbf{1}_{\stopping{\Gamma}=n}\,\mathrm{d}\probm\nonumber\\
&\le& \mbox{\S~By (\ref{eq:proof4:exponentiation}) and $X_{n}\ge 0$ \S} \nonumber\\
& & \sum_{n=0}^{\infty}\int {\left(1+\frac{\delta^2}{4}\cdot t^2\right)}^{-n} \cdot\mathbf{1}_{\stopping{\Gamma}=n}\,\mathrm{d}\probm\nonumber\\
&=& \sum_{n=0}^{\infty} {\left(1+\frac{\delta^2}{4}\cdot t^2\right)}^{-n} \cdot\probm\left(\stopping{\Gamma}=n\right) \nonumber\\
&= & \sum_{n=0}^{k-1} {\left(1+\frac{\delta^2}{4}\cdot t^2\right)}^{-n} \cdot\probm\left(\stopping{\Gamma}=n\right)\nonumber\\
& &~~{}+\sum_{n=k}^{\infty} {\left(1+\frac{\delta^2}{4}\cdot t^2\right)}^{-n} \cdot\probm\left(\stopping{\Gamma}=n\right)\nonumber\\
&\le & \left(1-\probm\left(\stopping{\Gamma}\ge k\right)\right)\nonumber\\
& & \qquad{}+{\left(1+\frac{\delta^2}{4}\cdot t^2\right)}^{-k}\cdot \probm\left(\stopping{\Gamma}\ge k\right)~~
\end{eqnarray}
for any $k\in\Nset$.
It follows that for all $k\in\Nset$,
\[
e^{-t\cdot\expv(X_0)}\le 1-\left(1-\left(1+\frac{\delta^2}{4}\cdot t^2\right)^{-k}\right)\cdot \probm\left( \stopping{\Gamma}\ge k\right)\enskip.
\]
Hence, for any $k\in\Nset$ and sufficiently small $t\in (0,\infty)$,
\[
\probm\left(\stopping{\Gamma}\ge k\right)\le \frac{1-e^{-t\cdot\expv(X_0)}}{1-\left(1+\frac{\delta^2}{4}\cdot t^2\right)^{-k}}\enskip.
\]
Then for sufficiently large $k\in\Nset$ with $t:=\frac{1}{\sqrt{k}}$,
\[
\probm\left(\stopping{\Gamma}\ge k\right)\le \frac{1-e^{-\frac{\expv(X_0)}{\sqrt{k}}}}{1-\left(1+\frac{\delta^2}{4}\cdot \frac{1}{k}\right)^{-k}}\enskip.
\]
Using the facts that $\lim\limits_{k\rightarrow\infty}(1+\frac{\delta^2}{4}\cdot\frac{1}{k})^{k}= e^{\frac{\delta^2}{4}}$ and $\lim\limits_{z\rightarrow 0^+}\frac{1-e^{-z}}{z}=1$, we have that the function
\[
k\mapsto\probm\left(\stopping{\Gamma}\ge k\right)\in \mathcal{O}\left(\frac{1}{\sqrt{k}}\right)\enskip.
\]
Since $\probm(\stopping{\Gamma}=\infty)=\lim\limits_{k\rightarrow\infty}\probm\left(\stopping{\Gamma}\ge k\right)$, one obtains immediately that $\probm(\stopping{\Gamma}=\infty)=0$ and $\probm(\stopping{\Gamma}<\infty)=1$.
\end{IEEEproof}

\begin{remark}
Compared with~\cite[Chapter~10.12]{probabilitycambridge}, Theorem~\ref{thm:supm} extends the result to arbitrary difference-bounded supermartingales.
Compared with~\cite[Theorem 4.1]{DBLP:journals/jcss/BrazdilKKV15}, Theorem~\ref{thm:supm} does not require the prerequisite that $\probm(\stopping{\Gamma}<\infty)=1$ and $\expv(\stopping{\Gamma})=\infty$.\hfill\IEEEQEDclosed
\end{remark}

\noindent{\textbf{Proposition}~\ref{prop:varcezero}.}
Let $\Gamma=\{X_n\}_{n\in\Nset_0}$ be a martingale adapted to a filtration $\{\mathcal{F}_n\}_{n\in\Nset_0}$ such that $X_0>0$ and $\condexpv{|X_{n+1}-X_n|}{\mathcal{F}_n}=0$ a.s. for all $n\in\Nset_0$.
Then $\probm\left(\stopping{\Gamma}=\infty\right)=1$.
\begin{IEEEproof}
Since $\condexpv{|X_{n+1}-X_n|}{\mathcal{F}_n}=0$ a.s., we have from the definition of conditional expectation that
\[
\int|X_{n+1}-X_n|\,\mathrm{d}\probm=\int\condexpv{|X_{n+1}-X_n|}{\mathcal{F}_n}\,\mathrm{d}\probm=0\enskip.
\]
Hence, it holds a.s. that for all $n\in\Nset_0$, $X_{n+1}=X_n$.
Thus, for all $k\in\Nset$,
\[
\probm\left(\stopping{\Gamma}\ge k\right)\ge\probm\left( \forall 0\le n\le k-1.\left(X_{n+1}=X_n\right)\right)=1.
\]
So, $\probm\left(\stopping{\Gamma}=\infty\right)=\lim\limits_{k\rightarrow\infty}\probm\left(\stopping{\Gamma}\ge k\right)=1$.
\end{IEEEproof}

\noindent\textbf{Theorem~\ref{thm:supmextended}.}
Consider any supermartingale $\Gamma=\{X_n\}_{n\in\Nset_0}$ adapted to a filtration $\{\mathcal{F}_n\}_{n\in\Nset_0}$ satisfying the following conditions:
\begin{compactenum}
\item $X_0$ is a constant random variable;
\item for all $n\in\Nset_0$, it holds for all $\omega$ that (i) $X_n(\omega)\ge 0$ and (ii) $X_n(\omega)=0$ implies  $X_{n+1}(\omega)=0$;
\item there exists a $\delta\in(0,\infty)$ such that for all $n\in\Nset_0$, it holds a.s. that $\condexpv{|X_{n+1}-X_n|}{\mathcal{F}_n}\ge \delta\cdot\mathbf{1}_{X_n>0}$.
\end{compactenum}
Then $\probm(\stopping{\Gamma}<\infty)=1$ and $k\mapsto\probm\left(\stopping{\Gamma}\ge k\right)\in \mathcal{O}\left(k^{-\frac{1}{6}}\right)$.
\begin{IEEEproof}
W.l.o.g., we assume that $X_0>0$. Let $\delta$ be given as in the statement of the theorem.
From
\[
\lim\limits_{k\rightarrow\infty}\frac{1-e^{-\frac{\expv(X_0)}{\sqrt{k}}}}{\frac{\expv(X_0)}{\sqrt{k}}}=1\mbox{ and } \lim\limits_{k\rightarrow\infty} \left(1+\frac{\delta^2}{16}\cdot \frac{1}{k}\right)^{-k}= e^{-\frac{\delta^2}{16}}\enskip,
\]
one can fix a constant natural number $N$ such that for all $k\ge N$,
\[
\frac{1-e^{-\frac{\expv(X_0)}{\sqrt{k}}}}{\frac{\expv(X_0)}{\sqrt{k}}}\le \frac{3}{2}\mbox{ and } 1-\left(1+\frac{\delta^2}{16}\cdot \frac{1}{k}\right)^{-k}\ge \frac{1-e^{-\frac{\delta^2}{16}}}{2}\enskip.
\]
Let
\[
C:= \frac{3}{2}\cdot \expv(X_0)\cdot \frac{2}{1-e^{-\frac{\delta^2}{16}}}\enskip.
\]
Choose a constant $c\in (0,1)$ such that
\[
\sum_{j=3}^{\infty} \frac{c^{j-2}}{j!}\le \frac{\delta^2}{16}\enskip.
\]
Note that from (E6), it holds a.s. for all $n$ that 
\begin{eqnarray*}
& &\condexpv{|X_{n+1}-X_n|}{\mathcal{F}_n}\\
&=&\condexpv{\mathbf{1}_{X_{n+1}< X_n}\cdot (X_{n}-X_{n+1})}{\mathcal{F}_n}\\
& &{}+\condexpv{\mathbf{1}_{X_{n+1}\ge X_n}\cdot (X_{n+1}-X_{n})}{\mathcal{F}_n}\\
&\ge& \mathbf{1}_{X_n>0}\cdot \delta\enskip.
\end{eqnarray*}
Moreover, from (E5), (E6) and definition of supermartingales, it holds a.s. that
\begin{eqnarray*}
& &\condexpv{X_{n+1}-X_n}{\mathcal{F}_n}\\
&=&\condexpv{\mathbf{1}_{X_{n+1}< X_n}\cdot (X_{n+1}-X_{n})}{\mathcal{F}_n}\\
& &{}+\condexpv{\mathbf{1}_{X_{n+1}\ge X_n}\cdot (X_{n+1}-X_{n})}{\mathcal{F}_n}\\
&=&-\condexpv{\mathbf{1}_{X_{n+1}< X_n}\cdot (X_{n}-X_{n+1})}{\mathcal{F}_n}\\
& &{}+\condexpv{\mathbf{1}_{X_{n+1}\ge X_n}\cdot (X_{n+1}-X_{n})}{\mathcal{F}_n}\\
&\le & 0\enskip.
\end{eqnarray*}
It follows that for all $n$, it holds a.s. that
\begin{equation}\label{eq:supmextended:delta}
\condexpv{\mathbf{1}_{X_{n+1}< X_n}\cdot (X_{n}-X_{n+1})}{\mathcal{F}_n}\ge \mathbf{1}_{X_n>0}\cdot \frac{\delta}{2}\enskip.
\end{equation}

Let $M$ be any real number satisfying $M> \max\{\expv(X_0), \sqrt[6]{N}\}$ and define the stopping time $R_M$ w.r.t $\{\mathcal{F}_n\}_{n\in\Nset_0}$ by
\[
R_M(\omega):=  \min\{n\mid X_{n}(\omega)\le 0\mbox{ or } X_{n}(\omega)\ge M\}
\]
where $\min\emptyset:=\infty$. Define the stochastic process $\Gamma'=\{X'_n\}_{n\in\Nset_0}$ adapted to $\{\mathcal{F}_n\}_{n\in\Nset_0}$ by:
\begin{equation}\label{eq:proof:primesupermartingale}
X'_n=X_n\wedge M~\mbox{ for all } n\in\Nset_0\enskip.
\end{equation}
It is clear that $\Gamma'$ is difference-bounded.
Below we prove that $\Gamma'$ is a supermartingale.
This can be observed from the following:
\begin{eqnarray*}
& &\condexpv{X'_{n+1}}{\mathcal{F}_n}-X'_n \\
&=& \mbox{\S~By~(E5), (E6) \S} \\
& &\condexpv{X'_{n+1}-X'_n}{\mathcal{F}_n} \\
&=& \mbox{\S~By~(E6) \S}\\
& & \condexpv{\mathbf{1}_{X_n>M}\cdot \left((X_{n+1}\wedge M) - M\right)}{\mathcal{F}_n} \\
& & \quad{}+\condexpv{\mathbf{1}_{X_n\le M}\cdot \left((X_{n+1}\wedge M)-X_n\right)}{\mathcal{F}_n}\\
&\le & \mbox{\S~By~(E8), (E10) \S}\\
& & \mathbf{1}_{X_n\le M}\cdot\condexpv{X_{n+1}-X_n}{\mathcal{F}_n}\\
&\le& 0\enskip.
\end{eqnarray*}
Hence $\Gamma'$ is a difference-bounded supermartingale.
Moreover, we have that the followings hold a.s. for all $n$:
\begin{eqnarray*}
& &\mathbf{1}_{0<X'_n< M}\cdot \condexpv{|X'_{n+1}-X'_n|}{\mathcal{F}_n} \\
&=& \mbox{\S~By~(E8) \S} \\
& & \condexpv{\mathbf{1}_{0<X'_n< M}\cdot |X'_{n+1}-X'_n|}{\mathcal{F}_n}\\
&\ge & \mbox{\S~By~(E10), (E6) \S} \\
& & \condexpv{\mathbf{1}_{0<X'_n< M}\cdot \mathbf{1}_{X'_{n+1}<X'_n}\cdot \left(X'_{n}-X'_{n+1}\right)}{\mathcal{F}_n} \\
&=& \condexpv{\mathbf{1}_{0<X'_n< M}\cdot \mathbf{1}_{X_{n+1}<X_n}\cdot \left(X_{n}-X_{n+1}\right)}{\mathcal{F}_n}\\
&=& \mbox{\S~By~(E8) \S} \\
& & \mathbf{1}_{0<X'_n< M}\cdot\condexpv{ \mathbf{1}_{X_{n+1}<X_n}\cdot \left(X_{n}-X_{n+1}\right)}{\mathcal{F}_n}\\
&\ge & \mbox{\S~By~(\ref{eq:supmextended:delta}) \S} \\
& & \mathbf{1}_{0<X'_n< M}\cdot \mathbf{1}_{X_n>0}\cdot \frac{\delta}{2}\\
&=& \mathbf{1}_{0<X'_n< M}\cdot \frac{\delta}{2}\enskip.
\end{eqnarray*}
Since $\condexpv{|X'_{n+1}-X'_n|}{\mathcal{F}_n}\ge 0$ a.s. (from (E10)), we obtain that a.s. 
\[
\condexpv{|X'_{n+1}-X'_n|}{\mathcal{F}_n}\ge \mathbf{1}_{0<X'_n< M}\cdot \frac{\delta}{2}\enskip.
\]
Hence, from (E13), it holds a.s. for all $n$ that 
\begin{eqnarray*}
\condexpv{(X'_{n+1}-X'_n)^2}{\mathcal{F}_n}&\ge &{\left(\condexpv{|X'_{n+1}-X'_n|}{\mathcal{F}_n}\right)}^2\\
&\ge& \mathbf{1}_{0<X'_n< M}\cdot \frac{\delta^2}{4}\enskip.
\end{eqnarray*}

Now define the discrete-time stochastic process $\{Y_n\}_{n\in\Nset_0}$ by
\[
Y_n:=\frac{e^{-t\cdot X'_n}}{\prod_{j=0}^{n-1} \condexpv{e^{-t\cdot \left(X'_{j+1}-X'_{j}\right)}}{\mathcal{F}_j}}
\]
where $t$ is an arbitrary real number in $(0,\frac{c}{M^3}]$.
Note that from difference-boundedness and (E10), $0<Y_n\le e^{n\cdot M\cdot t}$ a.s. for all $n\in\Nset_0$.
Then by the same analysis in (\ref{eq:proof4:martingale}), $\{Y_n\}_{n\in\Nset_0}$ is a martingale.
Furthermore, by similar analysis in (\ref{eq:proof4:exponentiation}), one can obtain that for every $n$, it holds a.s. that $0<X'_n< M$ implies
\begin{eqnarray*}
& &\expv\left(e^{-t\cdot \left(X'_{n+1}-X'_{n}\right)}\mid\mathcal{F}_n\right) \nonumber\\
&\ge & 1+\frac{t^2}{2}\cdot\condexpv{(X'_{n+1}-X'_{n})^2}{\mathcal{F}_n}- \sum_{j=3}^{\infty}\frac{{(M\cdot t)}^j}{{j}{!}}~~\\
&\ge & 1+\frac{t^2}{2}\cdot\frac{\delta^2}{4}- t^2\cdot \sum_{j=3}^{\infty}\frac{M^j\cdot t^{j-2}}{{j}{!}}~~\\
&\ge & 1+\frac{\delta^2}{8}\cdot t^2- t^2\cdot \sum_{j=3}^{\infty}\frac{M^{-2\cdot j+6}\cdot c^{j-2}}{{j}{!}}~~\\
&\ge & 1+\frac{\delta^2}{8}\cdot t^2- t^2\cdot \sum_{j=3}^{\infty}\frac{c^{j-2}}{{j}{!}}~~\\
&\ge & 1+\frac{\delta^2}{8}\cdot t^2- t^2\cdot \frac{\delta^2}{16}~~\\
&\ge & 1+\frac{\delta^2}{16}\cdot t^2\enskip.\nonumber
\end{eqnarray*}

Thus,
\begin{compactitem}
\item $|Y_{R_M\wedge n}|\le 1$ a.s. for all $n\in\Nset_0$, and
\item it holds a.s. that
\[
\left(\lim\limits_{n\rightarrow\infty} Y_{n\wedge R_M}\right)(\omega)=
\begin{cases}
0 & \mbox{if }R_M(\omega)=\infty\\
Y_{R_M(\omega)}(\omega) & \mbox{if }R_M(\omega)<\infty
\end{cases}\enskip.
\]
\end{compactitem}
Then from Optional Stopping Theorem (Item 1 of Theorem~\ref{thm:optstopping}), by letting $Y_\infty:=\lim\limits_{n\rightarrow\infty} Y_{n\wedge R_M}$
one has that
\[
\expv\left(Y_\infty\right)=\expv\left(Y_0\right)=e^{-t\cdot \expv(X_0)}\enskip.
\]
Moreover, one can obtain that
\begin{eqnarray}\label{eq:proof4:derivation2}
& &\expv\left(Y_\infty\right)\nonumber\\
&=& \mbox{\S~By Definition \S} \nonumber\\
& & \int Y_\infty\,\mathrm{d}\probm \nonumber\\
&=& \mbox{\S~By Linear Property of Lebesgue Integral \S} \nonumber\\
& &\int Y_\infty\cdot\mathbf{1}_{R_M=\infty}\,\mathrm{d}\probm+\int Y_\infty\cdot\mathbf{1}_{R_M<\infty}\,\mathrm{d}\probm\nonumber\\
&=& \mbox{\S~By Monotone Convergence Theorem~\cite[Chap. 6]{probabilitycambridge} \S} \nonumber\\
& & 0\cdot\probm\left(R_M=\infty\right)+\sum_{n=0}^{\infty}\int Y_\infty\cdot\mathbf{1}_{R_M=n}\,\mathrm{d}\probm\nonumber\\
&=& \sum_{n=0}^{\infty}\int Y_n\cdot\mathbf{1}_{R_M=n}\,\mathrm{d}\probm\nonumber\\
&\le& \mbox{\S~By $X'_{n}\ge 0$ \S} \nonumber\\
& & \sum_{n=0}^{\infty}\int {\left(1+\frac{\delta^2}{16}\cdot t^2\right)}^{-n} \cdot\mathbf{1}_{R_M=n}\,\mathrm{d}\probm\nonumber\\
&=& \sum_{n=0}^{\infty} {\left(1+\frac{\delta^2}{16}\cdot t^2\right)}^{-n} \cdot\probm\left(R_M=n\right) \nonumber\\
&= & \sum_{n=0}^{k-1} {\left(1+\frac{\delta^2}{16}\cdot t^2\right)}^{-n} \cdot\probm\left(R_M=n\right)\nonumber\\
& &~~{}+\sum_{n=k}^{\infty} {\left(1+\frac{\delta^2}{16}\cdot t^2\right)}^{-n} \cdot\probm\left(R_M=n\right)\nonumber\\
&\le & \left(1-\probm\left(R_M\ge k\right)\right)\nonumber\\
& & \qquad{}+{\left(1+\frac{\delta^2}{16}\cdot t^2\right)}^{-k}\cdot \probm\left(R_M\ge k\right)~~
\end{eqnarray}
for any $k\in\Nset$.
It follows that for all $k\in\Nset$,
\[
e^{-t\cdot\expv(X_0)}\le 1-\left(1-\left(1+\frac{\delta^2}{16}\cdot t^2\right)^{-k}\right)\cdot \probm\left( R_M\ge k\right)\enskip.
\]
Hence, for any $k\in\Nset$ and $t\in \left(0,\frac{c}{M^3}\right]$,
\[
\probm\left(R_M\ge k\right)\le \frac{1-e^{-t\cdot\expv(X_0)}}{1-\left(1+\frac{\delta^2}{16}\cdot t^2\right)^{-k}}\enskip.
\]
Then for any natural number $k\ge\frac{M^6}{c^2}$ with $t:=\frac{1}{\sqrt{k}}$,
\[
\probm\left(R_M\ge k\right)\le \frac{1-e^{-\frac{\expv(X_0)}{\sqrt{k}}}}{1-\left(1+\frac{\delta^2}{16}\cdot \frac{1}{k}\right)^{-k}}\enskip.
\]
In particular, we have that for all natural numbers $k\ge\frac{M^6}{c^2}$,
\[
\probm\left(R_M\ge k\right)\le C\cdot\frac{1}{\sqrt{k}}\enskip.
\]
Since $\probm(R_M=\infty)=\lim\limits_{k\rightarrow\infty}\probm\left(R_M\ge k\right)$, one obtains that $\probm(R_M=\infty)=0$ and $\probm(R_M<\infty)=1$.
By applying the third item of Optional Stopping Theorem (cf. Theorem~\ref{thm:optstopping}), one has that
$\expv(X_{R_M})\le \expv(X_0)$.
Thus, by Markov's Inequality,
\[
\probm(X_{R_M}\ge M)\le \frac{\expv(X_{R_M})}{M}\le \frac{\expv(X_0)}{M}\enskip.
\]

Now for any natural number $k$ such that $M:=\sqrt[6]{c^2\cdot k}>\max\{\expv(X_0),\sqrt[6]{N}\}$, we have
\begin{eqnarray*}
& &\probm(\stopping{\Gamma}\ge k) \\
&=&\probm(\stopping{\Gamma}\ge k\wedge X_{R_M}=0) + \probm(\stopping{\Gamma}\ge k\wedge X_{R_M}\ge M)\\
&=&\probm(R_M\ge k\wedge X_{R_M}=0) + \probm(\stopping{\Gamma}\ge k\wedge X_{R_M}\ge M)\\
&\le& \probm(R_M\ge k) + \probm(X_{R_M}\ge M)\\
&\le& \frac{C}{\sqrt{k}}+\frac{\expv(X_0)}{M} \\
&=& \frac{C}{\sqrt{k}}+\frac{\expv(X_0)}{\sqrt[6]{c^2\cdot k}}\enskip.
\end{eqnarray*}
It follows that $\probm(\stopping{\Gamma}<\infty)=1$ and $k\mapsto\probm\left(\stopping{\Gamma}\ge k\right)\in \mathcal{O}\left(k^{-\frac{1}{6}}\right)$.
\end{IEEEproof}

\noindent{\bf Lemma~\ref{thm:supm2}.}
Consider any difference-bounded supermartingale $\Gamma=\{X_n\}_{n\in\Nset_0}$ adapted to a filtration $\{\mathcal{F}_n\}_{n\in\Nset_0}$ satisfying that there exist $\delta\in(0,\infty)$ and $K\in\Nset$ such that the following conditions hold:
\begin{compactenum}
\item $X_0$ is a constant random variable;
\item for every $n\in\Nset_0$,  it holds for all $\omega$ that (i) $X_n(\omega)\ge 0$ and (ii) $X_n(\omega)=0$ implies  $X_{n+1}(\omega)=0$;
\item for every $n\in\Nset_0$, it holds a.s. that either $X_{n+1}\le X_n$ or $\condexpv{|X_{n+1}-X_n|}{\mathcal{F}_n}\ge \delta$;
\item for every $n\in\Nset_0$ it holds a.s. that there exists an $k$ such that (i) $n\le k< n+K$ and (ii) either $X_k=0$ or $\condexpv{|X_{k+1}-X_k|}{\mathcal{F}_k}\ge \delta$.
\end{compactenum}
Then $\probm(\stopping{\Gamma}<\infty)=1$ and $k\mapsto\probm\left(\stopping{\Gamma}\ge k\right)\in \mathcal{O}\left(\frac{1}{\sqrt{k}}\right)$.
\begin{IEEEproof}
Let $K,\delta$ be given as in the statement of the lemma. W.l.o.g, we assume that $X_0>0$.
Let $c\in (0,\infty)$ be such that for every $n\in\Nset_0$, it holds a.s. that $|X_{n+1}-X_n|\le c$.
Fix any sufficiently small real number $t\in (0,\infty)$ such that
\[
e^{c\cdot t}-(1+c\cdot t +\frac{1}{2}\cdot c^2\cdot t^2)\left(=\sum_{j=3}^\infty\frac{(c\cdot t)^j}{{j}{!}} \right)\le \frac{\delta^2}{4}\cdot t^2\enskip.
\]
Define the discrete-time stochastic process $\{Y_n\}_{n\in\Nset_0}$ as in (\ref{eq:proof4:definition}).
Note that from difference-boundedness, $0<Y_n\le e^{n\cdot c\cdot t}$ a.s. for all $n\in\Nset_0$.
Then from (\ref{eq:proof4:martingale}), $\{Y_n\}_{n\in\Nset_0}$ is a martingale.

Define $D_n$ to be the random variable $\condexpv{|X_{n+1}-X_n|}{\mathcal{F}_n}$ for $n\in\Nset_0$.
For every $n\in\Nset_0$, it holds a.s. that
\begin{eqnarray*}
& &\condexpv{e^{-t\cdot \left(X_{n+1}-X_{n}\right)}}{\mathcal{F}_n}\\
&=& \mbox{\S~By (E6)~\S} \\
& & \condexpv{\mathbf{1}_{X_{n+1}\le X_n}\cdot\mathbf{1}_{D_n<\delta}\cdot e^{-t\cdot \left(X_{n+1}-X_{n}\right)}}{\mathcal{F}_n} \\
& &{}+ \condexpv{\mathbf{1}_{D_n\ge\delta}\cdot e^{-t\cdot \left(X_{n+1}-X_{n}\right)}}{\mathcal{F}_n} \\
&=& \mbox{\S~By (E8)~\S} \\
& &\condexpv{\mathbf{1}_{X_{n+1}\le X_n}\cdot \mathbf{1}_{D_n<\delta}\cdot e^{-t\cdot \left(X_{n+1}-X_{n}\right)}}{\mathcal{F}_n} \\
& &{}+\mathbf{1}_{D_n\ge\delta}\cdot \condexpv{e^{-t\cdot \left(X_{n+1}-X_{n}\right)}}{\mathcal{F}_n}\enskip.
\end{eqnarray*}

Since
\[
\mathbf{1}_{X_{n+1}\le X_n}\cdot \mathbf{1}_{D_n<\delta}\cdot e^{-t\cdot \left(X_{n+1}-X_{n}\right)}\ge \mathbf{1}_{X_{n+1}\le X_n}\cdot \mathbf{1}_{D_n<\delta},
\]
one has from (E10) that a.s. 
\begin{eqnarray*}
&\condexpv{\mathbf{1}_{X_{n+1}\le X_n}\cdot \mathbf{1}_{D_n<\delta}\cdot e^{-t\cdot \left(X_{n+1}-X_{n}\right)}}{\mathcal{F}_n}\\
&\qquad\qquad\qquad\qquad\qquad\ge \condexpv{\mathbf{1}_{X_{n+1}\le X_n}\cdot \mathbf{1}_{D_n<\delta}}{\mathcal{F}_n}~.
\end{eqnarray*}
Moreover, from (\ref{eq:proof:convexity}) and (\ref{eq:proof4:exponentiation}), we obtain that a.s. 
\[
\mathbf{1}_{D_n\ge\delta}\cdot \condexpv{e^{-t\cdot \left(X_{n+1}-X_{n}\right)}}{\mathcal{F}_n}\ge \mathbf{1}_{D_n\ge\delta}\cdot\left(1+\frac{\delta^2}{4}\cdot t^2\right)\enskip.
\]
It follows from the third item in the statement of the theorem that for all $n\in\Nset_0$, it holds a.s. that
\begin{eqnarray*}
& &\condexpv{e^{-t\cdot \left(X_{n+1}-X_{n}\right)}}{\mathcal{F}_n} \\
&=&\mbox{\S~By (E6) \S} \\
& &\condexpv{\mathbf{1}_{X_{n+1}\le X_n}\cdot \mathbf{1}_{D_n<\delta}\cdot e^{-t\cdot \left(X_{n+1}-X_{n}\right)}}{\mathcal{F}_n} \\
& &{}+ \condexpv{\mathbf{1}_{D_n\ge\delta}\cdot e^{-t\cdot \left(X_{n+1}-X_{n}\right)}}{\mathcal{F}_n}\\
&=& \mbox{\S~By (E8) \S} \\
& &\condexpv{\mathbf{1}_{X_{n+1}\le X_n}\cdot \mathbf{1}_{D_n<\delta}\cdot e^{-t\cdot \left(X_{n+1}-X_{n}\right)}}{\mathcal{F}_n} \\
& &{}+\mathbf{1}_{D_n\ge\delta}\cdot \condexpv{e^{-t\cdot \left(X_{n+1}-X_{n}\right)}}{\mathcal{F}_n}\\
&\ge& \condexpv{\mathbf{1}_{X_{n+1}\le X_n}\cdot \mathbf{1}_{D_n<\delta}}{\mathcal{F}_n} \\
& &{}+\mathbf{1}_{D_n\ge\delta}\cdot\left(1+\frac{\delta^2}{4}\cdot t^2\right)\\
&=& \mbox{\S~By (E5), (E6) \S} \\
& &
\condexpv{\mathbf{1}_{X_{n+1}\le X_n}\cdot \mathbf{1}_{D_n<\delta}+\mathbf{1}_{D_n\ge\delta}\cdot\left(1+\frac{\delta^2}{4}\cdot t^2\right)}{\mathcal{F}_n} \\
&\ge& \mbox{\S~By (E10) \S} \\
& &
\condexpv{\mathbf{1}_{X_{n+1}\le X_n}\cdot \mathbf{1}_{D_n<\delta}+\mathbf{1}_{D_n\ge\delta}}{\mathcal{F}_n} \\
&=& \condexpv{1}{\mathcal{F}_n} \\
&=& 1~~~\mbox{\S~By (E5) \S} \enskip.
\end{eqnarray*}
It follows that for all $n\in\Nset_0$, $\condexpv{e^{-t\cdot \left(X_{n+1}-X_{n}\right)}}{\mathcal{F}_n}\ge 1$ a.s.
Hence, $|Y_n|\le 1$ a.s. for all $n\in\Nset_0$.
Furthermore, from the fourth item in the statement of theorem, it holds a.s. that $\stopping{\Gamma}\ge n$ implies
\[
\prod_{j=0}^{n-1} \condexpv{e^{-t\cdot \left(X_{j+1}-X_{j}\right)}}{\mathcal{F}_j}\ge {\left(1+\frac{\delta^2}{4}\cdot t^2\right)}^{\left\lfloor\frac{n}{K}\right\rfloor}\enskip.
\]
Thus, it holds a.s. that
\[
\left(\lim\limits_{n\rightarrow\infty} Y_{n\wedge \stopping{\Gamma}}\right)(\omega)=
\begin{cases}
0 & \mbox{if }\stopping{\Gamma}(\omega)=\infty\\
Y_{\stopping{\Gamma}(\omega)}(\omega) & \mbox{if }\stopping{\Gamma}(\omega)<\infty
\end{cases}\enskip.
\]
Then by Optional Stopping Theorem (the first item), by letting $Y_\infty:=\lim\limits_{n\rightarrow\infty} Y_{n\wedge\stopping{\Gamma}}$
one has that
\[
\expv\left(Y_\infty\right)=\expv\left(Y_0\right)=e^{-t\cdot \expv(X_0)}\enskip.
\]
Moreover, similar to (\ref{eq:proof4:derivation}), one can obtain that
\begin{eqnarray*}
\expv\left(Y_\infty\right)&=&\int Y_\infty\cdot\mathbf{1}_{\stopping{\Gamma}=\infty}\,\mathrm{d}\probm+\int Y_\infty\cdot\mathbf{1}_{\stopping{\Gamma}<\infty}\,\mathrm{d}\probm\\
&=& 0\cdot\probm\left(T=\infty\right)+\sum_{n=0}^{\infty}\int Y_\infty\cdot\mathbf{1}_{\stopping{\Gamma}=n}\,\mathrm{d}\probm\\
&=& \sum_{n=0}^{\infty}\int Y_n\cdot\mathbf{1}_{\stopping{\Gamma}=n}\,\mathrm{d}\probm\\
&\le& \sum_{n=0}^{\infty}\int {\left(1+\frac{\delta^2}{4}\cdot t^2\right)}^{-\left\lfloor \frac{n}{K}\right\rfloor} \cdot\mathbf{1}_{\stopping{\Gamma}=n}\,\mathrm{d}\probm\\
&=& \sum_{n=0}^{\infty} {\left(1+\frac{\delta^2}{4}\cdot t^2\right)}^{-\left\lfloor \frac{n}{K}\right\rfloor} \cdot\probm\left(\stopping{\Gamma}=n\right) \\
&= & \sum_{n=0}^{k-1} {\left(1+\frac{\delta^2}{4}\cdot t^2\right)}^{-\left\lfloor \frac{n}{K}\right\rfloor} \cdot\probm\left(\stopping{\Gamma}=n\right)\\
& &~~{}+\sum_{n=k}^{\infty} {\left(1+\frac{\delta^2}{4}\cdot t^2\right)}^{-\left\lfloor \frac{n}{K}\right\rfloor} \cdot\probm\left(\stopping{\Gamma}=n\right)\\
&\le & \left(1-\probm\left(\stopping{\Gamma}\ge k\right)\right)\\
& &\quad{}+{\left(1+\frac{\delta^2}{4}\cdot t^2\right)}^{-\left\lfloor \frac{k}{K}\right\rfloor}\cdot \probm\left(\stopping{\Gamma}\ge k\right)
\end{eqnarray*}
for any $k\in\Nset$.
It follows that for all $k\in\Nset$, 
\[
e^{-t\cdot\expv(X_0)}\le 1-\left(1-\left(1+\frac{\delta^2}{4}\cdot t^2\right)^{-\left\lfloor \frac{k}{K}\right\rfloor}\right)\cdot \probm\left(\stopping{\Gamma}\ge k\right)\enskip.
\]
Hence,
\[
\probm\left(\stopping{\Gamma}\ge k\right)\le \frac{1-e^{-t\cdot\expv(X_0)}}{1-\left(1+\frac{\delta^2}{4}\cdot t^2\right)^{-\left\lfloor \frac{k}{K}\right\rfloor}}\enskip.
\]
Then for sufficiently large $k\in\Nset$ with $t:=\frac{1}{\sqrt{k}}$,
\[
\probm\left(\stopping{\Gamma}\ge k\right)\le \frac{1-e^{-\frac{\expv(X_0)}{\sqrt{k}}}}{1-{\left(1+\frac{\delta^2}{4}\cdot \frac{1}{k}\right)}^{-\left\lfloor\frac{k}{K}\right\rfloor}}\enskip.
\]
Using the facts that $\lim\limits_{k\rightarrow\infty}(1+\frac{\delta^2}{4}\cdot\frac{1}{k})^{\left\lfloor\frac{k}{K}\right\rfloor}= e^{\frac{\delta^2}{4\cdot K}}$ and $\lim\limits_{z\rightarrow 0^+}\frac{1-e^{-z}}{z}=1$, we have that the function
\[
k\mapsto\probm\left(\stopping{\Gamma}\ge k\right)\in \mathcal{O}\left(\frac{1}{\sqrt{k}}\right)\enskip.
\]
Since $\probm(\stopping{\Gamma}=\infty)=\lim\limits_{k\rightarrow\infty}\probm\left(\stopping{\Gamma}\ge k\right)$, one obtains immediately that $\probm(\stopping{\Gamma}=\infty)=0$ and $\probm(\stopping{\Gamma}<\infty)=1$.
\end{IEEEproof}

\begin{definition}[Super-measure Functions]\label{def:supmfunc}
A \emph{super-measure} function is a function $h:\stackelems\rightarrow[0,\infty]$ satisfying that there exist $\zeta,\delta\in(0,\infty)$ such that for all stack elements $(\fn{f},\loc,\nu)$,
\begin{compactitem}
\item[(D1)] $\loc=\lout{\fn{f}}$ iff $h(\fn{f},\loc,\nu)=0$, and
\end{compactitem}
if $\loc\ne\lout{\fn{f}}$ and $h(\fn{f},\loc,\nu)<\infty$ then the followings hold:
\begin{compactitem}
\item[(D2)] if $\loc\in\alocs{f}\setminus\{\lout{\fn{f}}\}$ and $(\loc, u,\loc')$ is the only triple in $\transitions{\fn{f}}$ with source label $\loc$ and update function $u$, then it holds that
\begin{compactitem}
\item $\sum_{\mu\in \samples}\sampdpd(\mu)\cdot h\left(\fn{f},\loc',u(\nu,\mu)\right)\le h(\fn{f},\loc,\nu)$, and
\item (*) $\left|g(\mu)\right|\le\zeta$ for all $\mu\in\samples$, and
\item $\sum_{\mu\in\samples}\sampdpd(\mu)\cdot |g(\mu)|\ge\delta$\enskip,
\end{compactitem}
where $g(\mu):= h\left(\fn{f},\loc',u(\nu,\mu)\right)- h(\fn{f},\loc,\nu)$;
\item [(D3)] if $\loc\in\flocs{f}\setminus\{\lout{\fn{f}}\}$ and $(\loc,(\fn{g},v),\loc')$ is the only triple in $\transitions{\fn{f}}$ with source label $\loc$ and value-passing function $v$, then
\begin{compactitem}
\item $h\left(\fn{g},\lin{\fn{g}}, v(\nu)\right)+h(\fn{f},\loc',\nu)\le h(\fn{f},\loc,\nu)$, and
\item (*) $\left|h\left(\fn{g},\lin{\fn{g}}, v(\nu)\right)+h(\fn{f},\loc',\nu)- h(\fn{f},\loc,\nu)\right|\le\zeta$;
\end{compactitem}
\item[(D4)] if $\loc\in\clocs{f}\setminus\{\lout{\fn{f}}\}$ and $(\loc, \phi,\loc_1),(\loc, \neg\phi,\loc_2)$ are namely two triples in $\transitions{\fn{f}}$ with source label $\loc$ and propositional arithmetic predicate $\phi$, then
\begin{compactitem}
\item $\sum_{i=1}^2\mathbf{1}_{\nu\models\phi_i}\cdot h(\fn{f},\loc_i,\nu)\le h(\fn{f},\loc,\nu)$, and
\item (*) $|\left(\sum_{i=1}^2\mathbf{1}_{\nu\models\phi_i}\cdot h(\fn{f},\loc_i,\nu)\right)-h(\fn{f},\loc,\nu)|\le \zeta$
\end{compactitem}
where $\phi_1:=\phi$ and $\phi_2:=\neg\phi$;
\item[(D5)] if $\loc\in\dlocs{f}\setminus\{\lout{\fn{f}}\}$ and $(\loc, \star,\loc_1),(\loc, \star,\loc_2)$ are namely two triples in $\transitions{\fn{f}}$ with source label $\loc$, then
\begin{compactitem}
\item $\max\{h(\fn{f},\loc_1,\nu),h(\fn{f},\loc_2,\nu)\}\le h(\fn{f},\loc,\nu)$, and
\item (*) $\max_{i\in\{1,2\}}|h(\fn{f},\loc_i,\nu)-h(\fn{f},\loc,\nu)|\le \zeta$.
\end{compactitem}
\end{compactitem}
Conditions marked by (*) refer to difference-boundedness. \hfill\IEEEQEDclosed
\end{definition}

\begin{definition}\label{def:fnamelabelsets}
The sequence of sets $\{\Theta_n\}_{n\in\Nset_0}$ and the collection of numbers $\{K_{\fn{f},\loc}\}_{\fn{f}\in\fnames,\loc\in\locs{f}}$ are inductively defined through the procedure below:
\begin{compactitem}
\item initially, $\Theta_0:=\{(\fn{f},\loc)\mid \fn{f}\in\fnames, \loc\in\alocs{f}\cup\{\lout{\fn{f}}\}\}$ and  $K_{\fn{f},\loc}:=0$ for all $\fn{f}\in\fnames$ and $\loc\in\locs{f}$;
\item $\Theta_{n+1}:=\Theta_{n+1}^\mathrm{c}\cup \Theta_{n+1}^\mathrm{b}\cup \Theta_{n+1}^\mathrm{d}\cup\Theta_{n}$ where
\begin{compactitem}
\item $\Theta_{n+1}^\mathrm{c}$ is the set of all pairs $(\fn{f},\loc)$ such that (i) $\fn{f}\in\fnames$ and $\loc\in\flocs{f}\setminus\{\lout{\fn{f}}\}$, (ii) $(\loc, (\fn{g},v),\loc')$ is the only triple in $\transitions{\fn{f}}$ and (iii) $(\fn{f},\loc'),(\fn{g},\lin{\fn{g}})\in\Theta_{n}$, while $K_{\fn{f},\loc}:=K_{\fn{f},\loc'}+K_{\fn{g},\lin{\fn{g}}}+1$ for all such pairs outside $\Theta_{n}$, and
\item $\Theta_{n+1}^\mathrm{b}$ (resp. $\Theta_{n+1}^\mathrm{d}$) is the set of all pairs $(\fn{f},\loc)$ such that (i) $\fn{f}\in\fnames$ and $\loc\in\flocs{f}\setminus\{\lout{\fn{f}}\}$ (resp. $\loc\in\dlocs{f}\setminus\{\lout{\fn{f}}\}$), and (ii) $(\loc, \phi,\loc_1), (\loc, \neg\phi,\loc_2)$ (resp. $(\loc, \star,\loc_1), (\loc, \star,\loc_2)$) are namely the two triples in $\transitions{\fn{f}}$ and $(\fn{f},\loc_1),(\fn{f},\loc_2)\in\Theta_{n}$, while $K_{\fn{f},\loc}:=1+\max\{K_{\fn{f},\loc_1},K_{\fn{f},\loc_2}\}$ for all such pairs outside $\Theta_{n}$.
\end{compactitem}
\end{compactitem}
\hfill\IEEEQEDclosed
\end{definition}

\begin{remark}\label{rmk:fnamelabelsets}
Since there are finitely many function names and labels, there exists a $m\in\Nset_0$ such that $\Theta_{n}=\Theta_{m}$ for all $n\ge m$.
We let $m^*$ be the smallest among all those $m$'s. Then $\Theta_{n}=\Theta_{m^*}$ for all $n\ge m^*$.\hfill\IEEEQEDclosed
\end{remark}

By an easy inductive proof on the inductive construction of $\{\Theta_n\}_{n\in\Nset_0}$, it is straightforward to observe the following fact from our semantics.

\begin{lemma}\label{lemm:fnamelabelsets}
for all $n\in\Nset_0$ and for all histories $\rho$, there exists a $k\in\Nset$ such that
\begin{compactitem}
\item $n\le k<n+1+\max_{\fn{f}\in\fnames,\loc\in\locs{f}}K_{\fn{f},\loc}$, and
\item either (i) $\rho[k]=((\fn{f}',\loc,\nu')\cdot w,\mu)$ for some $\fn{f}'\in\fnames$,
$\loc\in\alocs{f'}$, $\nu'\in\val{\fn{f}'}$, $\mu\in\samples$ and configuration $w$, or (ii) $\rho[k]=(\varepsilon,\mu)$ for some $\mu\in\samples$.
\end{compactitem}
\hfill\IEEEQEDclosed
\end{lemma}

Below we prove Corollary~\ref{thm:supmfunc}.
For the sake of convenience, we temporarily define $\infty-\infty:=\infty$ in the proof 
(this situation will always happen with probability zero).

\noindent{\bf Corollary~\ref{thm:supmfunc}.}
If it holds that (i) $(\fn{f},\loc)\in \Theta_{m^*}$ for all $\fn{f}\in\fnames,\loc\in\locs{f}$ and (ii) there exists a super-measure function $h$ (for $W$),
then $\probm_\mathfrak{c}^\sigma (T<\infty)=1$ and $k\mapsto\probm_\mathfrak{c}^\sigma\left(T\ge k\right)\in \mathcal{O}\left(\frac{1}{\sqrt{k}}\right)$
for all schedulers $\sigma$ and non-terminal stack elements $\mathfrak{c}$ such that $h(\mathfrak{c})\in (0,\infty)$.
\begin{IEEEproof}
Let $h$ be any super-measure function with $\delta,\zeta$ given in Definition~\ref{def:supmfunc} and $\mathfrak{c}=(\fn{f}, \loc, \nu)$ be any non-terminal stack element such that $h(\fn{f}, \loc, \nu)<\infty$.
Let $\sigma$ be any scheduler.
Define the stochastic process $\Gamma=\{X_n\}_{n\in\Nset_0}$ adapted to $\{\mathcal{H}_n\}_{n\in\Nset_0}$
as in (\ref{eq:proof:xnprocess}) (under the probability measure $\probm_\mathfrak{c}^\sigma$).

We first prove that $\expv(|X_n|)<\infty$ for all $n$.
From (\ref{eq:proof:xnplusone}),
\[
X_{n+1}= \mathbf{1}_{\rvlen_n\ge 1}\cdot\left[D+Y^{n,\mathrm{a}}+Y^{n,\mathrm{c}}+Y^{n,\mathrm{b}}+Y^{n,\mathrm{d}}\right]
\]
where all relevant random variables are defined in the proof for Lemma~\ref{thm:soundness}.
Then from (\ref{eq:proof:xxp}) and (\ref{eq:proof:xpxpp}),
$X_{n+1}\le X'_{n+1}$
where
\[
X'_{n+1}= \mathbf{1}_{\rvlen_n\ge 1}\cdot\left[D+Y^{n,\mathrm{a}}+Y^{n,\mathrm{c}}+Y^{n,\mathrm{b}}+\widehat{Y}^{n,\mathrm{d}}\right]\enskip.
\]
Furthermore, from (\ref{eq:proof:xpxpp}) and (\ref{eq:proof:xppnplusone}), $\expv^\sigma_\mathfrak{c}(X'_{n+1})= \expv^\sigma_\mathfrak{c}(X''_{n+1})$
where
\[
X''_{n+1}= \mathbf{1}_{\rvlen_n\ge 1}\cdot\left[D+\widehat{Y}^{n,\mathrm{a}}+Y^{n,\mathrm{c}}+Y^{n,\mathrm{b}}+\widehat{Y}^{n,\mathrm{d}}\right]\enskip.
\]
Then, by conditions (D1)--(D5) for super-measure functions, one has that
\begin{eqnarray*}
X''_{n+1} &\le & \mathbf{1}_{\rvlen_n\ge 1}\cdot \Bigg[D+h\left(\rvfn_{n,0},~\rvlb_{n,0}, ~\rvval^{\rvfn_{n,0}}_{n,0}\right)\Bigg]\nonumber\\
&=& \mathbf{1}_{\rvlen_n\ge 1}\cdot X_n\nonumber\\
&=& \mathbf{1}_{\rvlen_n\ge 1}\cdot X_n+\mathbf{1}_{\rvlen_n=0}\cdot X_n\nonumber\\
&=& X_n\enskip.
\end{eqnarray*}
Hence, we obtain that
\[
\expv^\sigma_\mathfrak{c}(X_{n+1})\le\expv^\sigma_\mathfrak{c}(X'_{n+1})=\expv^\sigma_\mathfrak{c}(X''_{n+1})\le \expv^\sigma_\mathfrak{c}(X_n)\enskip.
\]
Thus by a straightforward induction on $n$, one has that $\expv^\sigma_\mathfrak{c}(X_n)\le \expv^\sigma_\mathfrak{c}(X_0)=h(\mathfrak{c})<\infty$ for all $n$.

Then we prove that $\Gamma$ is a supermartingale.
Actually, it follows directly from (\ref{eq:proof:xnplusonexpxpp}): for all $n$,
\[
\condexpv{X_{n+1}}{\mathcal{H}_n} \le X''_{n+1} \le X_n \mbox{ a.s.}
\]
Furthermore, we prove that $\Gamma$ is difference bounded.
From (\ref{eq:proof2:absdifference}) and (D1), we have
\[
|X_{n+1}-X_n|= \mathbf{1}_{\rvlen_n\ge 1}\cdot \left(\widetilde{Y}^{n,\mathrm{a}}+\widetilde{Y}^{n,\mathrm{c}} + \widetilde{Y}^{n,\mathrm{b}} + \widetilde{Y}^{n,\mathrm{d}}\right)\enskip.
\]
Hence by (D2)--(D5), $\Gamma$ is difference bounded.

Now we prove that for every $n\in\Nset_0$, it holds a.s. that either $X_{n+1}\le X_n$ or $\condexpv{|X_{n+1}-X_n|}{\mathcal{H}_n}\ge \delta$.
By our semantics, we have that
\begin{eqnarray*}
& & X_{n+1}-X_n\\
&=& \mathbf{1}_{\rvlen_n\ge 1}\cdot \left(Y^{n,\mathrm{a}}+Y^{n,\mathrm{c}} + Y^{n,\mathrm{b}} + Y^{n,\mathrm{d}}-\rvtop_n\right)  \\
&=& \mathbf{1}_{\rvlen_n\ge 1}\cdot \left(\widetilde{\mathtt{Y}}^{n,\mathrm{a}}+\widetilde{\mathtt{Y}}^{n,\mathrm{c}} + \widetilde{\mathtt{Y}}^{n,\mathrm{b}} + \widetilde{\mathtt{Y}}^{n,\mathrm{d}}\right)\\
&\le & \mathbf{1}_{\rvlen_n\ge 1}\cdot \widetilde{\mathtt{Y}}^{n,\mathrm{a}}+\mathbf{1}_{\rvlen_n\ge 1}\cdot\left(\widetilde{\mathtt{Y}}^{n,\mathrm{c}} + \widetilde{\mathtt{Y}}^{n,\mathrm{b}} + \widehat{\mathtt{Y}}^{n,\mathrm{d}}\right)
\end{eqnarray*}
where
\[
\widetilde{\mathtt{Y}}^{n,\mathrm{a}}:=\sum_{\mu\in\samples}\mathbf{1}_{\rvsam_{n+1}=\mu}\cdot \widetilde{\mathtt{Y}}^{n,\mathrm{a}}_{\mu}
\]
with
\[
\widetilde{\mathtt{Y}}^{n,\mathrm{a}}_{\mu}:=\sum_{\fn{f}\in\fnames}\sum_{\loc\in\alocs{f}\setminus\{\lout{\fn{f}}\}}\mathbf{1}_{\left(\rvfn_{n,0}, \rvlb_{n,0}\right)=
(\fn{f},\loc)}\cdot (Y^{n,\mathrm{a}}_{\fn{f},\loc,\mu}-\rvtop_n),
\]
and
\[
\widetilde{\mathtt{Y}}^{n,\mathrm{c}}:=\sum_{\fn{f}\in\fnames}\sum_{\loc\in\flocs{f}\setminus\{\lout{\fn{f}}\}}\mathbf{1}_{\left(\rvfn_{n,0}, \rvlb_{n,0}\right)=
(\fn{f},\loc)}\cdot (Y^{n,\mathrm{c}}_{\fn{f},\loc}-\rvtop_n),
\]
\[
\widetilde{\mathtt{Y}}^{n,\mathrm{b}}:=\sum_{\fn{f}\in\fnames}\sum_{\loc\in\clocs{f}\setminus\{\lout{\fn{f}}\}}\mathbf{1}_{\left(\rvfn_{n,0}, \rvlb_{n,0}\right)=(\fn{f},\loc)}\cdot (Y^{n,\mathrm{b}}_{\fn{f},\loc}-\rvtop_n),
\]
\[
\widetilde{\mathtt{Y}}^{n,\mathrm{d}}:=\sum_{\fn{f}\in\fnames}\sum_{\loc\in\dlocs{f}\setminus\{\lout{\fn{f}}\}}\mathbf{1}_{\left(\rvfn_{n,0}, \rvlb_{n,0}\right)=(\fn{f},\loc)}\cdot (Y^{n,\mathrm{d}}_{\fn{f},\loc}-\rvtop_n),
\]
\[
\widehat{\mathtt{Y}}^{n,\mathrm{d}}:=\sum_{\fn{f}\in\fnames}\sum_{\loc\in\dlocs{f}\setminus\{\lout{\fn{f}}\}}\mathbf{1}_{\left(\rvfn_{n,0}, \rvlb_{n,0}\right)=(\fn{f},\loc)}\cdot (\widehat{Y}^{n,\mathrm{d}}_{\fn{f},\loc}-\rvtop_n).
\]

Thus, from (\ref{eq:proof:absoluteconditional}) and (D1)--(D5), it holds a.s. that for all $n$,
\begin{compactitem}
\item either $X_{n+1}(\omega)=X_n(\omega)=0$,
\item or $\rvlb_{n,0}(\omega)\in L^{\rvfn_{n,0}(\omega)}\setminus \left(L^{\rvfn_{n,0}(\omega)}_{\mathrm{a}}\cup \lout{\rvfn_{n,0}(\omega)}\right)$ and $X_{n+1}(\omega)\le X_n(\omega)$,
\item or $\rvlb_{n,0}(\omega)\in L^{\rvfn_{n,0}(\omega)}_{\mathrm{a}}$ and $\condexpv{|X_{n+1}-X_n|}{\mathcal{H}_n}(\omega)\ge\delta$.
\end{compactitem}

Finally, since $(\fn{f},\loc)\in \Theta_{m^*}$ for all $\fn{f}\in\fnames,\loc\in\locs{f}$,
by letting $K:=\max_{\fn{f}\in\fnames,\loc\in\locs{f}} K_{\fn{f},\loc}$, we have that
for every $n\in\Nset_0$ it holds a.s. that there exists an $k$ such that (i) $n\le k< n+K$ and (ii) either $X_k=0$ or $\condexpv{|X_{k+1}-X_k|}{\mathcal{F}_k}\ge \delta$.
Now it follows from Lemma~\ref{thm:supm2} that
\[
\probm_\mathfrak{c}^\sigma(T<\infty)=\probm_\mathfrak{c}^\sigma(\stopping{\Gamma}<\infty)=1
\]
and
\[
k\mapsto\probm_\mathfrak{c}^\sigma(T\ge k)\in\mathcal{O}\left(\frac{1}{\sqrt{k}}\right)\enskip.
\]
\end{IEEEproof}

\begin{lemma}\label{thm:supm3}
Consider any supermartingale $\Gamma=\{X_n\}_{n\in\Nset_0}$ adapted to a filtration $\{\mathcal{F}_n\}_{n\in\Nset_0}$ satisfying that there exist $\delta\in(0,\infty)$ and $K\in\Nset$ such that the following conditions hold:
\begin{compactenum}
\item $X_0$ is a constant random variable;
\item for every $n\in\Nset_0$,  it holds for all $\omega$ that (i) $X_n(\omega)\ge 0$ and (ii) $X_n(\omega)=0$ implies  $X_{n+1}(\omega)=0$;
\item for every $n\in\Nset_0$, it holds a.s. that either $X_{n+1}\le X_n$ or $\condexpv{|X_{n+1}-X_n|}{\mathcal{F}_n}\ge \delta$;
\item for every $n\in\Nset_0$ it holds a.s. that there exists an $k$ such that (i) $n\le k< n+K$ and (ii) either $X_k=0$ or $\condexpv{|X_{k+1}-X_k|}{\mathcal{F}_k}\ge \delta$.
\end{compactenum}
Then $\probm(\stopping{\Gamma}<\infty)=1$ and $k\mapsto\probm\left(\stopping{\Gamma}\ge k\right)\in \mathcal{O}\left(k^{-\frac{1}{6}}\right)$.\hfill\IEEEQEDclosed
\end{lemma}
\begin{IEEEproof}
W.l.o.g., we assume that $X_0>0$. Let $K,\delta$ be given as in the statement of the lemma.
From
\[
\lim\limits_{k\rightarrow\infty}\frac{1-e^{-\frac{\expv(X_0)}{\sqrt{k}}}}{\frac{\expv(X_0)}{\sqrt{k}}}=1\mbox{ and } \lim\limits_{k\rightarrow\infty} \left(1+\frac{\delta^2}{16}\cdot \frac{1}{k}\right)^{-\left\lfloor\frac{k}{K}\right\rfloor}= e^{-\frac{\delta^2}{16\cdot K}}\enskip,
\]
one can fix a constant natural number $N$ such that for all $k\ge N$,
\[
\frac{1-e^{-\frac{\expv(X_0)}{\sqrt{k}}}}{\frac{\expv(X_0)}{\sqrt{k}}}\le \frac{3}{2}\mbox{ and } 1-\left(1+\frac{\delta^2}{16}\cdot \frac{1}{k}\right)^{-\left\lfloor\frac{k}{K}\right\rfloor}\ge \frac{1-e^{-\frac{\delta^2}{16\cdot K}}}{2}\enskip.
\]
Let
\[
C:= \frac{3}{2}\cdot \expv(X_0)\cdot \frac{2}{1-e^{-\frac{\delta^2}{16\cdot K}}}\enskip.
\]
Choose a constant $c\in (0,1)$ such that
\[
\sum_{j=3}^{\infty} \frac{c^{j-2}}{j!}\le \frac{\delta^2}{16}\enskip.
\]
Define $D_n:=\condexpv{|X_{n+1}-X_n|}{\mathcal{F}_n}$ for $n\in\Nset_0$.
Note that from (E6), it holds a.s. that 
\begin{eqnarray*}
D_n&=&\condexpv{\mathbf{1}_{X_{n+1}< X_n}\cdot (X_{n}-X_{n+1})}{\mathcal{F}_n}\\
& &{}+\condexpv{\mathbf{1}_{X_{n+1}\ge X_n}\cdot (X_{n+1}-X_{n})}{\mathcal{F}_n}
\end{eqnarray*}
Moreover, from (E5), (E6) and definition of supermartingales, it holds a.s. for all $n$ that 
\begin{eqnarray*}
& &\condexpv{X_{n+1}-X_n}{\mathcal{F}_n}\\
&=&\condexpv{\mathbf{1}_{X_{n+1}< X_n}\cdot (X_{n+1}-X_{n})}{\mathcal{F}_n}\\
& &{}+\condexpv{\mathbf{1}_{X_{n+1}\ge X_n}\cdot (X_{n+1}-X_{n})}{\mathcal{F}_n}\\
&=&-\condexpv{\mathbf{1}_{X_{n+1}< X_n}\cdot (X_{n}-X_{n+1})}{\mathcal{F}_n}\\
& &{}+\condexpv{\mathbf{1}_{X_{n+1}\ge X_n}\cdot (X_{n+1}-X_{n})}{\mathcal{F}_n}\\
&\le & 0\enskip.
\end{eqnarray*}
It follows that a.s. 
\[
\condexpv{\mathbf{1}_{X_{n+1}< X_n}\cdot (X_{n}-X_{n+1})}{\mathcal{F}_n}\ge
\frac{D_n}{2}\enskip.
\]

Let $M$ be any real number in $(\max\{\expv(X_0), \sqrt[6]{N}\},\infty)$ and define the stopping time $R_M$ w.r.t $\{\mathcal{F}_n\}_{n\in\Nset_0}$ by
\[
R_M(\omega):=  \min\{n\mid X_{n}(\omega)\le 0\mbox{ or } X_{n}(\omega)\ge M\}
\]
where $\min\emptyset:=\infty$. Define the stochastic process $\Gamma'=\{X'_n\}_{n\in\Nset_0}$ adapted to $\{\mathcal{F}_n\}_{n\in\Nset_0}$ as in (\ref{eq:proof:primesupermartingale}).
By the same analysis in the proof of Theorem~\ref{thm:supmextended}, $\Gamma'$ is a difference-bounded supermartingale.
Moreover, it holds a.s. for all $n$ that 
\begin{eqnarray*}
& &\mathbf{1}_{0<X'_n< M}\cdot \condexpv{|X'_{n+1}-X'_n|}{\mathcal{F}_n} \\
&=& \mbox{\S~By~(E8) \S} \\
& & \condexpv{\mathbf{1}_{0<X'_n< M}\cdot |X'_{n+1}-X'_n|}{\mathcal{F}_n}\\
&\ge & \mbox{\S~By~(E10) \S} \\
& & \condexpv{\mathbf{1}_{0<X'_n< M}\cdot \mathbf{1}_{X'_{n+1}<X'_n}\cdot \left(X'_{n}-X'_{n+1}\right)}{\mathcal{F}_n} \\
&=& \condexpv{\mathbf{1}_{0<X'_n< M}\cdot \mathbf{1}_{X_{n+1}<X_n}\cdot \left(X_{n}-X_{n+1}\right)}{\mathcal{F}_n}\\
&=& \mbox{\S~By~(E8) \S} \\
& & \mathbf{1}_{0<X'_n< M}\cdot\condexpv{ \mathbf{1}_{X_{n+1}<X_n}\cdot \left(X_{n}-X_{n+1}\right)}{\mathcal{F}_n}\\
&\ge & \mathbf{1}_{0<X'_n< M}\cdot \frac{D_n}{2}\enskip.
\end{eqnarray*}
Since $\condexpv{|X'_{n+1}-X'_n|}{\mathcal{F}_n}\ge 0$ a.s. (from (E10)), we obtain that
\[
\condexpv{|X'_{n+1}-X'_n|}{\mathcal{F}_n}\ge \mathbf{1}_{0<X'_n< M}\cdot \frac{D_n}{2}\enskip.
\]
Hence, from (E13), it holds a.s. that
\begin{eqnarray*}
\condexpv{(X'_{n+1}-X'_n)^2}{\mathcal{F}_n}&\ge &{\left(\condexpv{|X'_{n+1}-X'_n|}{\mathcal{F}_n}\right)}^2\\
&\ge& \mathbf{1}_{0<X'_n< M}\cdot \frac{D_n^2}{4}\enskip.
\end{eqnarray*}

Now define the discrete-time stochastic process $\{Y_n\}_{n\in\Nset_0}$ by
\[
Y_n:=\frac{e^{-t\cdot X'_n}}{\prod_{j=0}^{n-1} \condexpv{e^{-t\cdot \left(X'_{j+1}-X'_{j}\right)}}{\mathcal{F}_j}}\enskip.
\]
where $t$ is an arbitrary real number in $(0,\frac{c}{M^3}]$.
Note that from difference-boundedness, $0<Y_n\le e^{n\cdot M\cdot t}$ a.s. for all $n\in\Nset_0$.
Then by the same analysis in (\ref{eq:proof4:martingale}), $\{Y_n\}_{n\in\Nset_0}$ is a martingale.

By similar analysis in (\ref{eq:proof4:exponentiation}), it holds a.s. for all $n$ that $0<X'_n<M$ and $D_n\ge\delta$
implies
\begin{eqnarray*}
& &\expv\left(e^{-t\cdot \left(X'_{n+1}-X'_{n}\right)}\mid\mathcal{F}_n\right) \nonumber\\
&\ge & 1+\frac{t^2}{2}\cdot\condexpv{(X'_{n+1}-X'_{n})^2}{\mathcal{F}_n}- \sum_{j=3}^{\infty}\frac{{(M\cdot t)}^j}{{j}{!}}~~\\
&\ge & 1+\frac{t^2}{2}\cdot\frac{\delta^2}{4}- t^2\cdot \sum_{j=3}^{\infty}\frac{M^j\cdot t^{j-2}}{{j}{!}}~~\\
&\ge & 1+\frac{\delta^2}{8}\cdot t^2- t^2\cdot \sum_{j=3}^{\infty}\frac{M^{-2\cdot j+6}\cdot c^{j-2}}{{j}{!}}~~\\
&\ge & 1+\frac{\delta^2}{8}\cdot t^2- t^2\cdot \sum_{j=3}^{\infty}\frac{c^{j-2}}{{j}{!}}~~\\
&\ge & 1+\frac{\delta^2}{8}\cdot t^2- t^2\cdot \frac{\delta^2}{16}~~\\
&\ge & 1+\frac{\delta^2}{16}\cdot t^2\enskip.\nonumber
\end{eqnarray*}
Moreover, define random variables $V_1, V_2, V_3$ by:
\begin{compactitem}
\item $V_1:= \mathbf{1}_{0<X'_{n}<M}\cdot\mathbf{1}_{X_{n+1}\le X_n}\cdot \mathbf{1}_{D_n<\delta}$;
\item $V_2:= \mathbf{1}_{0<X'_{n}<M}\cdot \mathbf{1}_{D_n\ge\delta}$;
\item $V_3:= \mathbf{1}_{X'_{n}=0\vee X'_{n}=M}$. 
\end{compactitem}

Moreover, for every $n\in\Nset_0$, it holds a.s. that
\begin{eqnarray*}
& &\condexpv{e^{-t\cdot \left(X'_{n+1}-X'_{n}\right)}}{\mathcal{F}_n}\\
&=& \mbox{\S~By (E6)~\S} \\
& & \condexpv{V_1\cdot e^{-t\cdot \left(X_{n+1}-X_{n}\right)}}{\mathcal{F}_n} \\
& &{}+ \condexpv{V_2\cdot e^{-t\cdot \left(X'_{n+1}-X'_{n}\right)}}{\mathcal{F}_n} \\
& &{}+ \condexpv{V_3\cdot e^{-t\cdot \left(X'_{n+1}-X'_{n}\right)}}{\mathcal{F}_n} \\
&=& \mbox{\S~By (E8)~\S} \\
& &\condexpv{V_1\cdot e^{-t\cdot \left(X'_{n+1}-X'_{n}\right)}}{\mathcal{F}_n} \\
& &{}+V_2\cdot \condexpv{e^{-t\cdot \left(X'_{n+1}-X'_{n}\right)}}{\mathcal{F}_n}\\
& &{}+ \condexpv{V_3\cdot e^{-t\cdot \left(X'_{n+1}-X'_{n}\right)}}{\mathcal{F}_n} \\
&\ge& \mbox{\S~By (E10), (E6), (E1), (E5)~\S} \\
& &\condexpv{V_1+V_3}{\mathcal{F}_n}+V_2\cdot \left(1+\frac{\delta^2}{16}\cdot t^2\right)\\
&=& \mbox{\S~By (E5), (E6)~\S} \\
& &\condexpv{V_1+V_3+V_2\cdot \left(1+\frac{\delta^2}{16}\cdot t^2\right)}{\mathcal{F}_n} \\
&\ge& 1 \enskip.
\end{eqnarray*}

Thus,
\begin{compactitem}
\item $|Y_{R_M\wedge n}|\le 1$ a.s. for all $n\in\Nset_0$, and
\item it holds a.s. that
\[
\left(\lim\limits_{n\rightarrow\infty} Y_{n\wedge R_M}\right)(\omega)=
\begin{cases}
0 & \mbox{if }R_M(\omega)=\infty\\
Y_{R_M(\omega)}(\omega) & \mbox{if }R_M(\omega)<\infty
\end{cases}\enskip.
\]
\end{compactitem}
Then from Optional Stopping Theorem (Item 1 of Theorem~\ref{thm:optstopping}), by letting $Y_\infty:=\lim\limits_{n\rightarrow\infty} Y_{n\wedge R_M}$
one has that
\[
\expv\left(Y_\infty\right)=\expv\left(Y_0\right)=e^{-t\cdot \expv(X_0)}\enskip.
\]
Moreover, one can obtain that
\begin{eqnarray*}
& &\expv\left(Y_\infty\right)\nonumber\\
&=& \mbox{\S~By Definition \S} \nonumber\\
& & \int Y_\infty\,\mathrm{d}\probm \nonumber\\
&=& \mbox{\S~By Linear Property of Lebesgue Integral \S} \nonumber\\
& &\int Y_\infty\cdot\mathbf{1}_{R_M=\infty}\,\mathrm{d}\probm+\int Y_\infty\cdot\mathbf{1}_{R_M<\infty}\,\mathrm{d}\probm\nonumber\\
&=& \mbox{\S~By Monotone Convergence Theorem~\cite[Chap. 6]{probabilitycambridge} \S} \nonumber\\
& & 0\cdot\probm\left(R_M=\infty\right)+\sum_{n=0}^{\infty}\int Y_\infty\cdot\mathbf{1}_{R_M=n}\,\mathrm{d}\probm\nonumber\\
&=& \sum_{n=0}^{\infty}\int Y_n\cdot\mathbf{1}_{R_M=n}\,\mathrm{d}\probm\nonumber\\
&\le& \mbox{\S~By $X'_{n}\ge 0$ \S} \nonumber\\
& & \sum_{n=0}^{\infty}\int {\left(1+\frac{\delta^2}{16}\cdot t^2\right)}^{-\left\lfloor\frac{n}{K}\right\rfloor} \cdot\mathbf{1}_{R_M=n}\,\mathrm{d}\probm\nonumber\\
&=& \sum_{n=0}^{\infty} {\left(1+\frac{\delta^2}{16}\cdot t^2\right)}^{-\left\lfloor\frac{n}{K}\right\rfloor} \cdot\probm\left(R_M=n\right) \nonumber\\
&= & \sum_{n=0}^{k-1} {\left(1+\frac{\delta^2}{16}\cdot t^2\right)}^{-\left\lfloor\frac{n}{K}\right\rfloor} \cdot\probm\left(R_M=n\right)\nonumber\\
& &~~{}+\sum_{n=k}^{\infty} {\left(1+\frac{\delta^2}{16}\cdot t^2\right)}^{-\left\lfloor\frac{n}{K}\right\rfloor} \cdot\probm\left(R_M=n\right)\nonumber\\
&\le & \left(1-\probm\left(R_M\ge k\right)\right)\nonumber\\
& & \qquad{}+{\left(1+\frac{\delta^2}{16}\cdot t^2\right)}^{-\left\lfloor\frac{k}{K}\right\rfloor}\cdot \probm\left(R_M\ge k\right)~~
\end{eqnarray*}
for any $k\in\Nset$.
It follows that for all $k\in\Nset$,
\[
e^{-t\cdot\expv(X_0)}\le 1-\left(1-\left(1+\frac{\delta^2}{16}\cdot t^2\right)^{-\left\lfloor\frac{k}{K}\right\rfloor}\right)\cdot \probm\left( R_M\ge k\right)\enskip.
\]
Hence, for any $k\in\Nset$ and $t\in \left(0,\frac{c}{M^3}\right]$,
\[
\probm\left(R_M\ge k\right)\le \frac{1-e^{-t\cdot\expv(X_0)}}{1-\left(1+\frac{\delta^2}{16}\cdot t^2\right)^{-\left\lfloor\frac{k}{K}\right\rfloor}}\enskip.
\]
Then for any natural number $k\ge\frac{M^6}{c^2}$ with $t:=\frac{1}{\sqrt{k}}$,
\[
\probm\left(R_M\ge k\right)\le \frac{1-e^{-\frac{\expv(X_0)}{\sqrt{k}}}}{1-\left(1+\frac{\delta^2}{16}\cdot \frac{1}{k}\right)^{-\left\lfloor\frac{k}{K}\right\rfloor}}\enskip.
\]
In particular, we have that for all natural numbers $k\ge\frac{M^6}{c^2}$,
\[
\probm\left(R_M\ge k\right)\le C\cdot\frac{1}{\sqrt{k}}\enskip.
\]
Since $\probm(R_M=\infty)=\lim\limits_{k\rightarrow\infty}\probm\left(R_M\ge k\right)$, one obtains that $\probm(R_M=\infty)=0$ and $\probm(R_M<\infty)=1$.
By applying the third item of Optional Stopping Theorem (cf. Theorem~\ref{thm:optstopping}), one has that
$\expv(X_{R_M})\le \expv(X_0)$.
Thus, by Markov's Inequality,
\[
\probm(X_{R_M}\ge M)\le \frac{\expv(X_{R_M})}{M}\le \frac{\expv(X_0)}{M}\enskip.
\]

Now for any natural number $k$ such that $M:=\sqrt[6]{c^2\cdot k}>\max\{\expv(X_0),\sqrt[6]{N}\}$, we have
\begin{eqnarray*}
& &\probm(\stopping{\Gamma}\ge k) \\
&=&\probm(\stopping{\Gamma}\ge k\wedge X_{R_M}=0) + \probm(\stopping{\Gamma}\ge k\wedge X_{R_M}\ge M)\\
&=&\probm(R_M\ge k\wedge X_{R_M}=0) + \probm(\stopping{\Gamma}\ge k\wedge X_{R_M}\ge M)\\
&\le& \probm(R_M\ge k) + \probm(X_{R_M}\ge M)\\
&\le& \frac{C}{\sqrt{k}}+\frac{\expv(X_0)}{M} \\
&=& \frac{C}{\sqrt{k}}+\frac{\expv(X_0)}{\sqrt[6]{c^2\cdot k}}\enskip.
\end{eqnarray*}
It follows that $\probm(\stopping{\Gamma}<\infty)=1$ and $k\mapsto\probm\left(\stopping{\Gamma}\ge k\right)\in \mathcal{O}\left(k^{-\frac{1}{6}}\right)$.
\end{IEEEproof}

\section{Details for Several Examples}\label{app:examples}

\noindent\textbf{Example~\ref{ex:special:nonnegativity}.}
In Definition~\ref{def:rsupm}, the Non-negativity condition is necessary; in other words,  it is necessary having  $X_{\stopping{\Gamma}}=0$ rather than $X_{\stopping{\Gamma}}\le 0$ when $\stopping{\Gamma}<\infty$.
This can be observed as follows.
Consider the discrete-time stochastic processes $\{X_n\}_{n\in\Nset_0}$ and $\Gamma=\{Y_n\}_{n\in\Nset_0}$ given as follows:
\begin{compactitem}
\item the random variables $X_0,\dots,X_n,\dots$ are independent, $X_0$ is the random variable with constant value $\frac{1}{2}$ and each $X_n$ ($n\ge 1$) satisfies that $\probm\left(X_n=1\right)=e^{-\frac{1}{n^2}}$ and $\probm\left(X_n=-4\cdot n^2\right)=1-e^{-\frac{1}{n^2}}$;
\item $Y_n:=\sum_{j=0}^{n}X_j$ for $n\ge 0$.
\end{compactitem}
Let the filtration $\{\mathcal{F}_n\}_{n\in\Nset_0}$ be given such that each $\mathcal{F}_n$ is the $\sigma$-algebra generated by $X_0,\dots,X_n$ (i.e., the smallest $\sigma$-algebra that makes  $X_0,\dots,X_n$ measurable).
It is straightforward to see that every $Y_n$ is integrable and $\mathcal{F}_n$-measurable, and every $X_{n+1}$ is independent of $\mathcal{F}_n$.
Thus for $n\ge 0$,
we have that (cf. properties for conditional expectation in Appendix~\ref{app:condexpv})
\begin{eqnarray*}
\condexpv{Y_{n+1}}{\mathcal{F}_n}&=& \condexpv{Y_{n}+X_{n+1}}{\mathcal{F}_n}\\
\mbox{(by (E6), (E5))} &=& Y_{n}+\condexpv{X_{n+1}}{\mathcal{F}_n}\\
\mbox{(by (E9))} &=& Y_{n}+\expv\left(X_{n+1}\right)\\
&=& Y_{n}+\left(e^{-\frac{1}{(n+1)^2}} - 4\cdot \frac{1-e^{-\frac{1}{(n+1)^2}}}{\frac{1}{(n+1)^2}}\right)\\
&\le& Y_{n}+1-4\cdot \left(1-e^{-1}\right)\\
&\le& Y_n-1.52~~~,
\end{eqnarray*}
where the first inequality is obtain by the fact that the function $x\mapsto \frac{1-e^{-x}}{x}$ is decreasing over $(0,\infty)$. Hence, $\{Y_n\}_{n\in\Nset_0}$ satisfies the Ranking Condition. However, since $Y_n<0$ once $X_n=-4\cdot n^2$, 
$\probm\left(\stopping{\Gamma}> n\right)=\prod_{j=1}^{n}e^{-\frac{1}{j^2}}$.
It follows directly that
$\probm\left(\stopping{\Gamma}=\infty\right)=\lim\limits_{n\rightarrow\infty}\probm\left(\stopping{\Gamma}> n\right)=e^{-\frac{\pi^2}{6}}>0$\enskip.\hfill\IEEEQEDclosed

\noindent\textbf{Example~\ref{ex:special:cboundedness}.}
The conditional difference-boundedness in Proposition~\ref{prop:cbrsupm} cannot be dropped.
Consider the family $\{Y_n\}_{n\in\Nset_0}$ of independent random variables defined as follows:
$Y_0:=3$ and each $Y_n$ ($n\ge 1$) satisfies that
$\probm\left(Y_n=2^{n-1}\right)=\frac{1}{2}$ and $\probm\left(Y_n=-2^{n-1}-2\right)=\frac{1}{2}$.
Let the stochastic process $\Gamma=\{X_n\}_{n\in\Nset_0}$ be inductively defined by: $X_0:=Y_0$ and for all $n\in\Nset_0$,
\[
X_{n+1}:=\mathbf{1}_{X_n>0}\cdot\left(X_n+Y_{n+1}\right) \enskip.
\]
Note that once $n\ge 1$ and $Y_n(\omega)=-2^{n-1}-2$, $X_n(\omega)=0$ and $\stopping{\Gamma}(\omega)\le n$ as $3+\sum_{k=1}^{n-1} 2^{k-1}=2^{n-1}+2$.
Let $\{\mathcal{F}_n\}_{n\in\Nset_0}$ be the filtration such that each $\mathcal{F}_n$ is the smallest $\sigma$-algebra that makes all $Y_0,\dots,Y_n$ measurable, so that $\Gamma$ is adapted to $\{\mathcal{F}_n\}_{n\in\Nset_0}$.
Then, one has that for all $n\in\Nset_0$ (cf. properties for conditional expectation in Appendix~\ref{app:condexpv}),
\begin{eqnarray*}
& &\condexpv{X_{n+1}}{\mathcal{F}_n}-X_n \\
& = & \condexpv{\mathbf{1}_{X_n>0}\cdot\left(X_n+Y_{n+1}\right)}{\mathcal{F}_n} -X_n\\
&=& \condexpv{\mathbf{1}_{X_n>0}\cdot X_n+\mathbf{1}_{X_n>0}\cdot Y_{n+1}}{\mathcal{F}_n} -X_n\\
& = & \mbox{\S~By (E6) \S} \\
& & \condexpv{\mathbf{1}_{X_n>0}\cdot X_n}{\mathcal{F}_n}+ \condexpv{\mathbf{1}_{X_n>0}\cdot Y_{n+1}}{\mathcal{F}_n}-X_n\\
& = & \mbox{\S~By (E5), (E8) \S} \\
& & \mathbf{1}_{X_n>0}\cdot X_n+\mathbf{1}_{X_n>0}\cdot\condexpv{Y_{n+1}}{\mathcal{F}_n}-X_n \\
& = & \mbox{\S~By (E9) and $X_n\ge 0$ \S} \\
& &
\mathbf{1}_{X_n>0}\cdot\left[2^{n-1}\cdot\frac{1}{2}-\left(2^{n-1}+2\right)\cdot\frac{1}{2}\right] \\
&=& -\mathbf{1}_{X_n>0} \enskip.
\end{eqnarray*}
Hence $\{X_n\}_{n\in\Nset_0}$ is a ranking supermartingale and for all $n$ and $\omega$, $X_n(\omega)=0$ implies $X_{n+1}(\omega)=0$.
However,
\[
\expv(\stopping{\Gamma})=\sum_{n=1}^\infty n\cdot \probm\left(\stopping{\Gamma}=n\right)=\sum_{n=1}^\infty \frac{n}{2^n}=2,
\]
which implies that $\expv(\stopping{\Gamma})<\frac{\expv(X_0)}{1}=3$. \hfill\IEEEQEDclosed

\noindent\textbf{Example~\ref{ex:special:noconcentration}.}
In general, the difference-boundedness condition cannot be dropped in Theorem~\ref{thm:concentration}.
Fix any $\alpha\in (1,\infty)$ and consider the family $\{Y_n\}_{n\in\Nset_0}$ of independent random variables defined as follows:
$Y_0:=3$ and each $Y_n$ ($n\ge 1$) satisfies that
$\probm\left(Y_n=2\right)=\frac{n^\alpha}{(n+1)^\alpha}$ and $\probm\left(Y_n=-2\cdot n-1\right)=1-\frac{n^\alpha}{(n+1)^\alpha}$.
Let the stochastic process $\Gamma=\{X_n\}_{n\in\Nset_0}$ be inductively defined by: $X_0:=Y_0$ and for all $n\in\Nset_0$, $X_{n+1}:=\mathbf{1}_{X_n>0}\cdot\left(X_n+Y_{n+1}\right)$~.
Note that once $n\ge 1$ and $Y_n(\omega)=-2\cdot n-1$, $X_n(\omega)=0$ and $\stopping{\Gamma}(\omega)\le n$ as $3+\sum_{k=1}^{n-1} 2=2\cdot n+1$.
Let $\{\mathcal{F}_n\}_{n\in\Nset_0}$ be the filtration such that each $\mathcal{F}_n$ is the smallest $\sigma$-algebra that makes all $Y_0,\dots,Y_n$ measurable.
Then, one has that for all $n\in\Nset_0$ (cf. properties for conditional expectation in Appendix~\ref{app:condexpv}),
\begin{eqnarray*}
& &\condexpv{X_{n+1}}{\mathcal{F}_n}-X_n \\
& = & \condexpv{\mathbf{1}_{X_n>0}\cdot\left(X_n+Y_{n+1}\right)}{\mathcal{F}_n} -X_n\\
&=& \condexpv{\mathbf{1}_{X_n>0}\cdot X_n+\mathbf{1}_{X_n>0}\cdot Y_{n+1}}{\mathcal{F}_n} -X_n\\
& = & \mbox{\S~By (E6) \S} \\
& & \condexpv{\mathbf{1}_{X_n>0}\cdot X_n}{\mathcal{F}_n}+ \condexpv{\mathbf{1}_{X_n>0}\cdot Y_{n+1}}{\mathcal{F}_n}-X_n\\
& = & \mbox{\S~By (E5), (E8) \S} \\
& & \mathbf{1}_{X_n>0}\cdot X_n+\mathbf{1}_{X_n>0}\cdot\condexpv{Y_{n+1}}{\mathcal{F}_n}-X_n \\
& = & \mbox{\S~By (E9) and $X_n\ge 0$ \S} \\
& &
\mathbf{1}_{X_n>0}\cdot\Bigg[2\cdot\frac{(n+1)^\alpha}{(n+2)^\alpha} \\
& & \quad{}-(2\cdot n+3)\cdot \frac{{(n+2)}^\alpha-{(n+1)}^\alpha}{{(n+2)}^\alpha}\Bigg] \\
&\le& \mbox{\S~By Lagrange's Mean-Value Theorem~\S} \\
& &
\mathbf{1}_{X_n>0}\cdot\Bigg[2\cdot\frac{(n+1)^\alpha}{(n+2)^\alpha} \\
& & \quad{}-(2\cdot n+3)\cdot \frac{\alpha\cdot {(n+1)}^{\alpha-1}}{{(n+2)}^\alpha}\Bigg] \\
&\le&
\mathbf{1}_{X_n>0}\cdot\left[2\cdot\frac{(n+1)^\alpha}{(n+2)^\alpha}-2\cdot (n+1)\cdot \frac{\alpha\cdot {(n+1)}^{\alpha-1}}{{(n+2)}^\alpha}\right] \\
&=& \mathbf{1}_{X_n>0}\cdot\left(2-2\cdot\alpha\right)\cdot \frac{(n+1)^\alpha}{(n+2)^\alpha} \\
&\le & -(2\cdot\alpha - 2)\cdot {\left(\frac{1}{2}\right)}^\alpha\cdot \mathbf{1}_{X_n>0} \enskip.
\end{eqnarray*}
Hence $\{X_n\}_{n\in\Nset_0}$ is a ranking supermartingale and for all $n$ and $\omega$, $X_n(\omega)=0$ implies $X_{n+1}(\omega)=0$.
However, since $\stopping{\Gamma}(\omega)=n$ once $Y_n(\omega)=-2\cdot n-1$,
one can calculate exactly that for $n\ge 1$,
$\probm(\stopping{\Gamma}> n)=\prod_{k=1}^n \frac{k^\alpha}{{(k+1)}^\alpha}=\frac{1}{{(n+1)}^\alpha}$\enskip.
Hence, $\{X_n\}_{n\in\Nset_0}$ does not admit exponential decrease of tail probabilities.
\hfill\IEEEQEDclosed

\noindent\textbf{Example~\ref{ex:special:supmoptimal}.}
Consider the family $\{Y_n\}_{n\in\Nset_0}$ of independent random variables defined as follows:
$Y_0:=1$ and each $Y_n$ ($n\ge 1$) satisfies that
$\probm\left(Y_n=1\right)=\frac{1}{2}$ and $\probm\left(Y_n=-1\right)=\frac{1}{2}$.
Let the stochastic process $\Gamma=\{X_n\}_{n\in\Nset_0}$ be inductively defined by: $X_0:=Y_0$ and for all $n\in\Nset_0$, $X_{n+1}:=\mathbf{1}_{X_n>0}\cdot\left(X_n+Y_{n+1}\right)$~.
Choose the filtration $\{\mathcal{F}_n\}_{n\in\Nset_0}$ such that every $\mathcal{F}_n$ is the smallest $\sigma$-algebra that makes $Y_0,\dots, Y_n$ measurable.
Then $\Gamma$ models the classical symmetric random walk.
Since a.s.
\begin{eqnarray*}
 \condexpv{X_{n+1}-X_n}{\mathcal{F}_n} &=&\condexpv{\mathbf{1}_{X_n>0}\cdot Y_{n+1}}{\mathcal{F}_n} \\
&=& \mbox{\S~By (E8), (E9) \S} \\
& & \mathbf{1}_{X_n>0}\cdot\expv\left(Y_{n+1}\right)\\
&=& 0 \enskip,
\end{eqnarray*}
$\Gamma$ is a difference-bounded martingale. Moreover, $X_n>0$ implies $\condexpv{|X_{n+1}-X_n|}{\mathcal{F}_n}=1$ a.s.:
\begin{eqnarray*}
\condexpv{\left|X_{n+1}-X_n\right|}{\mathcal{F}_n} & = & \condexpv{\mathbf{1}_{X_n>0}\cdot |Y_{n+1}|}{\mathcal{F}_n} \\
& = & \mbox{\S~By (E8) \S} \\
& & \mathbf{1}_{X_n>0}\cdot\condexpv{|Y_{n+1}|}{\mathcal{F}_n} \\
& = & \mbox{\S~By (E9) \S} \\
& &
\mathbf{1}_{X_n>0}\enskip.
\end{eqnarray*}
From Theorem~\ref{thm:supm}, one obtains that $\probm(\stopping{\Gamma}<\infty)=1$ and $k\mapsto\probm\left(\stopping{\Gamma}\ge k\right)\in \mathcal{O}\left(\frac{1}{\sqrt{k}}\right)$.
In~\cite[Theorem 4.1]{DBLP:journals/jcss/BrazdilKKV15}, it is shown (in the context of pBPA) that $k\mapsto\probm\left(\stopping{\Gamma}\ge k\right)\in \Omega\left(\frac{1}{\sqrt{k}}\right)$.
Hence the tail bound in Theorem~\ref{thm:supm} is optimal. \hfill\IEEEQEDclosed

\noindent\textbf{\bf Example~\ref{ex:special:positivity}.}
Consider the discrete-time stochastic process $\{X_n\}_{n\in\Nset_0}$ such that all $X_n$'s are independent, $X_0=1$ and every $X_n$ ($n\ge 1$) observes the two-point distribution such that $\probm\left(X_n=2^{-n+1}\right)=\probm\left(X_n=-2^{-n+1}\right)=\frac{1}{2}$.
Choose the filtration $\{\mathcal{F}_n\}_{n\in\Nset_0}$ such that every $\mathcal{F}_n$ is the smallest $\sigma$-algebra that makes $X_0,\dots, X_n$ measurable.
Let the stochastic process $\Gamma=\{Y_n\}_{n\in\Nset_0}$ be inductively defined by: $Y_0:=X_0$ and for all $n\in\Nset_0$,
$Y_{n+1}:=\mathbf{1}_{Y_n>0}\cdot\left(Y_n+X_{n+1}\right)$.
Since
\begin{eqnarray*}
 \condexpv{Y_{n+1}-Y_n}{\mathcal{F}_n} &=&\condexpv{\mathbf{1}_{Y_n>0}\cdot X_{n+1}}{\mathcal{F}_n} \\
&=& \mbox{\S~By (E8), (E9) \S} \\
& & \mathbf{1}_{Y_n>0}\cdot\expv\left(X_{n+1}\right)\\
&=& 0 \enskip,
\end{eqnarray*}
$\Gamma$ is a difference-bounded martingale. Moreover, it holds a.s. that
\begin{eqnarray*}
 \condexpv{\left|Y_{n+1}-Y_n\right|}{\mathcal{F}_n} &=&\condexpv{\mathbf{1}_{Y_n>0}\cdot|X_{n+1}|}{\mathcal{F}_n} \\
&=& \mbox{\S~By (E8), (E9) \S} \\
& & \mathbf{1}_{Y_n>0}\cdot\expv\left(|X_{n+1}|\right)\\
&=& 2^{-n}\cdot \mathbf{1}_{Y_n>0} \enskip.
\end{eqnarray*}
However, $\probm\left(\stopping{\Gamma}=\infty\right)=\frac{1}{2}$ as whether $\stopping{\Gamma}=\infty$ or not relies only on $X_1$.\hfill\IEEEQEDclosed

\end{document}